\documentclass[aps,prc,amsmath,superscriptaddress,showpacs,twocolumn,captions=nooneline]{revtex4-1} 
\usepackage{graphicx,amsmath,amssymb,bm,color,inputenc,fontenc,mathrsfs}
\usepackage{epstopdf}
\usepackage{float}
\usepackage[caption = false]{subfig}
\usepackage{comment}
\usepackage{bm} 
\usepackage[dvipsnames,usenames]{xcolor}

\begin{document}

\title[title]{Exact solution of multi-angle quantum many-body collective neutrino flavor oscillations}
\author{Ermal Rrapaj}
 \email{ermalrrapaj@gmail.com}
\affiliation{Department of Physics, University of California, Berkeley, CA 94720, USA}
 \affiliation{School of Physics and Astronomy, University of Minnesota, Minneapolis, MN 55455, USA}

\begin{abstract}
 I study the flavor evolution of a dense neutrino gas by considering vacuum contributions, matter effects and neutrino self-interactions. Assuming a system of two flavors in
  a uniform matter background, the time evolution of the many-body system in discretized momentum space is computed.
  The multi-angle neutrino-neutrino interactions
 are treated exactly and compared to both the single-angle approximation and mean field calculations. 
 The mono-energetic two neutrino beam scenario is solved analytically. 
 I proceed to solve flavor oscillations for mono-energetic cubic lattices and quadratic lattices of two energy levels. In addition I study various configurations of twelve, sixteen, and twenty neutrinos.
 I find that when all neutrinos are initially of the same flavor, all methods agree. When both flavors are present, I find collective oscillations and flavor equilibration develop in the many body treatment but 
 not in the mean field method. This difference persists in dense matter with tiny mixing angle and it can be ascribed to non-negligible flavor polarization correlations being present. 
 Entanglement entropy is significant in all such cases. The relevance for supernovae or neutron stars mergers is contingent upon the value of the normalization
 volume $V$ and the large $N$ dependence of the timescale associated with oscillations.
 In future work, I intend study this dependence using larger lattices and also include anti-neutrinos.
\end{abstract}
\maketitle

\section{Introduction}
Neutrinos are some of the most abundant particles found in nature, produced during the early universe \cite{Kostelecky:1993, Abazajian:2002, Ho:2005, Steigman:2012, Follin:2015}, 
from stars like the sun during their lifetime~\cite{Davis:1968, Bahcall:1989},
and in copious amounts during core collapse supernovae~\cite{Hirata:1987, Bionta:1987, Alekseev:1987, Qian:1994, Qian:1995, Pastor:2002, Balantekin:2005, Fuller:2006}. 
From solar, atmospheric, and reactor experiments it is known that they are massive, 
and the mass eigenstates are different from the weak interaction eigenstates~\cite{Fukuda:1998mi, Kouchner:2014, Kudenko:2017}. As a result, 
neutrinos oscillate between their three flavors during their propagation in vacuum~\cite{Gava:2008}.
For instance, the solar neutrino problem, the discrepancy between the electron neutrino flux emitted from the sun and observed on earth, was a long standing problem~\cite{Bowles:1993,Haxton:1995}.
The resolution came from the modified dispersion relation in matter which induces resonant flavor conversions commonly referred to as the Mikheyev-Smirnov-Wolfenstein (MSW) 
effect~\cite{Wolfestein:1978,Mikheev:1987,Bellerive:2016}. Neutrino--matter interactions induce a self-energy which is the origin of this effect. 

In this work I study flavor oscillations under the influence of the self-energy induced by neutrino-neutrino interactions, called collective oscillations.
This type of oscillations has been studied extensively in the literature \cite{Duan:2007,Dasgupta:2008cd,Gava:2009,Duan:2009,Balantekin:2009dy,
Dasgupta:2010cd,Duan:2010,Raffelt:2010,Raffelt:2011,Duan:2011,Cherry:2012zw}.
In core-collapse supernovae and merging of neutron stars a very large number of neutrinos 
are produced~\cite{Mirizzi:2015eza}. In such situations even the average energies of different flavors are different since only electron neutrinos participate in charged-current weak interactions.
Neutrino-neutrino scattering, being between particles of the same type, is of a different nature than neutrino-matter scattering, and gives rise to forward scattering terms in the many-body Hamiltonian which 
contribute to oscillations \cite{Pantaleone:1992, Pantaleone:1992-2}. These terms are dependent only on the angle between neutrinos and couple neutrinos of different energies making flavor
evolution a rather intricate many-body problem.
Neutrino transport and flavor oscillations are active areas of research, and no exact treatment has yet been provided. Due to the amount of energy neutrinos carry during a core collapse supernova, 
they can have a significant impact on the explosion. For instance, the kinetic energy of the material ejected is only about 1\% of the neutrino energy~\cite{Hix:2014,Janka:2017vlw}. If the majority of neutrinos are of electron flavor, they can be
re-absorbed by matter and provide the energy required for the explosion. Knowing the flavor evolution is crucial in understanding the role neutrinos play in this environment. 

The mean field method was first proposed in \cite{Pantaleone:1992, Pantaleone:1992-2} and
has been widely applied subsequently. The time evolution equations in this method become nonlinear 
and analytical solutions are available only in special cases. However, the method needs to be compared to exact solutions to test its region of validity.

 Seminal work on the role of 
neutrino-neutrino interactions in flavor oscillations has been conducted in ~\cite{Friedland:2003, Bell:2003, Friedland:2003eh}. 
However, at the time, only a part of the full Hamiltonian was considered, that is a simplified version of neutrino-neutrino interactions, which here I call single-angle approximation,
without the presence of the mass term which causes flavor mixing. 

Later an algebraic approach was developed which retains the many-body nature of the Hamiltonian and where two body operators are not replaced by effective one body counterparts \cite{Balantekin:2006}. In this approach, the many-body system is formulated as an SU(2)
coherent state path integral of two flavors (SU(3) for three flavors). This approach, as demonstrated in this article, is particularly useful for finding exact solutions and constants of motion.
An alternative approximation to the mean field is to replace the angular dependence in neutrino-neutrino interactions by the average angle but keep the two body term; this is commonly referred to as the 
single-angle approximation. This simplified Hamiltonian can be studied within the Richardson-Gaudin framework~\cite{Richardson:1966,Gaudin:1976} and one can solve for the eigenvalues~\cite{Cervia:2019nzy,Patwardhan:2019zta}.
Using the single-angle approximation, one can calculate the spectral split for a large number of neutrinos~\cite{Birol:2018}.
However, there is no \textit{a priori} physical motivation for replacing the angular dependence of neutrino-neutrino interactions by an average value. This averaging replaces the complicated angular structure in the 
two-body neutrino interactions with the squared total angular momentum (in flavor space) of the system, which is not present in the original Hamiltonian. This is the Casimir operator~\cite{Georgi:1982} of the SU(2) algebra and commutes 
with the SU(2) generators, thus it introduces symmetries not present in the original Hamiltonian. 
From this perspective, the single-angle approximation also needs to be tested for validity.

In this work, I solve the many-body multi-angle Hamiltonian and compare to the single-angle approximation and the multi-angle mean field method.
As a first step, I assume a simplified scenario of electron neutrinos $\nu_e$, and an additional flavor which can be considered as a superposition of tau and muon 
neutrinos denoted by $\nu_x$, in matter at constant density. In addition, I discretize the momentum space, which allows for the treatment of the many-body problem on the lattice.
This appears to be one of the first attempts at an exact treatment of neutrino-neutrino interactions on the lattice. The role of anti-neutrinos is postponed for future works.
In addition, time evolution for large lattices is computationally expensive for classical computers as the degrees of freedom increase exponentially with the number of lattice points.
Hence, I focus on smaller lattices.
The Hamiltonian for two neutrino flavors is an ideal candidate for quantum computing given its resemblance to the Ising model~\cite{Lidar:1997}. This technology allows one to 
overcome the limitations of classical computers and explore much larger lattices than currently available. I plan to explore this research direction in future work. 

Section \ref{sec:alg} describes the operator algebra required for a treatment of neutrino oscillations in a discretized momentum lattice. Then, section \ref{sec:latt} studies
the equations of motion in all the three methods outlined here. Throughout this work, the time unit chosen for plotting results is $\triangle t = \mu_0^{-1}=(\frac{G_F}{2\sqrt{2}V})^{-1}$.
In section \ref{sec:twobeam} a system of two neutrino beams is analyzed, and in sections
\ref{sec:cubic} and \ref{sec:quadratic} cubic and two energy level quadratic systems are studied respectively. In section~\ref{sec:12} a system of twelve neutrinos is studied in vacuum and dense matter.
In section~\ref{sec:16} two different configurations of sixteen neutrinos in dense matter are studied. I conclude the analysis with a system of twenty neutrinos in section~\ref{sec:20}.
Section~\ref{sec:ee} studies the entropy of entanglement of each neutrino in the 
many-body wave-function.
The findings are summarized and conclusions are drawn in section~\ref{sec:conc}.

\section{two flavor oscillations as a lattice of SU(2) algebras}
\label{sec:alg}
By introducing creation and annihilation operators for one neutrino 
with three momentum $\bold{p}$, the generators of an SU(2) algebra can be written~\cite{Balantekin:2006, Pehlivan:2011, Balantekin:2018}: 
\begin{eqnarray}
J^+(\bold{p}) &=& a_e^\dagger(\bold{p}) a_x(\bold{p}), \> \> \>
J^-(\bold{p})=a_x^\dagger(\bold{p}) a_e(\bold{p}), \nonumber \\
J^z(\bold{p}) &=& \frac{1}{2}\left(a_e^\dagger(\bold{p})a_e(\bold{p})-a_x^\dagger(\bold{p})a_x(\bold{p})
\right). \label{su2}
\end{eqnarray}
where $a_e^\dagger(\bold{p})$ is the creation operator of an electron flavor neutrino and $a_x^\dagger(\bold{p})$ of a neutrino of flavor x, which can be thought of as a superposition of 
muon and tau neutrinos.
Indeed, there is an SU(2) algebra associated with each momentum value, all commuting with each other. 
To be mathematically rigorous, I discretize the momenta using a box quantization so that I get an SU(2) algebra instead of the usual current algebra.  
The sum of these operators over all possible values of momenta also generates a global SU(2) algebra. 
The Hamiltonian is comprised of three contributions: the mass (vacuum) term, matter interactions, and neutrino self-interactions. 
I start with the first two terms, for which one needs to pick either the mass or the flavor basis to express them. I opted to work in the flavor basis,  
\begin{equation}
\begin{split}
 H_{\nu} =&  \sum_{\bold{p}}  \> \omega_p  \bold{B} \cdot \bold{J}(\bold{p}) +  \sqrt{2} G_F \sum_{\bold{p}} 
\> N_e(\bold{p}) \>  J^z(\bold{p})\\
\label{msw}
\end{split}
\end{equation}
where $\bold{B}= \sin\left({2\theta}\right) \hat{x} - \cos \left({2\theta}\right) \hat{z}$, $\omega_p=\delta m^2/(2p)$, and $\sin \theta=0.297$ is the mixing angle between the mass 
basis and the flavor basis in vacuum~\cite{Tanabashi:2018}.
In Eq.~(\ref{msw}), the first term represents neutrino mixing and the second one 
represents neutrino forward scattering off the background matter.
The electron density $N_e$ in the second term of Eq.~(\ref{msw}) is inside the summation since electron 
densities encountered by neutrinos traveling in different directions can be different, but for simplicity from this point forward it is assumed to be constant. While this assumption is obviously 
unacceptable over large distances, here I focus on oscillations that happen over very short ranges for which the approximation should hold.
Neutrino-neutrino forward-scattering contributions, which are the basis of collective oscillations, are described by the Hamiltonian 
\begin{equation}
\begin{split}
H_{\nu \nu} =& \sqrt{2} \frac{G_F}{V} \sum_{\bold{p}\bold{q}}  \>  (1-\cos\vartheta_{\bold{p}\bold{q}}) \> \bold{J}(\bold{p}) \cdot \bold{J}(\bold{q})\\
\end{split}
\label{nunu}
\end{equation}
where $\vartheta_{\bold{p}\bold{q}}$ is the angle between neutrino momenta $\bold{p}$ and $\bold{q}$ and $V$ 
is the normalization volume.  The $(1-\cos\vartheta_{\bold{p}\bold{q}})$ term in the integral above 
means that neutrinos traveling in the same direction do not forward scatter off each other. Contributions from collisions are proportional to $G_F^2$ and are sub-leading with respect to forward-scattering.
The total Hamiltonian is $H = H_{\nu} + H_{\nu \nu}$. In writing $H_{\nu \nu}$ neutrino masses were set to zero.

\section{Many-body Hamiltonian in momentum lattice}
\label{sec:latt}
For practical computations, the SU(2) algebra operators are represented by Pauli matrices, $\bold{J}_{\bold{p}} \equiv \frac{1}{2} \bold{\sigma}_{\bold{p}}$.
For notational convenience, $\alpha_{\text{vac}}=2\theta$. Matter effects are equivalent to a modified neutrino oscillation frequency and mixing angle,
\begin{equation}
\begin{split}
&\frac{\sin(\alpha_{\text{vac}})}{\sin \left(\alpha_{p} \right)} = \frac{\overline{\omega}_p}{\omega_p}\\
&=  \ \sqrt{\sin^2(\alpha_{\text{vac}})+\left(\cos (\alpha_{\text{vac}})  - \frac{\sqrt{2} G_FN_e}{\omega_p}\right)^2}.\\
\end{split}
\label{eq:matter_effect}
\end{equation}
The Hamiltonian described in the previous section can be written as a sum of operators at different lattice points.
In the Heisenberg picture, operators evolve in time due to their commutation relation with the Hamiltonian,
\begin{equation*}
  d_t O = i[H,O]
\end{equation*}
I start with the global SU(2) generators $J^{k}=\sum_{\bold{p}}J^{k}_{\bold{p}}$,
\begin{equation}
 \begin{split}
  d_t \bold{J} =  \sum_{\bold{p}} \overline{\omega}_{p} \left(\bold{\overline{B}}_p \times \bold{J}_{\bold{p}}\right),
 \end{split}
 \label{eq:dtJk}
\end{equation}
where $\bold{\overline{B}}_p=\sin\left(\alpha_p\right) \hat{x} - \cos \left(\alpha_p\right) \hat{z}$ denotes the one body contribution with a mixing angle modified in the presence of matter.
Only the mass term is responsible for flavor oscillations in either basis, as the neutrino self interaction term  commutes with $J^k$. For a complete analysis the reader is referred to~\cite{Pehlivan:2011, Balantekin:2018}.
One can also explore what happens to a particular lattice point (direction),
\begin{equation}
 \begin{split}
   d_t \bold{J}_{\bold{p}}
 =&  \overline{\omega}_{p}  \left(\bold{\overline{B}_p} \times \bold{J}_{\bold{p}}\right) 
 + \mu \sum_{\bold{q}} (1-\cos\vartheta_{\bold{p},\bold{q}}) \left(\bold{J}_{\bold{q}}\times \bold{J}_{\bold{p}}\right).\\
 \end{split}
 \label{eq:dtjkp}
\end{equation}
As I explore in this work, they lead to fast collective oscillations which can be observed by looking at oscillations of individual lattice points.
Equations~(\ref{eq:dtJk}) and~(\ref{eq:dtjkp}) are derived in appendix~\ref{ap:dtjk}.
As this is one of the first lattice treatments of neutrino oscillations it is important to compare to the approximations commonly used in literature. 
\subsection*{Mean field approximation}
Neutrino flavor oscillations are commonly studied in the mean field approximation, 
in which products of two one body operators are approximated by the product of a one body operator and the expectation of the other.
While this approximation greatly simplifies calculations, its validity is questionable.
The underlying assumption upon which it is based is that two or higher many particle correlations are too small to have a significant impact. 
In this work I find that when this assumption fails, also the mean field method fails to provide the correct flavor evolution in time.
For the moment, let us describe the method. I will return upon this issue later in this article.
For the Hamiltonian under consideration, the mean field treatment is given as follows,
\begin{equation*}
 \begin{split}
  H_{\text{MF}} 
  =& \sum_{\bold{p}\bold{q}}  \frac{\overline{\omega}_p}{N_{\nu}} \left[  \sin\left( {\alpha}_p \right) J^x_{\bold{p}}  I_{\bold{q}} 
  -\cos \left( {\alpha}_p \right) J^z_{\bold{p}} I_{\bold{q}} \right] \\
  +& \sum_{\bold{p}\bold{q}} \mu (1-\cos \vartheta_{\bold{p}\bold{q}}) \left( \bold{P}_{\bold{p}} \cdot \bold{J}_{\bold{q}} \right), \\
 \end{split}
\end{equation*}
where $ \bold{P}_{\bold{p}} =\langle  \bold{J}_{\bold{p}}  \rangle$.
The time evolution of the SU(2) algebra generators is,
\begin{equation}
 \begin{split}
   d_t \bold{J}_p =& \overline{\omega}_p \bold{\overline{B}}_p \times \bold{J}_{\bold{p}}\ +  \mu \sum_{\bold{q}} (1-\cos \vartheta_{\bold{p}\bold{q}}) \bold{P}_{\bold{q}} \times \bold{J}_{\bold{p}}.
 \end{split}
 \label{eq:mf1}
\end{equation}
As the expectation value of the polarization $\bold{P}_{\bold{q}}$ also evolves in time, an additional equation of motion is needed for completeness and consistency,
\begin{equation}
 d_t \bold{P}_{\bold{p}}= \overline{\omega}_p \bold{\overline{B}}_p \times \bold{P}_{\bold{p}}\ +  \mu \sum_{\bold{q}} (1-\cos \vartheta_{\bold{p}\bold{q}}) \bold{P}_{\bold{q}} \times \bold{P}_{\bold{p}}.
 \label{eq:mf2}
\end{equation}
By solving this equation one studies neutrino oscillations in the mean field approximation.
As I show in the next sections, the underlying assumption for this approximation is valid when only one flavor is initially present. I find that when two flavors are initially present, fast 
collective oscillations are obtained in the exact treatment but do not appear in the mean field method. In the recent literature, linear instability analysis has been employed in collective flavor oscillations and fast conversions have been 
found~\cite{Banerjee:2011, Sarikas:2012, Saviano:2012, Mirizzi:2012, Izaguirre:2017, Azari:2019jvr}. For these fast conversions to develop, among many other criteria, anti-neutrinos must be present.
An additional criterion for the two beam scenario is that the angle between the two neutrino modes must be acute~\cite{Dasgupta:2017oko}.
In my analysis fast oscillations develop without the need for anti-neutrinos (for all lattices considered here) and regardless of the angle in the two beam case, which is quite an interesting difference.
I also compare to the single-angle approximation on the lattice. For the mean field method I perform a multi-angle treatment.

\section{Analytical solution for two neutrino beams}
\label{sec:twobeam}
If two neutrino beams with the same energy but different directions are considered, one can solve the system exactly by making use of the Dirac basis~\cite{garfken67:math}
\begin{equation*}
 \begin{split}
 \sigma_{\bold{p}\bold{q}}^{\mu \nu}= \sigma_{\bold{p}}^{\mu} \otimes \sigma_{\bold{q}}^{\nu},\
 \sigma^{\mu}=\{I_2,\sigma^x, \sigma^y, \sigma^z \}.
 \end{split}
\end{equation*}
The subscripts denote the lattice points associated with the Pauli matrices.
Any four dimensional matrix $M$ can be decomposed in this orthogonal basis as follows,
\begin{equation*}
\begin{split}
 M =& \sum_{\mu \nu}c^{\mu \nu} \sigma^{\mu \nu}, \ c^{\mu \nu}= \frac{1}{4}\text{Tr}[M \sigma^{\mu \nu}].
\end{split}
\end{equation*}
The Hamiltonian for this system is
\begin{equation*}
 \begin{split}
  H=&H_{\bold{1}\bold{1}}+H_{\bold{2}\bold{1}}+H_{\bold{1}\bold{2}}+H_{\bold{2}\bold{2}}\\
  \end{split}
  \end{equation*}
 And each term is given in terms of the elements of the  Dirac basis, 
  \begin{equation*}
  \begin{split}
  H_{\bold{p}\bold{q}}=& \frac{\overline{\omega}}{2} \left[\sin(\alpha)\sigma_{\bold{p}\bold{q}}^{1 0} - \cos(\alpha) \sigma_{\bold{p}\bold{q}}^{30}\right] \\
  +& \frac{\mu}{4} (1-\cos \vartheta_{\bold{p},\bold{q}})
  \left[ \sigma_{\bold{p}\bold{q}}^{11}+ \sigma_{\bold{p}\bold{q}}^{22}+ \sigma_{\bold{p}\bold{q}}^{33}\right].\\
 \end{split}
\end{equation*}
Since both beams have the same energy, $\overline{\omega}_{\bold{1}}=\overline{\omega}_{\bold{2}}=\overline{\omega}$, $\alpha_1=\alpha_2=\alpha$, and $\bold{\overline{B}}_{\bold{1}}=\bold{\overline{B}}_{\bold{2}}=\bold{\overline{B}}$.
Any operator can be expressed as a function of time, $O(t)=e^{i H t}Oe^{-i H t}$.
In appendix~\ref{ap:twobeams}, the decomposition of the time evolution operator $e^{-i H t}$ in the Dirac basis is provided. 
Once the evolution operator is known, the  generators $(J^k=J^k_{\bold{1}}+J^k_{\bold{2}})$
of the SU(2) algebra can be calculated at any moment in time (appendix~\ref{ap:twobeams}),
\begin{equation*}
\begin{split}
J^x(t)=& \frac{1}{2} \left(\cos ^2(\frac{\overline{\omega}}{2} t)-\cos (2 \alpha ) \sin ^2(\frac{\overline{\omega}}{2} t)\right) (\sigma^{01}_{\bold{1}\bold{2}}+\sigma^{10}_{\bold{1}\bold{2}})\\
+& \frac{1}{2}\cos (\alpha ) \sin (\overline{\omega} t)  (\sigma^{02}_{\bold{1}\bold{2}}+\sigma^{20}_{\bold{1}\bold{2}})\\
-& \frac{1}{2} \sin (2 \alpha ) \sin ^2(\frac{\overline{\omega}}{2} t)(\sigma^{03}_{\bold{1}\bold{2}}+\sigma^{30}_{\bold{1}\bold{2}}), \\
J^z(t)=& \frac{1}{2} \sin (\alpha ) \sin (\overline{\omega} t)(\sigma^{02}_{\bold{1}\bold{2}}+\sigma^{20}_{\bold{1}\bold{2}})\\
-&\frac{1}{2} \sin (2 \alpha ) \sin ^2(\frac{\overline{\omega}}{2} t)(\sigma^{01}_{\bold{1}\bold{2}}+\sigma^{10}_{\bold{1}\bold{2}})\\
+&\frac{1}{2} \left(\cos (2 \alpha ) \sin ^2(\frac{\overline{\omega}}{2} t)+\cos ^2(\frac{\overline{\omega}}{2} t)\right)(\sigma^{03}_{\bold{1}\bold{2}}+\sigma^{30}_{\bold{1}\bold{2}}). \\
\end{split}
\end{equation*}
From the time dependence of these operators one can verify  that $\bold{\overline{B}} \cdot \bold{J}(t)$ is a constant of motion,
\begin{equation*}
 \begin{split}
  \bold{\overline{B}} \cdot \bold{J}(t)=& \sin \left(\alpha \right) J^x(t) -\cos \left(\alpha \right) J^z(t)\\
  =& \frac{1}{2} \sin \left(\alpha \right) \big( \sigma^{01}_{\bold{1}\bold{2}} + \sigma^{10}_{\bold{1}\bold{2}} \big)
  - \frac{1}{2} \cos \left(\alpha \right) \big( \sigma^{03}_{\bold{1}\bold{2}} + \sigma^{30}_{\bold{1}\bold{2}} \big).
 \end{split}
\end{equation*}
As collective oscillations appear by looking at individual points (directions) in the momenta lattice, I focus on point $\bold{1}$ 
and study the time evolution of the polarization for different initial conditions,
\begin{equation}
 \begin{split}
  &\langle {\nu_e}_{\bold{1}}, {\nu_e}_{\bold{2}} | {J_Z}_{\bold{1}} | {\nu_e}_{\bold{1}}, {\nu_e}_{\bold{2}} \rangle (t)=
  -\langle {\nu_x}_{\bold{1}}, {\nu_x}_{\bold{2}} | {J_Z}_{\bold{1}} | {\nu_x}_{\bold{1}}, {\nu_x}_{\bold{2}} \rangle (t)\\
  &=\frac{\cos (2 \alpha ) \sin ^2(\frac{\overline{\omega}}{2} t)+\cos ^2(\frac{\overline{\omega}}{2} t)}{2}.\\
 \end{split}
 \label{eq:eema}
\end{equation}
and,
\begin{equation}
 \begin{split}
 &\frac{\langle {\nu_e}_{\bold{1}}, {\nu_x}_{\bold{2}} | {J_Z}_{\bold{1}} | {\nu_e}_{\bold{1}}, {\nu_x}_{\bold{2}} \rangle (t)}
  {\langle {\nu_e}_{\bold{1}}, {\nu_e}_{\bold{2}} | {J_Z}_{\bold{1}} | {\nu_e}_{\bold{1}}, {\nu_e}_{\bold{2}} \rangle (t)}
  =-\frac{\langle {\nu_x}_{\bold{1}}, {\nu_e}_{\bold{2}} | {J_Z}_{\bold{1}} | {\nu_x}_{\bold{1}}, {\nu_e}_{\bold{2}} \rangle (t)}
  {\langle {\nu_e}_{\bold{1}}, {\nu_e}_{\bold{2}} | {J_Z}_{\bold{1}} | {\nu_e}_{\bold{1}}, {\nu_e}_{\bold{2}} \rangle (t)}\\
  &= \cos \left(2  \mu (1-\cos\vartheta_{\bold{1}\bold{2}}) t \right). \\
 \end{split}
 \label{eq:2nu}
\end{equation}
From these equations I find that the frequency of oscillations is proportional to $\overline{\omega}$  when only one flavor is present and 
proportional to $\mu$ when both flavors are present. Collective oscillations do not depend on 
the energy scale or the mixing angle, provided $\alpha \neq 0$. In addition, the angle between the two neutrinos does not prevent oscillations from developing 
as long as they are not parallel to each other. 
To understand the difference between these two initial conditions I study the flavor polarization correlations.
Here I define the correlation between two operators as 
$C(O_i,O_j, |\Psi \rangle)=\langle \Psi| O_i O_j| \Psi \rangle - \langle \Psi|O_i| \Psi \rangle \langle \Psi|O_j|\Psi \rangle$.

For the two beam scenario I find
\begin{equation}
 \begin{split}
  C(J^z_{\bold{1}},J^z_{\bold{2}}, |\nu_e \nu_e \rangle ) =& C(J^z_{\bold{1}},J^z_{\bold{2}}, |\nu_x \nu_x \rangle) = 0,\\
  \frac{C(J^z_{\bold{1}},J^z_{\bold{2}}, | \nu_e \nu_x \rangle)}{\langle \nu_e \nu_x | J^z_1 J^z_2| \nu_e \nu_x \rangle} 
  =& \frac{C(J^z_{\bold{1}},J^z_{\bold{2}}, | \nu_x \nu_e \rangle )}{\langle \nu_x \nu_e | J^z_1 J^z_2| \nu_x \nu_e \rangle} \\
  =& \sin^2(2\mu (1 - \cos \vartheta_{\bold{1}\bold{2}})t).
 \end{split}
\end{equation}
From these correlations I expect the mean field approximation to match the exact solution for initial conditions where only one flavor is present, and maximal disagreement 
when the net polarization is 0 (equal amounts of each flavor).

It is worth pointing out that when both neutrinos have the same initial flavor, the wavefunction is an eigenstate of the two body term where both $|\nu_e\nu_e\rangle$ and $|\nu_x\nu_x\rangle$ have the same eigenvalue. As a result, the time evolution is initially impacted by the one body term only. And since both neutrinos have the same frequency $\omega$, they will precess around $\bold{B}$ in a synchronized fashion.
Thus, they will be a superposition of pure flavor eigenstates $|\nu_e\nu_e\rangle$ and $|\nu_x\nu_x\rangle$, and the two body term will play no role. If the initial wavefunction has both flavors present, it is not an eigenstate of the two body term and the time evolution will depend on both terms in the Hamiltonian.
\subsection{Single-angle approximation}
The average angle for the two beam scenario is $\langle \cos \vartheta \rangle =(1+\cos \vartheta_{\bold{1}\bold{2}})/2$. 
The respective time evolution of the expectation value of ${J_Z}_{\bold{1}}$ is identical to the multi-angle results for initial conditions of only electron or x flavor.
When both flavors are present the results are different,
\begin{equation}
 \begin{split}
  \frac{\langle {\nu_e}_{\bold{1}}, {\nu_x}_{\bold{2}} | {J_Z}_{\bold{1}} | {\nu_e}_{\bold{1}}, {\nu_x}_{\bold{2}} \rangle (t)}
  {\langle {\nu_e}_{\bold{1}}, {\nu_e}_{\bold{2}} | {J_Z}_{\bold{1}} | {\nu_e}_{\bold{1}}, {\nu_e}_{\bold{2}} \rangle (t)}
  =& \cos \left(\mu (1-\cos\vartheta_{\bold{1}\bold{2}}) t \right).\\
 \end{split}
 \label{eq:eesa}
\end{equation}
Correlations vanish for a single flavor and differ for mixed flavor initial conditions,
\begin{equation*}
 \begin{split}
  \frac{C(J^z_{\bold{1}},J^z_{\bold{2}}, \nu_e \nu_x)}{\langle \nu_e \nu_x | J^z_1 J^z_2| \nu_e \nu_x \rangle} 
  =& \frac{C(J^z_{\bold{1}},J^z_{\bold{2}}, \nu_x \nu_e)}{\langle \nu_x \nu_e | J^z_1 J^z_2| \nu_x \nu_e \rangle}\\
  =& \sin^2(\mu (1-\cos \vartheta_{\bold{1}\bold{2}})t).
 \end{split}
\end{equation*}
The attentive reader might be puzzled as to why I am comparing multi and single-angle calculations
for a system of two particles where indeed there is only one angle present between them. As shown in appendix~\ref{ap:twobeams}, the single-angle approximation allows for neutrinos to interact with themselves,
differently from the original Hamiltonian. Even for a small system of two particles this causes a difference with exact results. 
This difference seems mainly quantitative rather than qualitative at this stage. 
However, as the number of particles increases, so does the number of angles between them, and one needs to compare with multi-angle results.

\subsection{Mean field approximation}
\label{subec:mf}
The total polarization and difference of polarizations are defined as
\begin{equation*}
 \begin{split}
  \bold{P}_{\text{tot}}  =& \bold{P}_{\bold{1}}+\bold{P}_{\bold{2}},\\
  \bold{P}_{\text{diff}} =& \bold{P}_{\bold{1}}-\bold{P}_{\bold{2}}.
 \end{split}
\end{equation*}
This simplifies calculations as equations of motion partially decouple,
\begin{equation}
 \begin{split}
  d_t \bold{P}_{\text{tot}} =& \overline{\omega} \bold{\overline{B}} \times \bold{P}_{\text{tot}},\\
  d_t \bold{P}_{\text{diff}} =& \left( \overline{\omega} \bold{\overline{B}} + \mu (1-\cos \vartheta_{\bold{p}\bold{q}}) \bold{P}_{\text{tot}} \right) \times \bold{P}_{\text{diff}}.
 \end{split} 
 \label{eq:ttotdiff}
\end{equation}
The first equation has the analytical solution,
\begin{equation*}
 \begin{split}
  \bold{P}_{\text{tot}}(t)=& \bold{\overline{B}} \times \bold{P}_{0} +  \sin \left(\overline{\omega } t \right) \left(\bold{\overline{B}}\cdot \bold{P}_{0}\right) \bold{\overline{B}}\\
  -& \cos \left(\overline{\omega } t \right) \bold{\overline{B}} \times (\bold{\overline{B}} \times \bold{P}_{0}).
 \end{split}
\end{equation*}
where $\bold{P}_0=\bold{P}_{\text{tot}}(t=0)$. More details can be found in appendix~\ref{ap:ptot}. The second equation is solved numerically for generic initial conditions.
However, here I am interested in two specific cases: two neutrino beams of one flavor ($|\nu_e\nu_e\rangle$) and one neutrino of each flavor ($|\nu_e \nu_x \rangle$).

\subsubsection*{Case: $|\nu_e\nu_e\rangle$}
Initially both neutrino beams are of electron flavor,
\begin{equation*}
 \begin{split}
  &\bold{P}_{\text{tot}}(t=0)=2\bold{P}_{\bold{1}}(t=0)=\hat{z},\\
  &\bold{P}_{\text{diff}}=\bold{0}.
 \end{split}
\end{equation*}
Since the time evolution equations are of first order, once one knows the initial conditions, the solutions can be found,
\begin{equation*}
 \begin{split}
 \bold{P}_{\text{diff}}=&\bold{0}, \\
 P^z_{\text{tot}}=& \left(\cos ^2(\alpha )+\sin ^2(\alpha ) \cos (\overline{\omega} t)\right).
 \end{split}
\end{equation*}
The expectation values for individual polarizations are,
\begin{equation}
 \begin{split}
  J^{z}{}_{\bold{1}}(t)= J^{z}{}_{\bold{2}}(t) =\frac{1}{2} \left(\cos ^2(\alpha )+\sin ^2(\alpha ) \cos (\overline{\omega} t)\right).\\
 \end{split}
 \label{eq:mf2p}
\end{equation}
By trigonometric identities one can prove that Eqs.~(\ref{eq:mf2p}) and~(\ref{eq:eema}) are identical. In this case all methods agree, there are no correlations present, and the frequency of oscillations depends on $\overline{\omega}$
but not on $\mu$.

\subsubsection*{Case: $|\nu_e\nu_x\rangle$}
If the initial wave-function has equal amounts of electron neutrinos $\nu_e$ and $x$ neutrinos $\nu_x$, 
\begin{equation*}
 \begin{split}
  &\bold{P}_{\text{tot}}(t=0)=\bold{0},\\
  &\bold{P}_{\text{diff}}=2\bold{P}_{\bold{1}}(t=0)=\hat{z}.
 \end{split}
\end{equation*}
The total polarization is a constant of motion and the difference of polarizations depends only on the mass term,
\begin{equation}
 \begin{split}
  &d_t \bold{P}_{\text{diff}} = \overline{\omega} \bold{\overline{B}} \times \bold{P}_{\text{diff}}. \\
 \end{split}
 \label{eq:s0}
\end{equation}
Equation~(\ref{eq:s0}) is identical to Eq.~(\ref{eq:ttotdiff}) in the previous section and, thus, has the same solution.
The expectation values for individual polarizations are,
\begin{equation}
 \begin{split}
  J^{z}{}_{\bold{1}}(t)=& -J^{z}{}_{\bold{2}}(t)= \frac{1}{2} \left(\cos ^2(\alpha )+\sin ^2(\alpha ) \cos (\overline{\omega} t)\right).
 \end{split}
 \label{eq:mf2pex}
\end{equation}
The time evolution for individual modes depends only on the mass term, in sharp contrast with the result from the exact solution.
For initial conditions without net polarization the mean field method shows no collective oscillations, which is quite interesting, 
as from the previous section, this is the scenario where correlations are present.

\subsection{Numerical results for $|\nu_e\nu_x\rangle$}
\begin{figure*}[ht]
 \subfloat[$ {\omega}/2 = \mu_0,\ \mu/4 = \mu_0, \ \cos \vartheta_{\bold{1}\bold{2}}= \frac{1}{2}$]{\includegraphics[height=0.2\linewidth,width=0.45\linewidth]{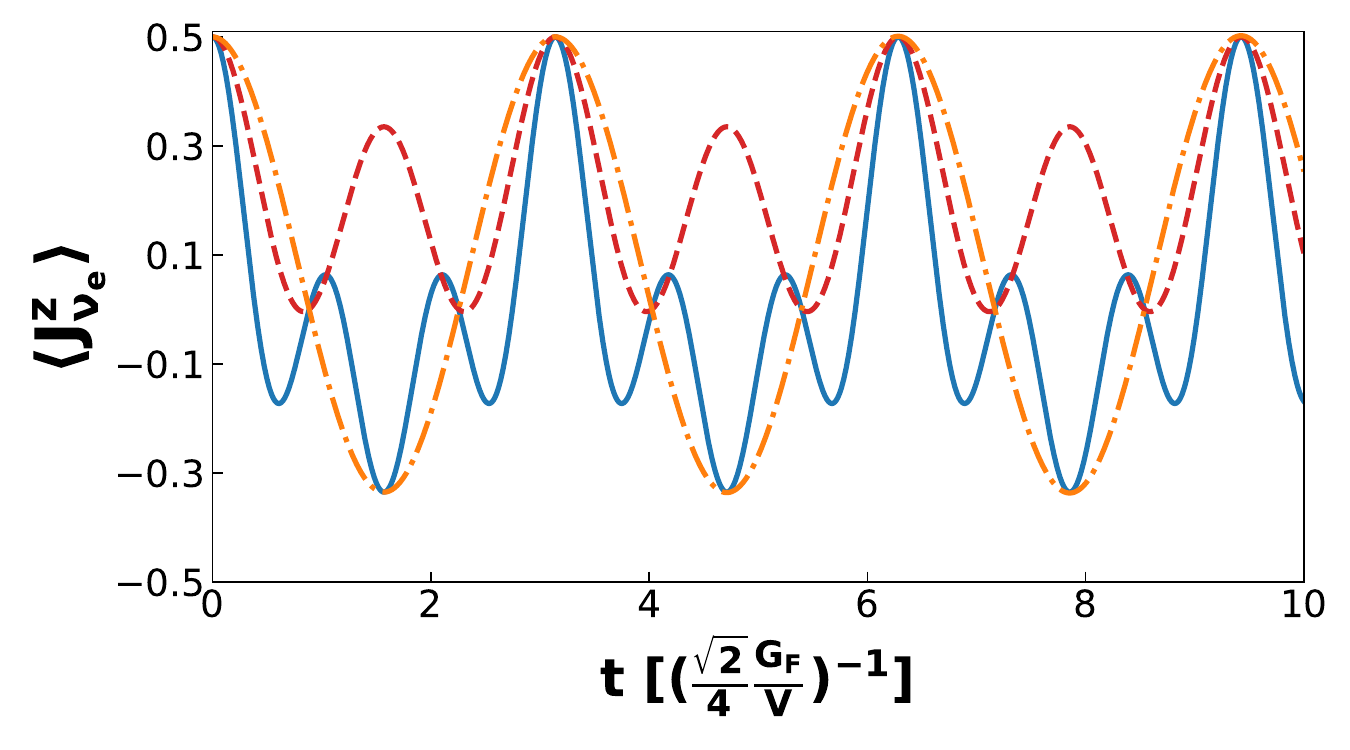}}
 \subfloat[$ {\omega}/2= \mu_0, \ \mu/4 =  \mu_0, \ \cos \vartheta_{\bold{1}\bold{2}}=- \frac{1}{2}$]{\includegraphics[height=0.2\linewidth,width=0.45\linewidth]{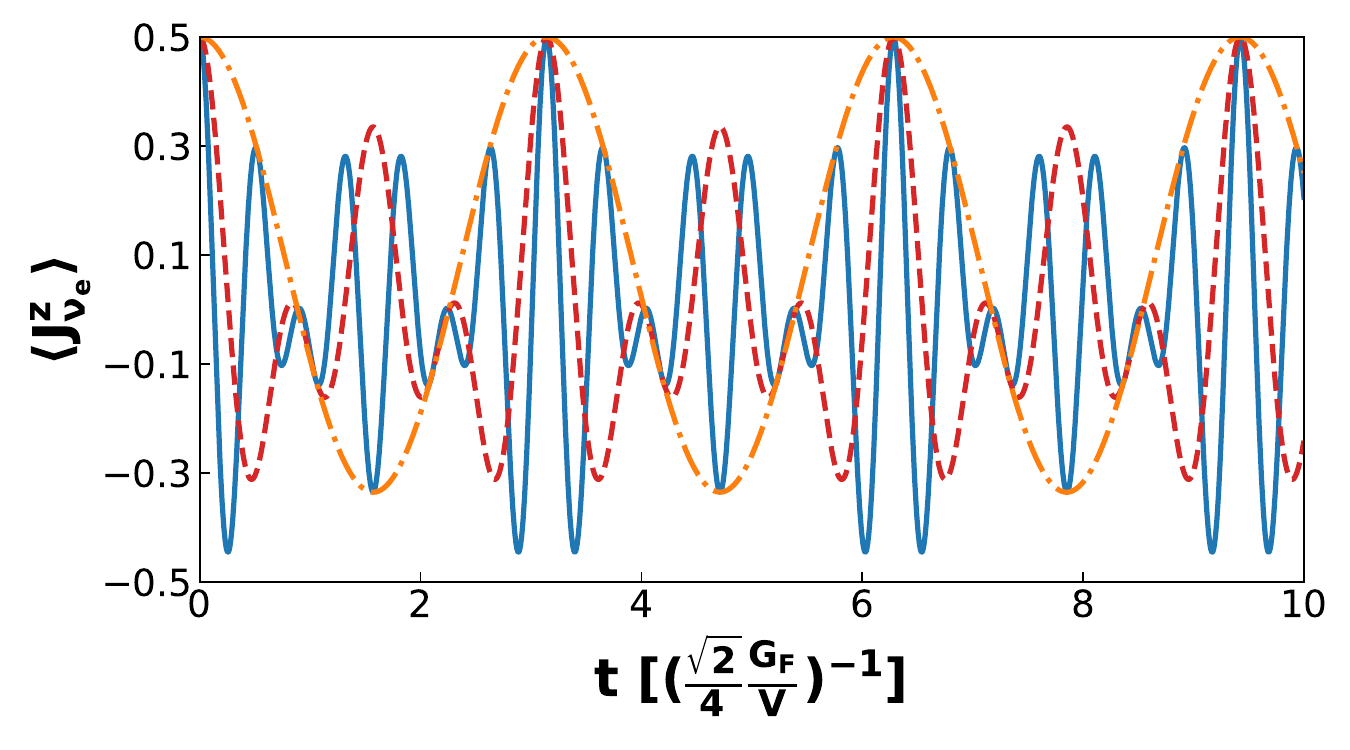}}\\
 \subfloat[$ {\omega}/2= \mu_0/5, \ \mu/4 = \mu_0, \ \cos \vartheta_{\bold{1}\bold{2}}= \frac{1}{2}$]{\includegraphics[height=0.2\linewidth,width=0.45\linewidth]{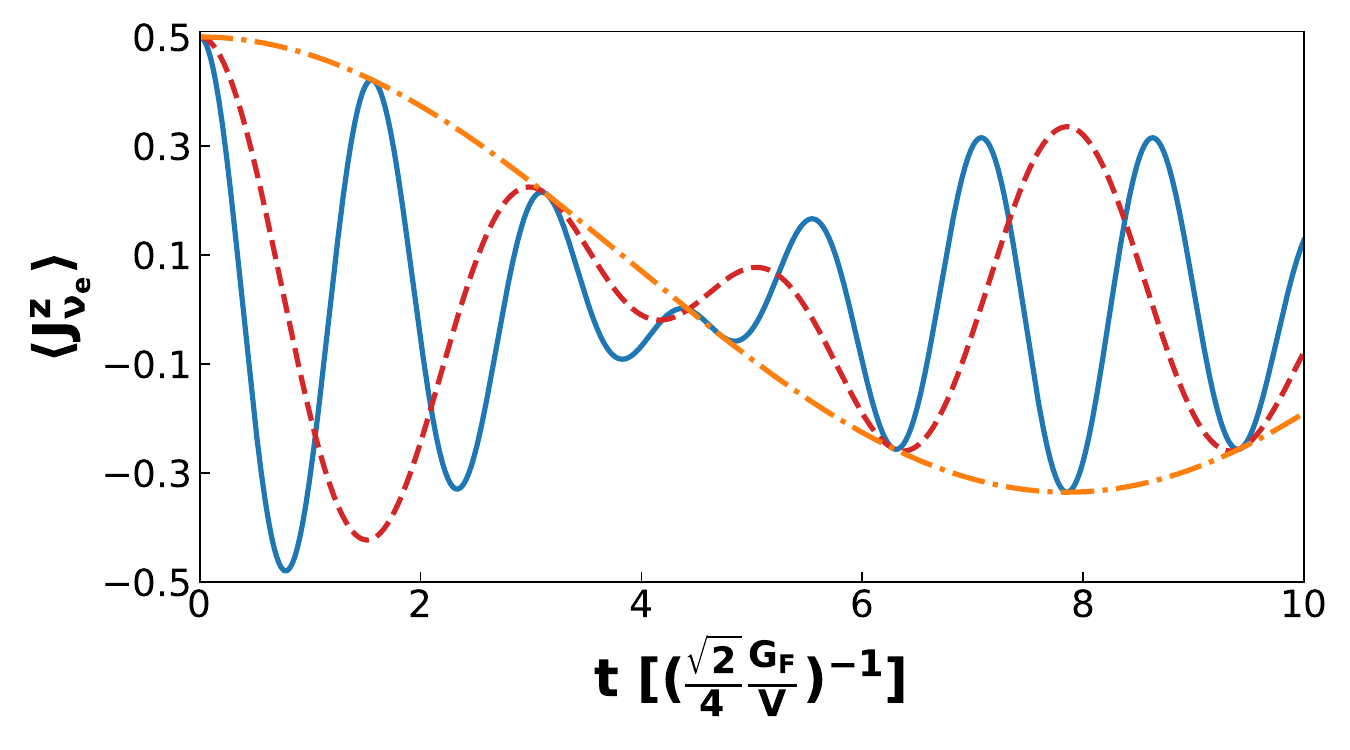}}
 \subfloat[$ {\omega}/2= \mu_0/5, \ \mu/4 = \mu_0, \cos \vartheta_{\bold{1}\bold{2}}=- \frac{1}{2}$]{\includegraphics[height=0.2\linewidth,width=0.45\linewidth]{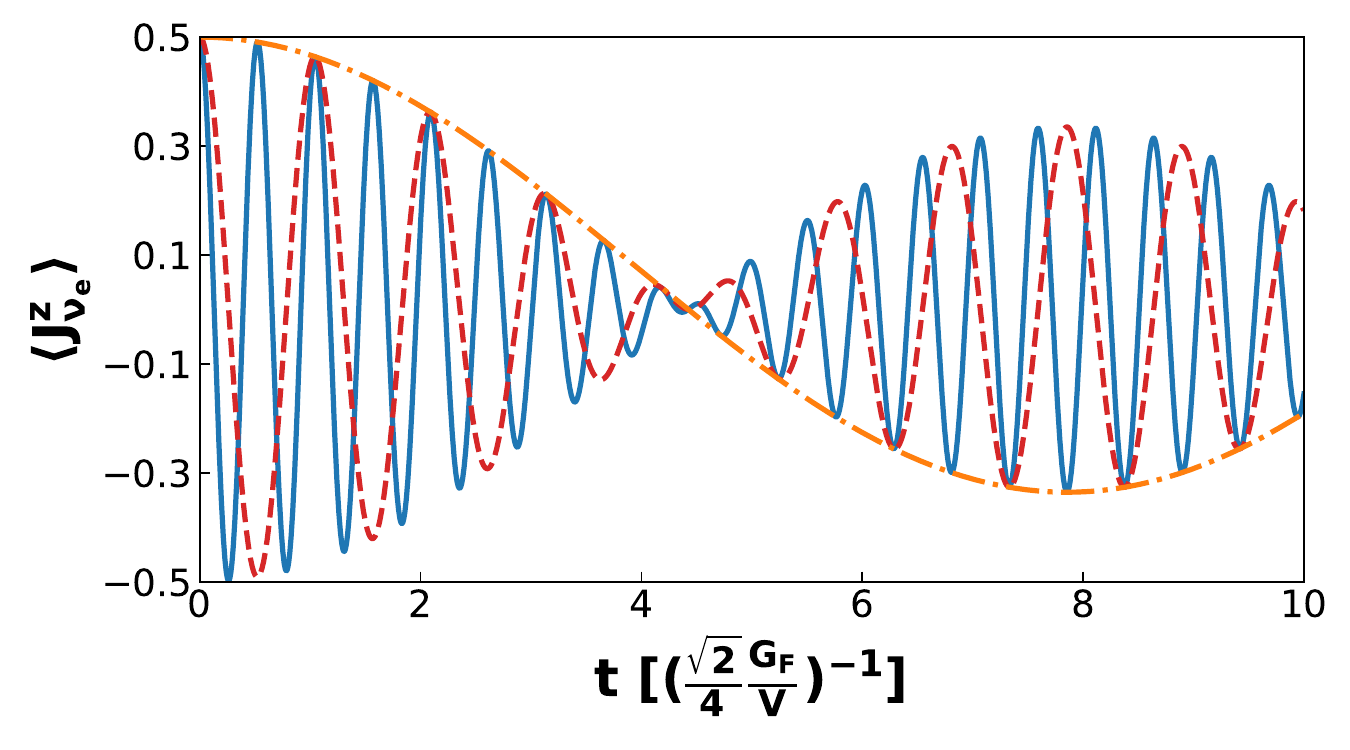}}\\
 \caption{Multi-angle (solid blue line), 
 single-angle (dashed red line), and mean field (dashed dotted orange line) calculations of the flavor polarization as function of time for two beam neutrinos 
 with   $\cos \vartheta_{\bold{1}\bold{2}}=\pm \frac{1}{2}$ and initial wave-function $|\Psi_0\rangle= | \nu_e \nu_x \rangle$ in vacuum from equations~(\ref{eq:2nu}),~(\ref{eq:eesa}) and~(\ref{eq:mf2pex}) respectively. }
 \label{fig:210}
\end{figure*}
To simplify the numerical implementation of the equations in the previous section, $\mu_0=\sqrt{2}\frac{G_F}{V}$ is set as the energy (inverse time) unit.
I focus on mode  $1$, and assume the initial wave-function is $|\Psi_0\rangle = | {\nu_e}_{\bold{1}} {\nu_x}_{\bold{2}} \rangle$.
Figure~\ref{fig:210} assumes a vacuum mixing angle between electrons and $x$.
By varying the two couplings and the angle between the two modes one can see that fast collective flavor oscillations develop in vacuum. As Eq.~(\ref{eq:2nu}) shows, the mass term $\omega$
is responsible for the small oscillation frequency, and the high frequency is linearly proportional to $\mu$.
In other words, these are frequency modulated oscillations where 
${\omega}_p$ plays the role of the original frequency and $\mu$ is the modulating frequency. Oscillations develop regardless of the sign of $\cos\left(\alpha\right)$.
However, the mean field calculation fails to find any collective oscillatory behavior and only displays the usual vacuum oscillation with frequency $\omega_p$.

 \begin{figure*}[ht]
 \subfloat[$\overline{\omega}/2= \mu_0, \ \mu/4 = \mu_0$]{\includegraphics[height=0.2\linewidth,width=0.45\linewidth]{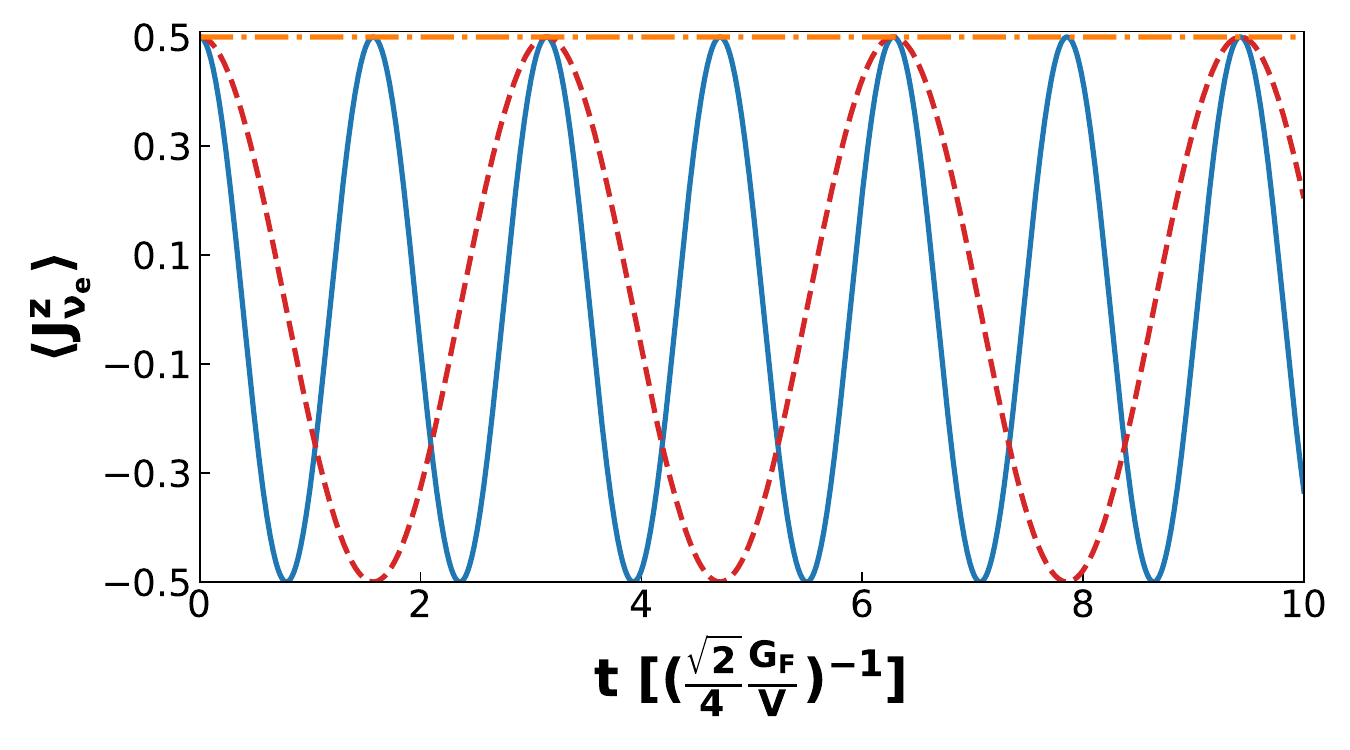}}
 \subfloat[$\overline{\omega}/2= \mu_0/5, \ \mu/4 = \mu_0$]{\includegraphics[height=0.2\linewidth,width=0.45\linewidth]{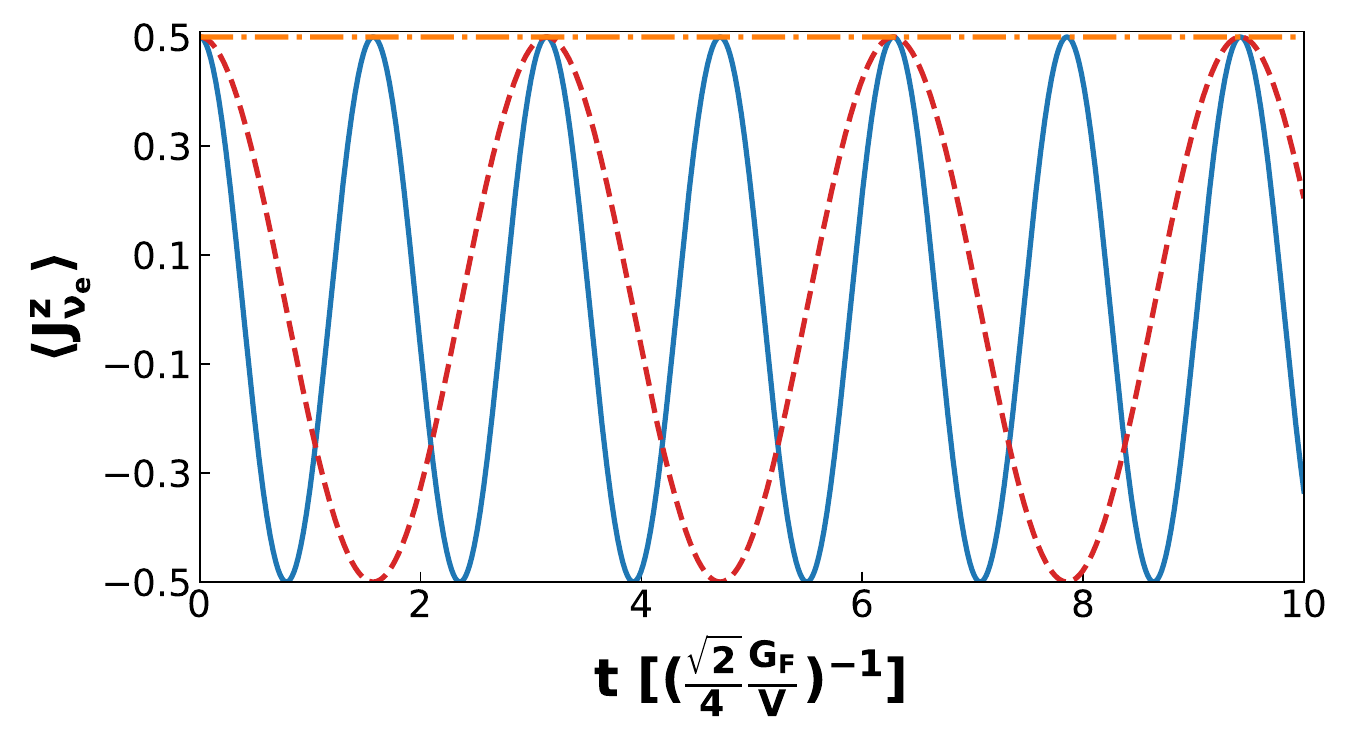}}\\
 \caption{Multi-angle (solid blue line), 
 single-angle (dashed red line), and mean field (dashed dotted orange line) calculations of the flavor polarization as function of time for two beam neutrinos 
 with  initial wave-function $|\Psi_0\rangle= | \nu_e \nu_x \rangle$ in medium where $\cos \vartheta_{\bold{1}\bold{2}}=\frac{1}{2}$ and $\alpha = 0.01$ from equations~(\ref{eq:2nu}),~(\ref{eq:eesa}) and~(\ref{eq:mf2pex}) respectively.}
 \label{fig:matter2p}
\end{figure*}  
Next I turn my attention to flavor oscillations in the presence of matter. While a detailed study is beyond the scope of this work, I focus on the adiabatic approximation (constant matter density).
In other words, I study length scales for which matter density does not vary drastically. As Eq.~(\ref{eq:matter_effect}) shows, the effective mixing angle in dense media is much smaller than in vacuum
and the effective mass coupling increases. In Fig.~\ref{fig:matter2p} I used 
$\alpha=0.01$ as an illustration. Collective oscillations develop for both multi and single-angle calculations while the mean field approximation 
shows no oscillations at all. It seems that as long as there is some mixing, no matter how small the mixing angle, fast oscillations develop and the frequency of oscillations in this case depends only 
on $\mu$.
The explanation for the mean field result can be found by analyzing Eq.~(\ref{eq:s0}). If the mixing angle is very small and there is no net polarization initially, $\bold{P}_{\text{tot}}(0)=\bold{0}$ and $\bold{P}_{\text{diff}}(0) = \hat{z}$.
Then, the rate of change of the polarization difference is
\begin{equation*}
\begin{split}
  d_t \bold{P}_{\text{diff}} \approx& -\overline{\omega}_p \hat{z} \times \bold{P}_{\text{diff}} = \bold{0}.
\end{split}
 \end{equation*}
 This is a rather interesting finding as collective oscillations in dense media could have quite dramatic implications for core collapse supernovae and neutron star binary mergers.
 One can perform an order of magnitude estimate of the ratio between the neutrino mean free path from charged reactions and for collective oscillations.
 To estimate the ratio of these two length scales, I assume the normalizing volume in the two body term to be $V\propto n^{-1}_{\nu}$,
 \begin{equation*}
  \begin{split}
   \frac{\lambda_{\nu_e + n \rightarrow p +e}}{\lambda_{\text{osc}}}\propto & \frac{\sqrt{2}G_F n_{\nu}}{G_F^{2}\omega^2n_b}=\frac{\sqrt{2}n_{\nu}}{G_F \omega^2 n_B}\\
   =&1.2 \times 10^{3} \left(\frac{n_{\nu}/n_B}{10^{-6}}\right) \left(\frac{10\ \text{MeV}}{\omega}\right)^2.
  \end{split}
\end{equation*}
This result seems to suggest that once collective oscillation develop they will effectively equilibrate neutrino flavors between collisions with matter. 
However, the result depends on the numerical value of $V$, which I discuss in section~\ref{sec:inf_two_body}.
I also consider the case for which the initial wave-function is only electron flavor. From Eqs.~(\ref{eq:eema}) and~(\ref{eq:eesa}) the single-angle approximation gives the same result
as the multi-angle treatment and they both show ordinary vacuum oscillations. In addition, no correlations are present, and the mean field approximation result in Eq.~(\ref{eq:mf2p}) 
agrees with the exact solution.  

\section{Cubic Lattice results}
\label{sec:cubic}
\begin{figure}[ht]
\includegraphics[height=0.5\linewidth,width=0.75\linewidth]{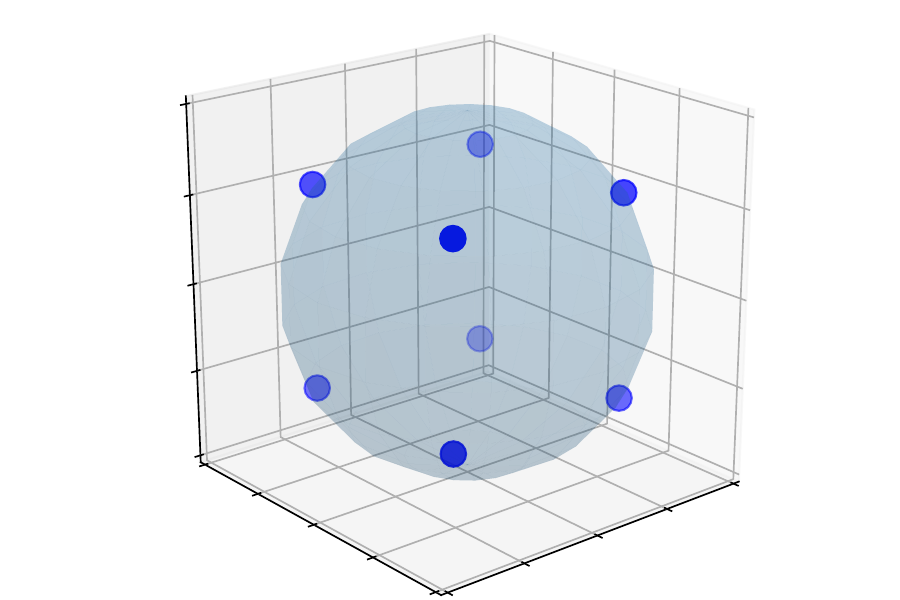}
\caption{Cubic lattice structure of eight neutrino momenta which share the same magnitude.}
\label{fig:8p}
\end{figure}

\begin{figure}[ht]
 \subfloat[Flavor polarization]
 {\includegraphics[height=0.45\linewidth,width=\linewidth]{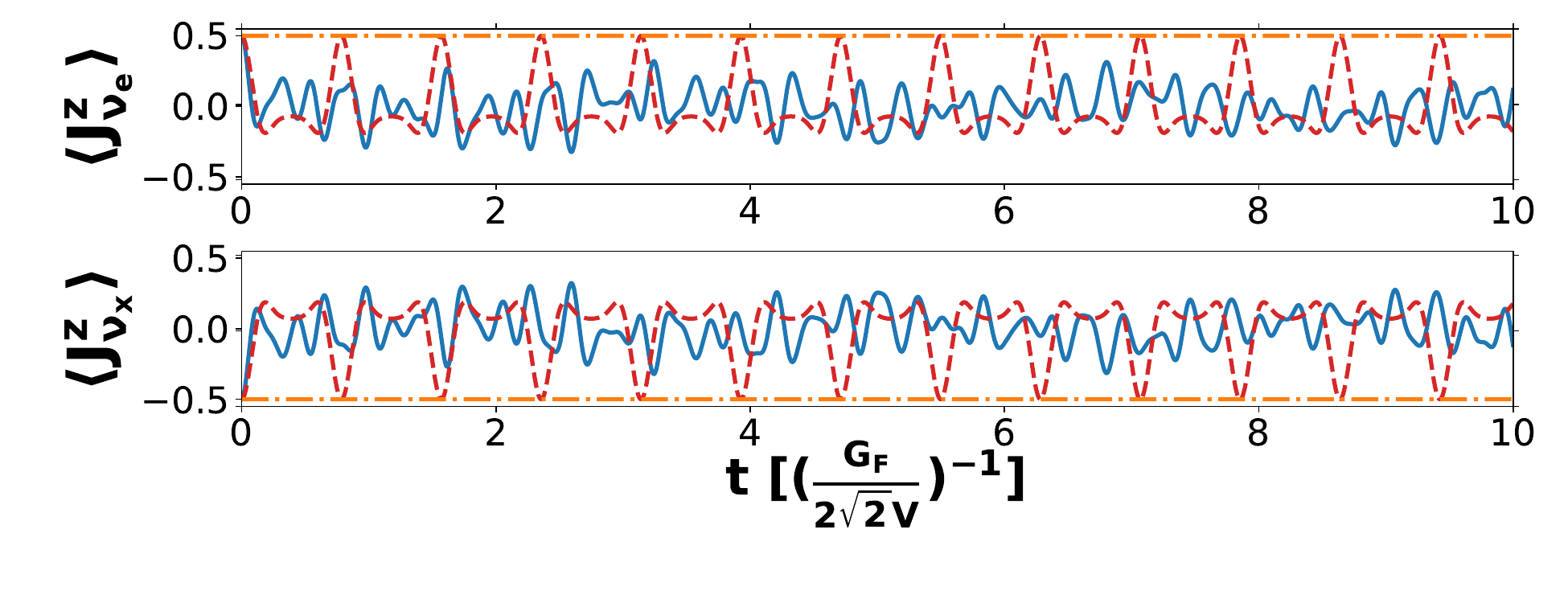}}\\
 \subfloat[Correlations]{\includegraphics[height=0.35\linewidth,width=\linewidth]{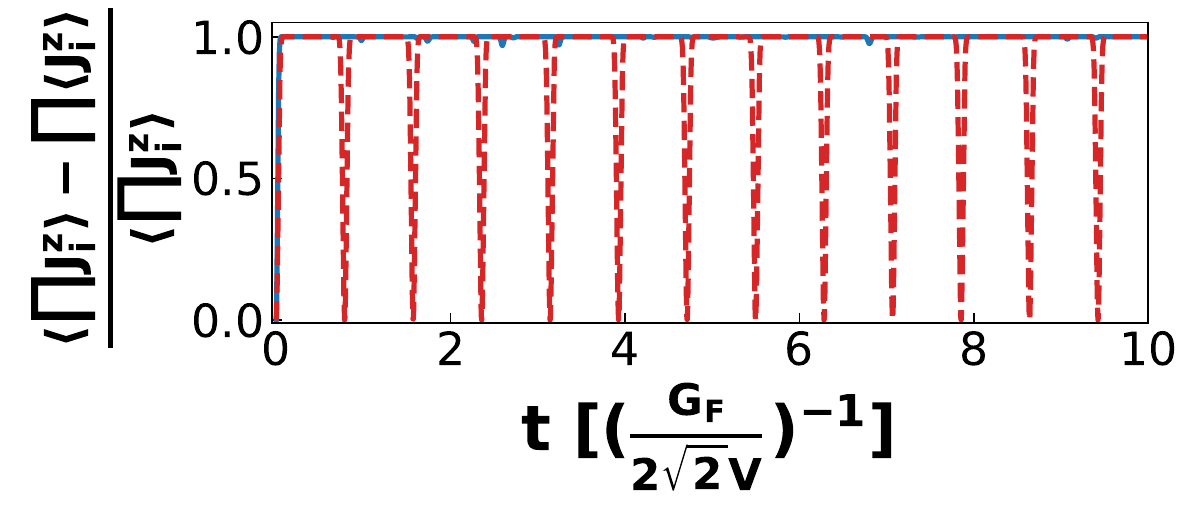}}
 \caption{Multi-angle (solid blue line), 
 single-angle (dashed red line), and mean field (dashed dotted orange line) flavor polarization and correlations for the lattice in Fig.~\ref{fig:8p}
 and initial wave-function $|\Psi_0\rangle= | \nu_e^{(4)} \nu_{x}^{(4)}\rangle$ in medium with $ {\omega}/2= \mu = \mu_0$ and $\alpha_p=0.01$.  These results were obtained based on eq.~(\ref{eq:psit}).}
 \label{fig:4e4x}
 \end{figure}

While the results in the previous section were obtained analytically, for systems of more than two neutrinos I proceed to solve numerically for the wavefunction as function of time,
\begin{equation}
 | \Psi(t)\rangle=e^{-iHt}| \Psi_0\rangle.
 \label{eq:psit}
\end{equation}
In this section I consider a three dimensional cube as depicted in Fig.~\ref{fig:8p}. The basis used in this case is a direct product of eight Pauli matrices which allows 
for the exact computation of the time evolution operator.
I study oscillations in matter with an equal number of electron and $x$ flavor in the initial wave-function. Figure~\ref{fig:4e4x} 
shows the correlation of all points and the polarizations for two lattice points with initial flavors $\nu_e$ and $\nu_x$ respectively for a mixing angle $\alpha_p=0.01$. 
The mean field shows no flavor evolution, while the many-body methods depict rapid flavor equilibration. In the single-angle approximation, as correlations change between maximal and minimal values, so does the 
flavor mixing. The multi-angle calculation shows strong correlations, and the flavor polarization oscillates close to zero as a result. 
The presence of correlations explains the difference between exact calculations and mean field similarly to the two beam case.
As before, correlations disappear when all neutrinos start with the same flavor and all methods agree.
I verified numerically that $\bold{\overline{B}_p}\cdot \bold{J}$ is a constant of motion. 
Flavor evolution in the vacuum with no net initial polarization and alternatively with only one flavor initially is provided in 
appendix~\ref{app:cubic}.

\subsection*{Initial conditions with partial polarization}
\begin{figure}[ht]
 \subfloat[$ {\omega}/2= \mu_0, \ \mu/4 = \mu_0$]{\includegraphics[height=0.45\linewidth,width=\linewidth]{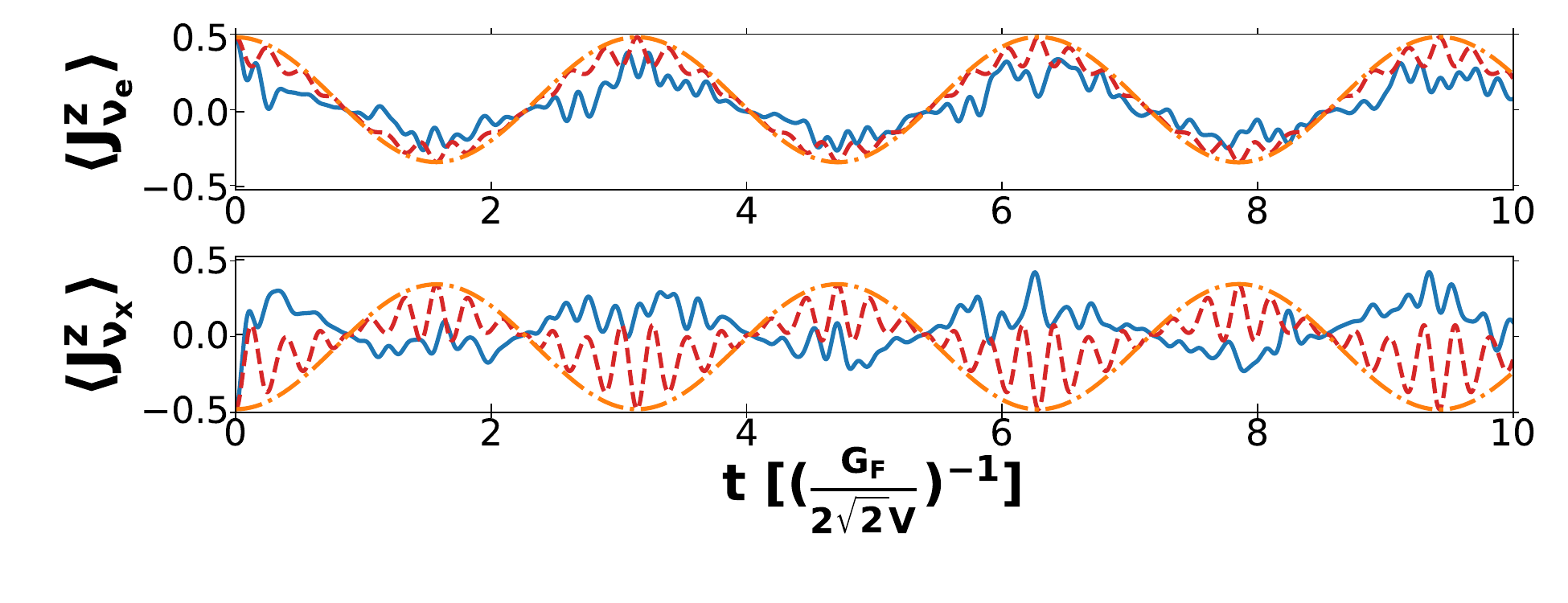}}\\
 \subfloat[$ {\omega}/2= \mu_0/5, \ \mu/4 = \mu_0$]{\includegraphics[height=0.45\linewidth,width=\linewidth]{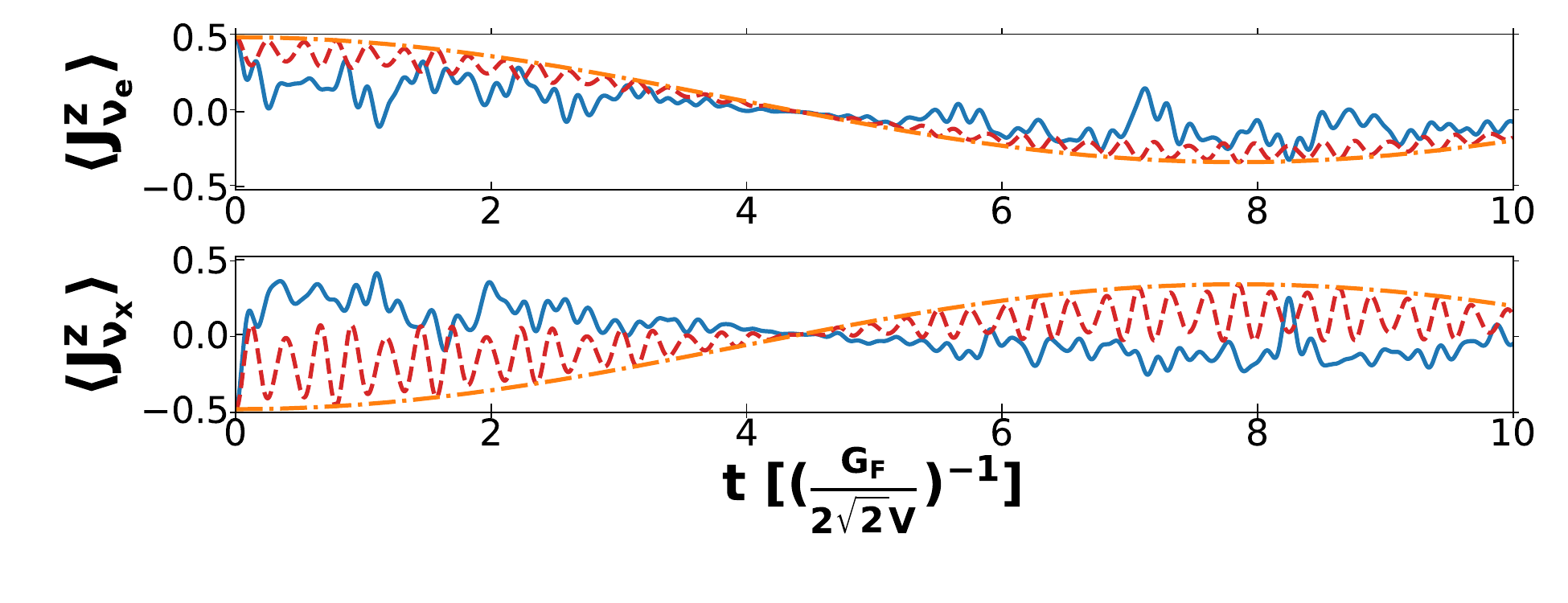}}\\
 \caption{Multi-angle (solid blue line), 
 single-angle (dashed red line), and mean field (dashed dotted orange line) calculations of the flavor polarization as function of time for the cubic lattice in Fig.~\ref{fig:8p}
 and initial wave-function $|\Psi_0\rangle= | \nu_e^{(6)} \nu_{x}^{(2)}\rangle$ in vacuum.   These results were obtained based on eq.~(\ref{eq:psit}).}
 \label{fig:6e2x}
\end{figure}

In the previous sections I have studied case when all neutrinos start with one flavor, or both flavors are present in the same amount.
However, an intermediate scenario is needed where there is some flavor polarization in the initial
wave-function, but not all neutrinos are of the same type. As an illustration, I performed the flavor evolution for the cubic lattice with an initial wave-function of six electron neutrinos and
two x neutrinos, $| \Psi_0 \rangle =| \nu_e^{(6)}\nu_x^{(2)}\rangle$. 
Figure~\ref{fig:6e2x} displays the flavor evolution for two lattice points which start oscillations
with $\nu_e$ and $\nu_x$ respectively.
Collective oscillations still develop but are not as pronounced as in the previous cases. Similar to previous results, the mean field method does not show any collective behavior.
In addition, both the single-angle and mean field approximations display oscillations with a contrary phase with respect to the exact solution when it comes to the flavor of lower concentration.
In the multi-angle lattice treatment, the $x$ flavor quickly synchronizes in phase with the electron flavor, but this does not happen for the two other methods. 

Since in this section I have studied systems of neutrinos with the same frequency $\omega_p$, a few remarks are needed. Firstly, from eq.~(\ref{eq:mf2}) one can conclude that initially the cross products vanish since the polarizations are aligned along the z-axis. As all neutrinos have the same frequency, in the mean field approximation they will precess around B with same frequency, remaining parallel, and thus the two body term will never impact the flavor evolution regardless of particle number. On the other hand, in the many body case the result is qualitatively different. The mean field initial state with both flavors present is not an eigenstate of the two body interaction in eq.~(\ref{nunu}), and its time evolution will be affected by this term in the Hamiltonian as shown in figs.~(\ref{fig:4e4x}) and~(\ref{fig:6e2x}).
\section{Two Energy Levels: quadratic lattice}
\label{sec:quadratic}
\begin{figure}[ht]
\centering
\includegraphics[scale=0.22]{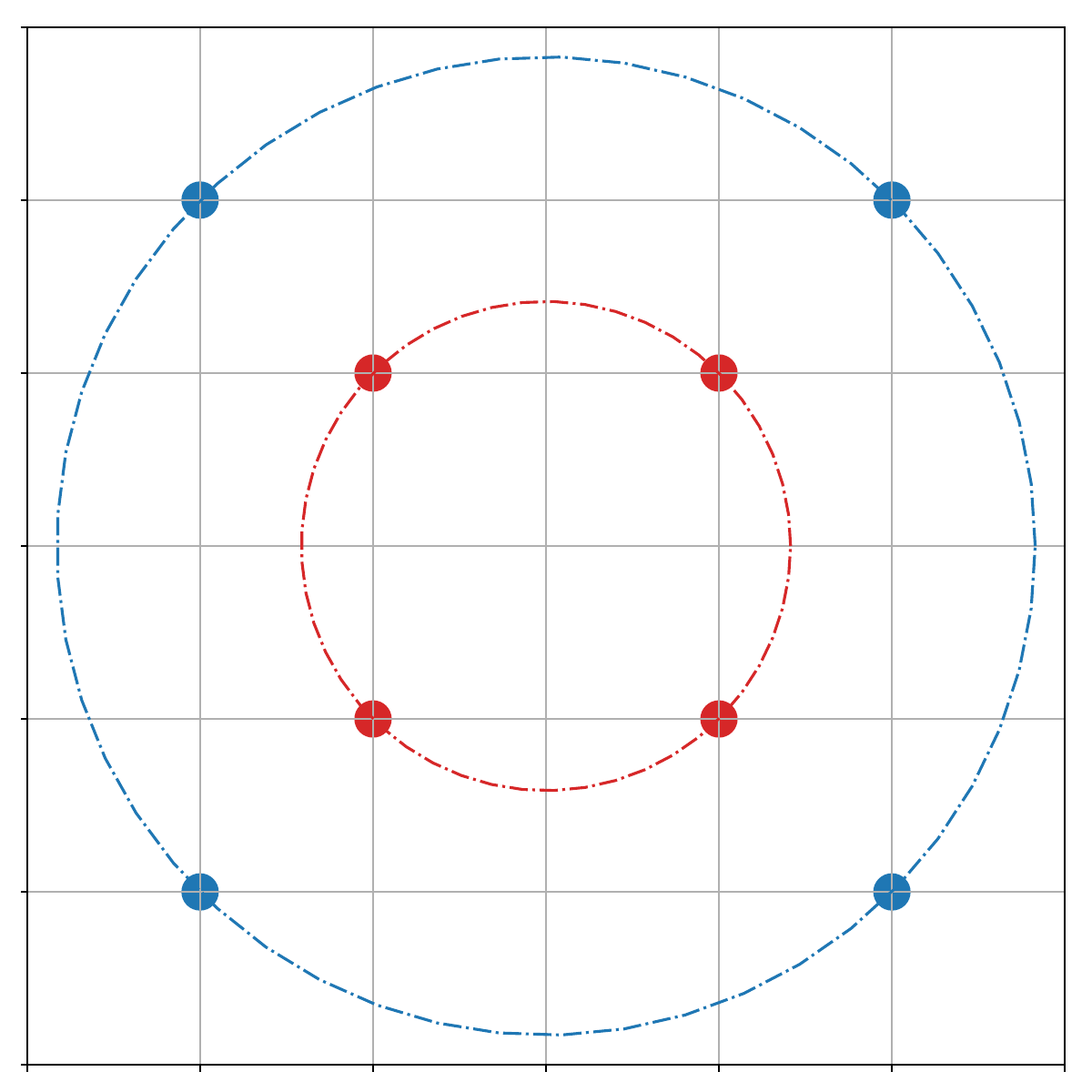}
\caption{Two quadratic lattice structures in momentum space. The smaller lattice is initially of electron flavor and the larger lattice of x flavor.}
\label{fig:44lattice}
\end{figure}
So far mono-energetic flavor oscillations have been considered. Due to their interactions with matter during a core collapse supernova, 
electron neutrinos decouple in lower density regions with respect to other flavors and the resulting energy spectrum is lower~\cite{Bruenn:1987,Tamborra:2012,Janka:2017vlw}. 
To account for this difference in spectrum, in this section I work with 
two quadratic lattices of different momenta values as shown in Fig.~\ref{fig:44lattice}. I pick a momentum magnitude ratio of three as a conservative representation of the mean energy ratio 
found in numerical simulations~\cite{Bruenn:1987,Janka:2017vlw}.
I denote by $\omega_p$  and $\omega_{3p}$ for the frequencies respectively.

\subsection{Initial Conditions without polarization}
\begin{figure}[ht]
 \subfloat [Flavor polarization]
 {\includegraphics[height=0.45\linewidth,width=\linewidth]{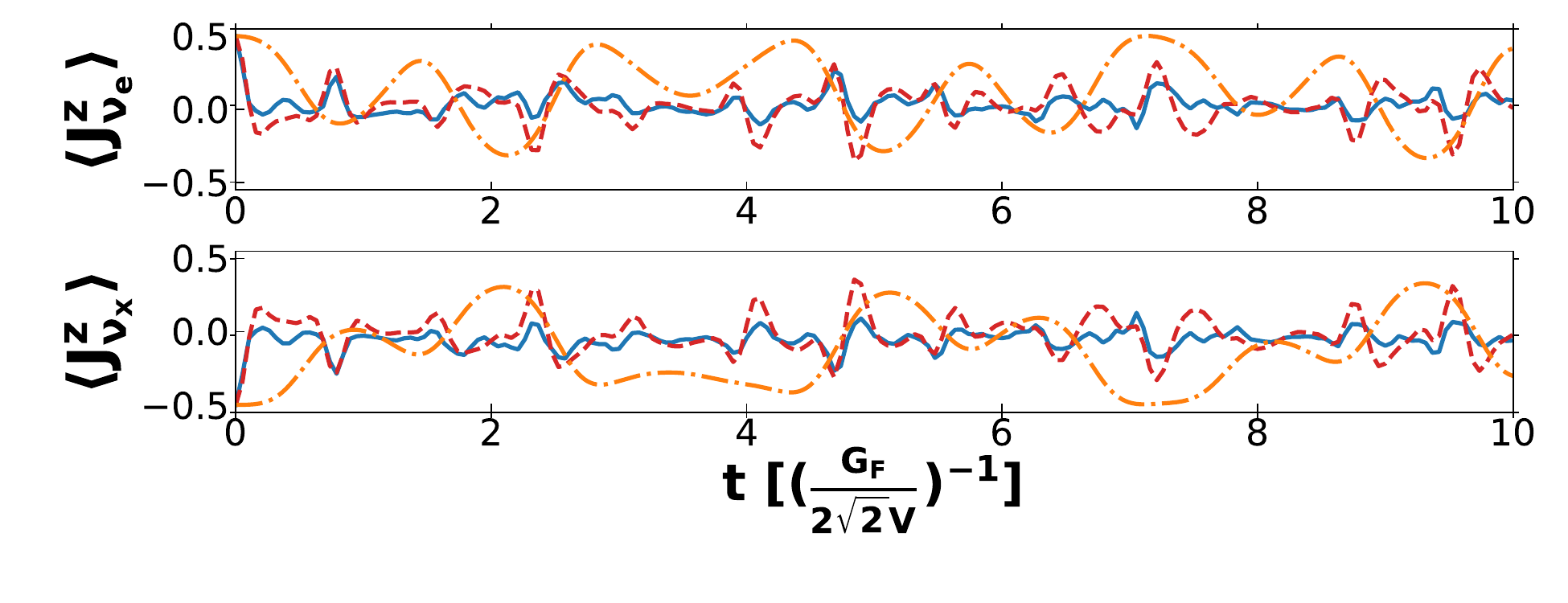}}\\
 \subfloat[Correlations]{\includegraphics[height=0.35\linewidth,width=\linewidth]{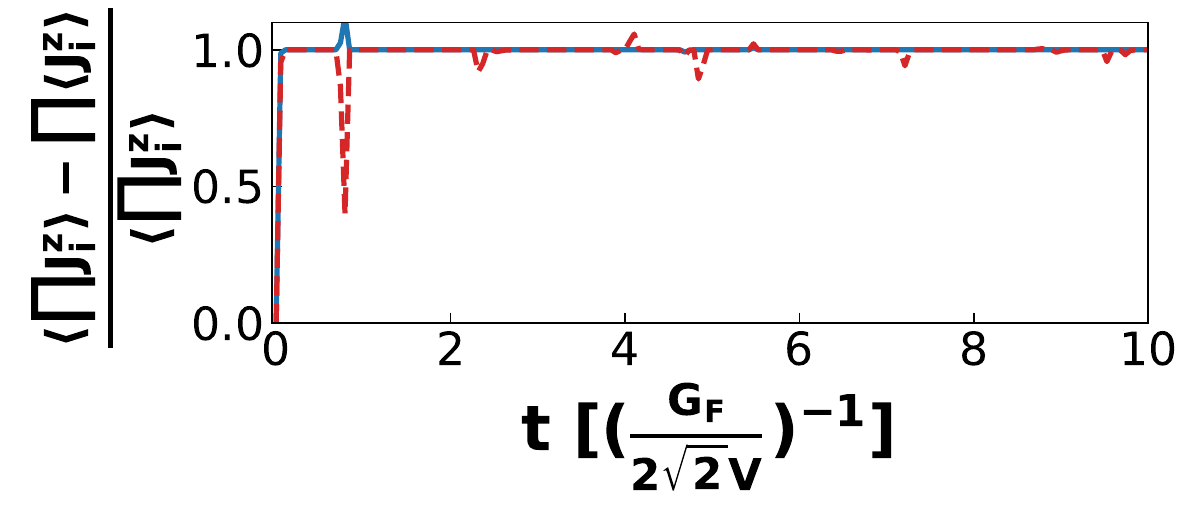}}
 \caption{Multi-angle (solid blue line), 
 single-angle (dashed red line), and mean field (dashed dotted orange line) flavor polarization and correlations for the lattice in Fig.~\ref{fig:44lattice}
 and initial wave-function $|\Psi_0\rangle= | \nu_e^{(4)} \nu_{x}^{(4)}\rangle$ in vacuum with $ {\omega}_p/2= \mu/4 = \mu_0$.   These results were obtained based on eq.~(\ref{eq:psit}).}
 \label{fig:4410_1}
 \end{figure}

 Qualitatively, oscillations agree with previous results in this work. The presence of two energy levels is seen in the pattern of oscillations which is rather complicated as shown
 in Fig.~\ref{fig:4410_1}. Polarization correlations among all neutrinos are also displayed in this figure.

\subsection{Initial Conditions with partial polarization}
I verified that if the initial wave-function contains only one neutrino flavor, no collective oscillations occur and the mean field agrees with the lattice calculation. 
In addition, correlations vanish as well.

\begin{figure}[ht]
 \includegraphics[height=0.65\linewidth,width=\linewidth]{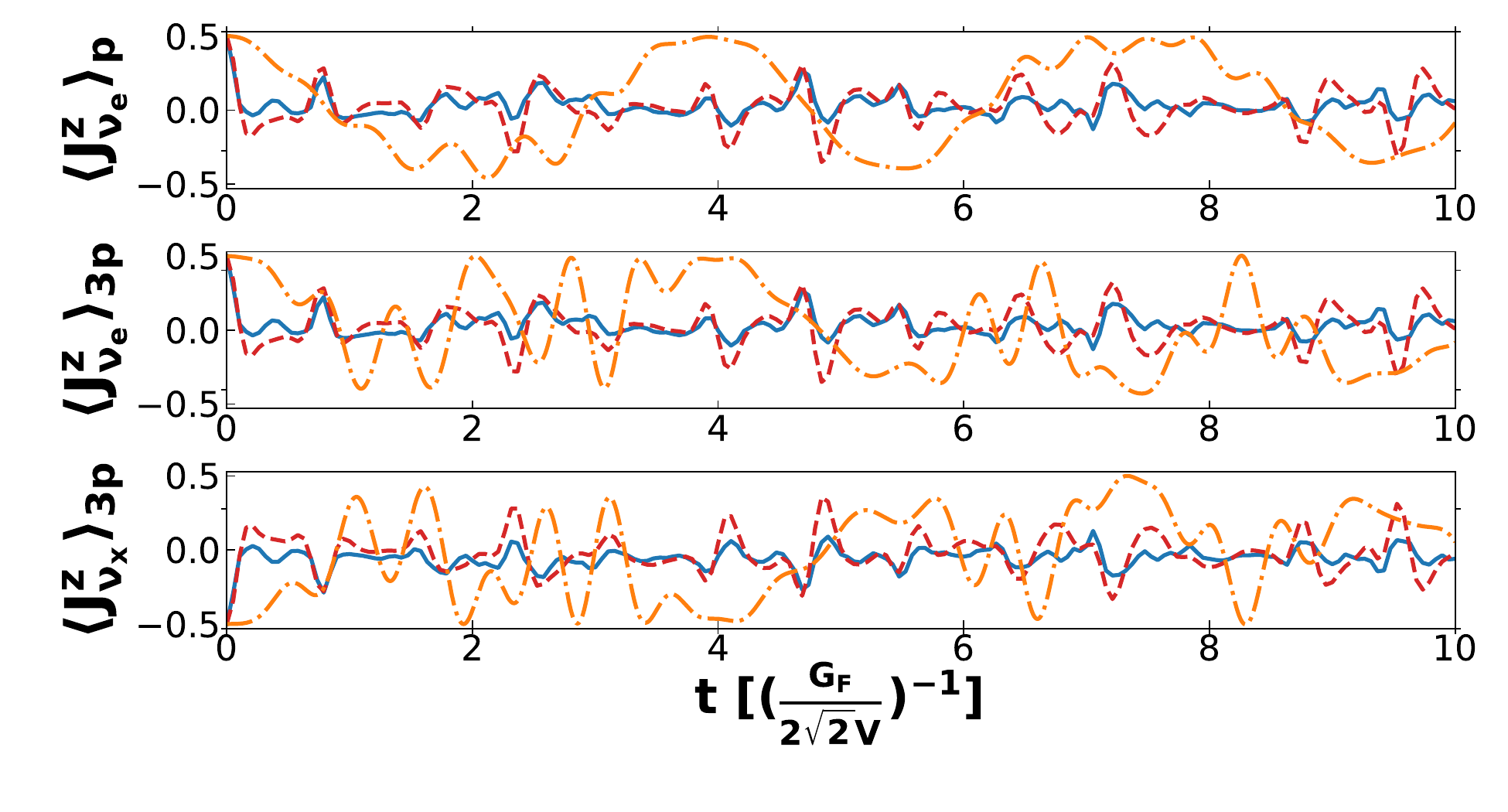}
 \caption{Multi-angle (solid blue line), 
 single-angle (dashed red line), and mean field (dashed dotted orange line) flavor polarization for the lattice in Fig.~\ref{fig:44lattice}
 and initial wave-function $|\Psi_0\rangle= | \nu_e^{(6)} \nu_{x}^{(2)}\rangle$ in vacuum with $ {\omega}_p/2= \mu/4 = \mu_0$.   These results were obtained based on eq.~(\ref{eq:psit}).}
 \label{fig:6210_1}
 \end{figure}

\begin{figure}[ht]
 \includegraphics[height=0.35\linewidth,width=\linewidth]{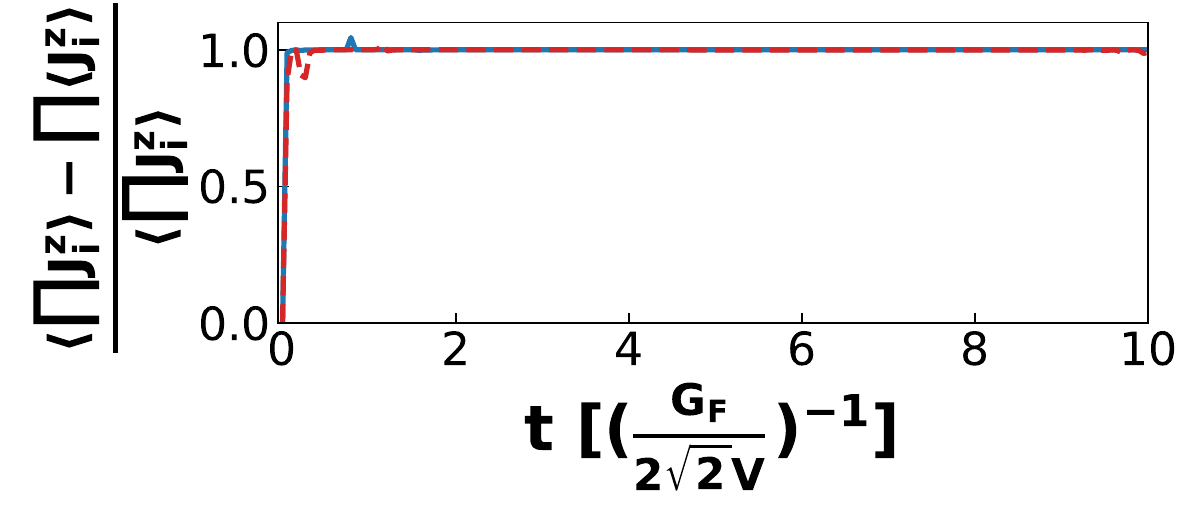}
 \caption{Multi-angle (solid blue line), 
 single-angle (dashed red line) correlations for the lattice in Fig.~\ref{fig:44lattice}
 and initial wave-function $|\Psi_0\rangle= | \nu_e^{(6)} \nu_{x}^{(2)}\rangle$ in vacuum with $ {\omega}_p/2= \mu/4 = \mu_0$.   These results were obtained based on eq.~(\ref{eq:psit}).}
 \label{fig:6210_2}
 \end{figure}

\begin{figure}[ht]
\includegraphics[height=0.65\linewidth,width=\linewidth]{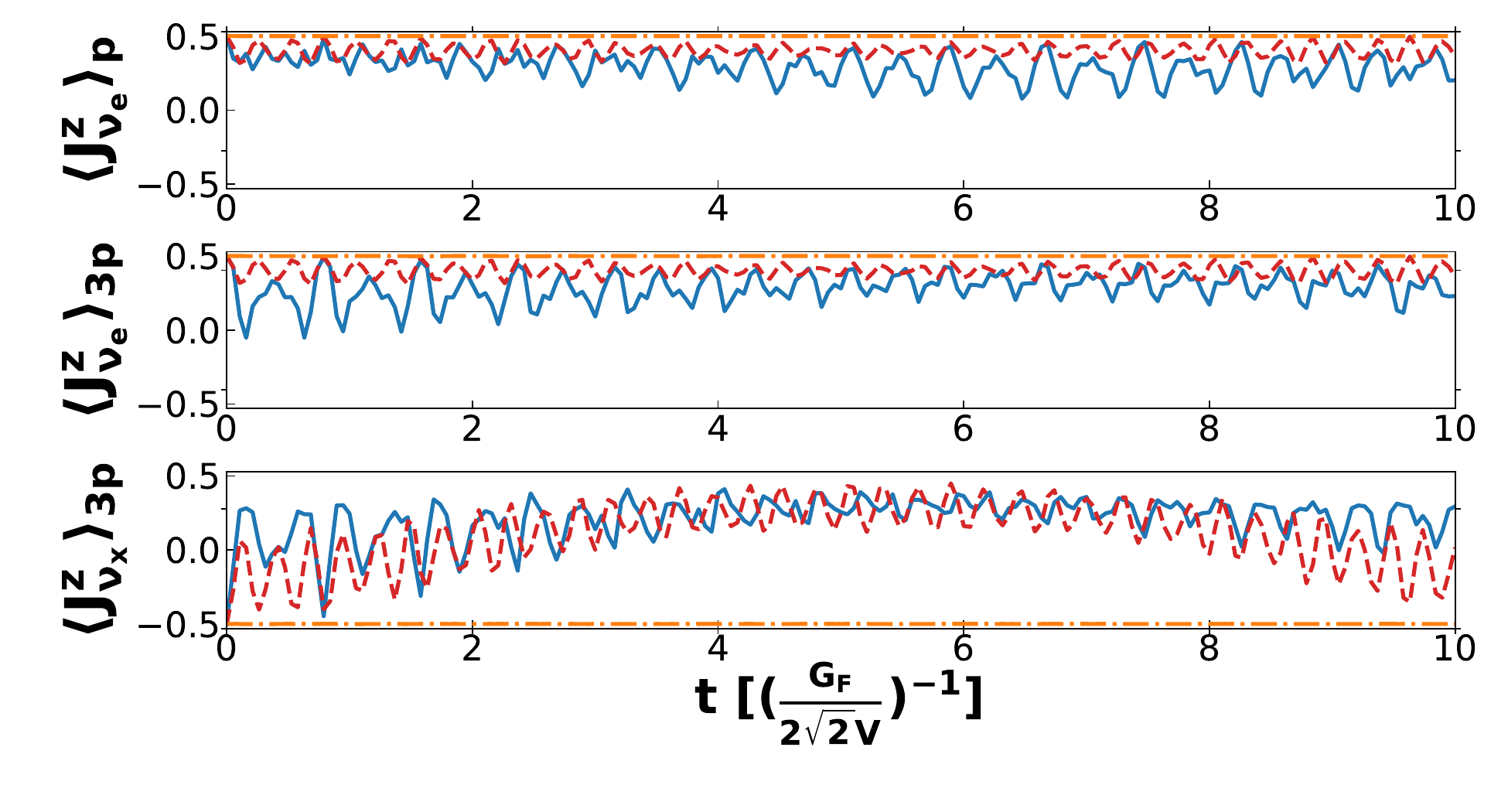}
 \caption{Multi-angle (solid blue line), 
 single-angle (dashed red line), and mean field (dashed dotted orange line) flavor polarization for the lattice in Fig.~\ref{fig:44lattice}
 and initial wave-function $|\Psi_0\rangle= | \nu_e^{(6)} \nu_{x}^{(2)}\rangle$ in medium. These results were obtained based on eq.~(\ref{eq:psit}).}
 \label{fig:6e2xm}
 \end{figure}

Next I consider a mixture of 75\% electron flavor, $| \Psi_0 \rangle =| \nu_e^{(6)}\nu_x^{(2)}\rangle$ like in the case of the cubic lattice. From Fig.~\ref{fig:6210_1} one can 
see that collective oscillations develop and from fig.~\ref{fig:6210_2} that correlations are present. Lattice points 4 and 8 show flavor oscillations for the larger lattice with $\nu_x$ and $\nu_e$ flavors initially; lattice point 1 is 
a representative from the smaller lattice.
The amplitude of collective oscillations is less pronounced and it oscillates close to zero. The mean field still differs from the exact solution.
As a final check, I study oscillations in a dense medium in Fig.~\ref{fig:6e2xm}. The effective frequencies increase while the mixing angles decrease resulting in $\alpha_p/\alpha_{3p}\approx 3, \ \overline{\omega}_{p}/\overline{\omega}_{3p}\approx1$ with $\alpha_p=0.01$ and $\omega_p/2=\mu/4=\mu_0$. The mean field shows no oscillations for any of the neutrinos, while the multi and single-angle develop collective oscillations. This is more evident for the x flavor neutrinos in Fig.~\ref{fig:6e2xm}. This confirms the presence of correlations among neutrinos of different energies even in a dense medium. The two energy level neutrino systems in the following sections have the same momenta values in vacuum and respective frequencies and mixing angles in medium as described here.
 \begin{figure}[ht]
 \includegraphics[height=0.35\linewidth,width=\linewidth]{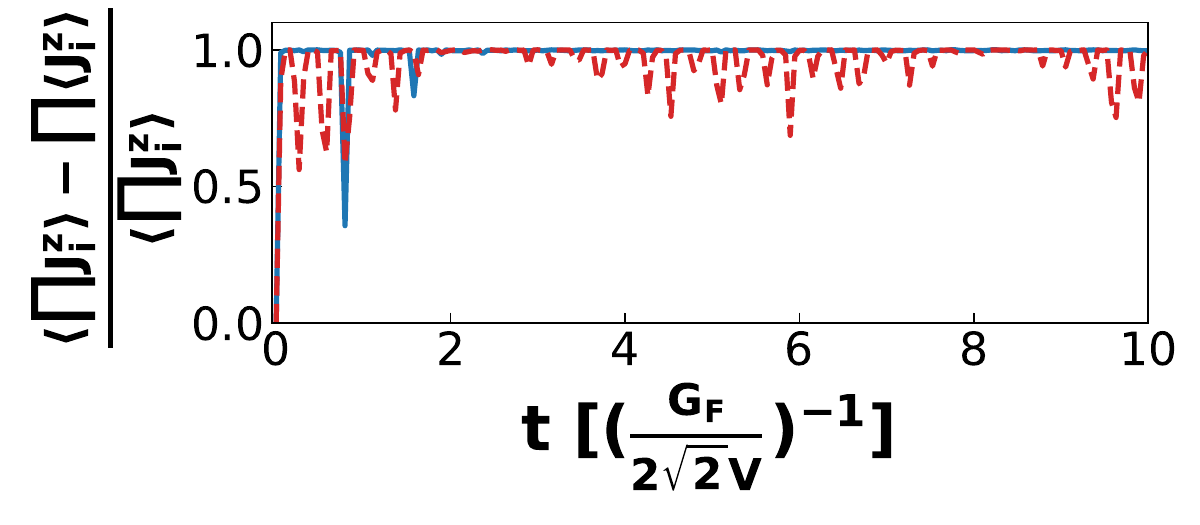}
 \caption{Multi-angle (solid blue line), 
 single-angle (dashed red line) correlations for the lattice in Fig.~\ref{fig:44lattice}
 and initial wave-function $|\Psi_0\rangle= | \nu_e^{(6)} \nu_{x}^{(2)}\rangle$ in medium. These results were obtained based on eq.~(\ref{eq:psit}).}
 \label{fig:6e2xm_2}
 \end{figure}

 \section{System of twelve neutrinos}
 \label{sec:12}
 \begin{figure}[ht]
 \includegraphics[scale=0.5]{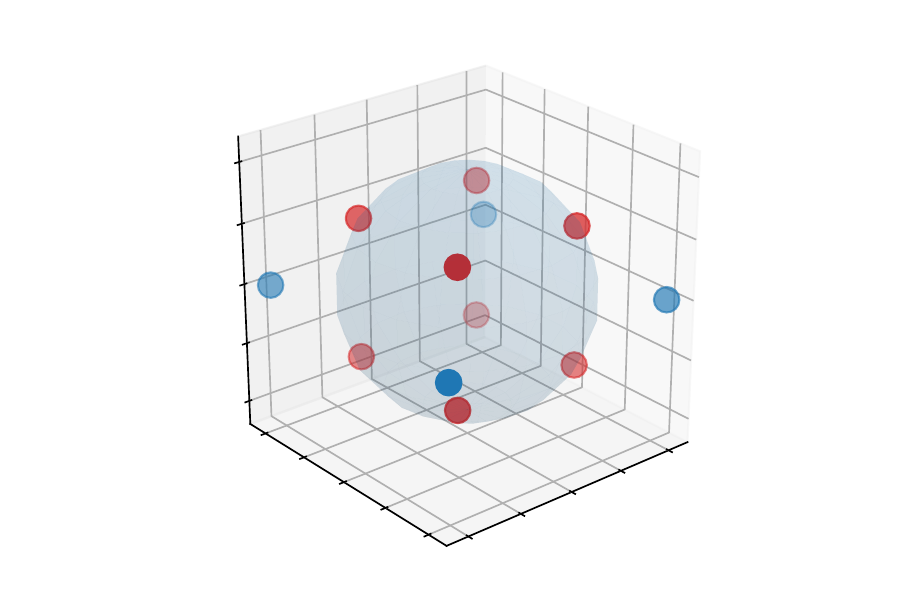}
 \label{fig:84lattice}
 \caption{Cubic lattices with initially electron neutrinos (smaller lattice) and quadratic lattice of x flavor neutrinos (greater lattice).}
\end{figure}
 In this section I consider a system of twelve neutrinos with two thirds initially of electron flavor. The configuration chosen is depicted in fig.~\ref{fig:84lattice}. Electron neutrinos
 have a momentum magnitude three times smaller than x flavor neutrinos following the same rationale of the previous section. 
 
 \begin{figure}[ht]
 \subfloat [Flavor polarization]
 {\includegraphics[height=0.45\linewidth,width=\linewidth]{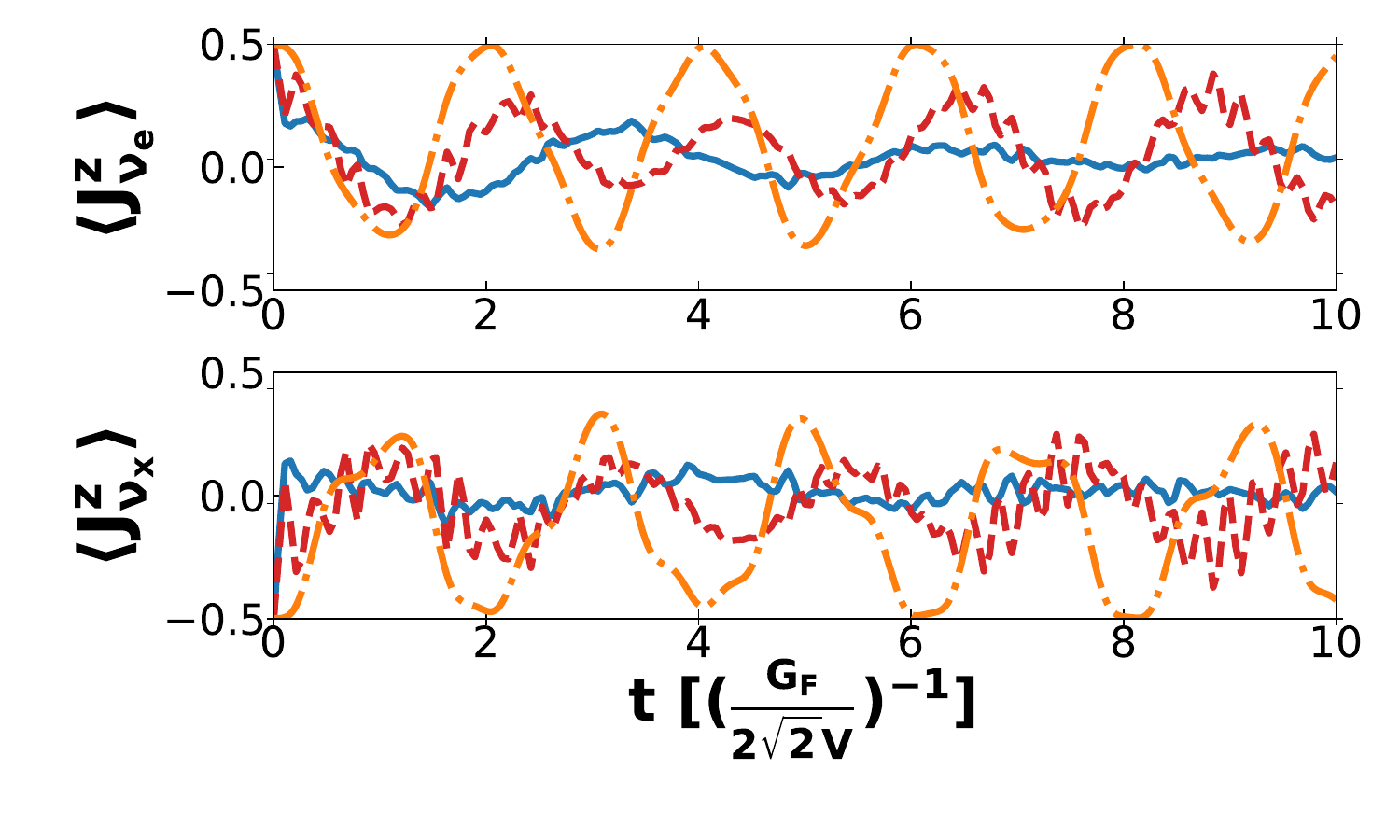}}\\
 \subfloat[Correlations]{\includegraphics[height=0.35\linewidth,width=\linewidth]{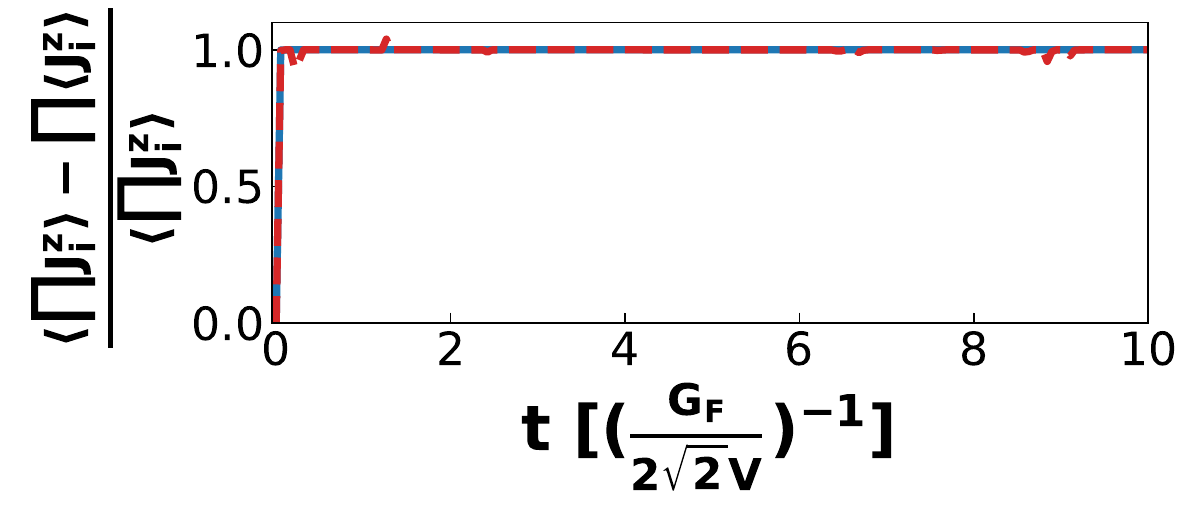}}
 \caption{Multi-angle (solid blue line), 
 single-angle (dashed red line), and mean field (dashed dotted orange line) flavor polarization and correlations for the lattice in Fig.~\ref{fig:84lattice}
 and initial wave-function $|\Psi_0\rangle= | \nu_e^{(8)} \nu_{x}^{(4)}\rangle$ in vacuum with $ {\omega}/2= \mu/4 = \mu_0$.   These results were obtained based on eq.~(\ref{eq:psit}).}
 \label{fig:8410}
 \end{figure}
 
\begin{figure}[ht]
 \subfloat [Flavor polarization]
 {\includegraphics[height=0.45\linewidth,width=\linewidth]{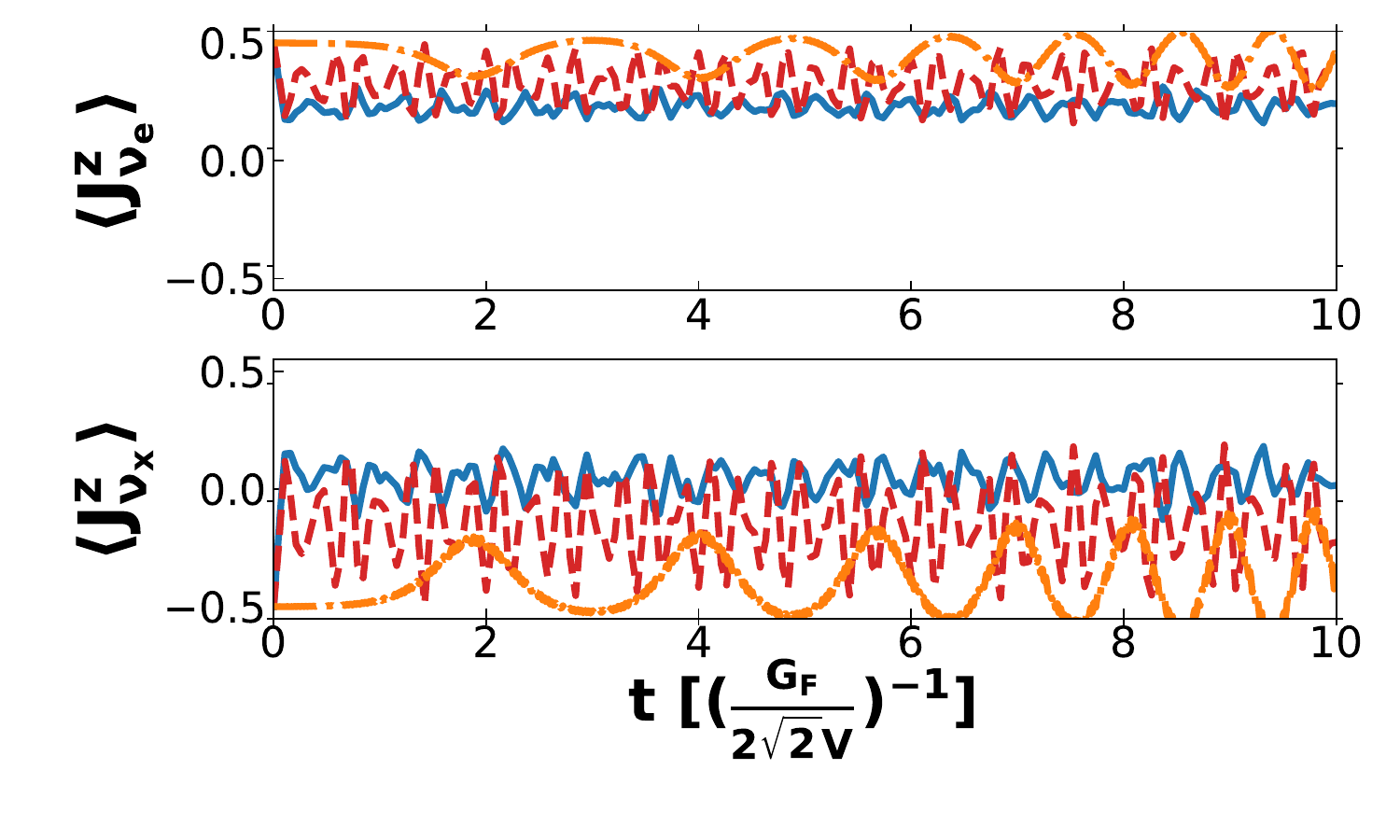}}\\
 \subfloat[Correlations]{\includegraphics[height=0.35\linewidth,width=\linewidth]{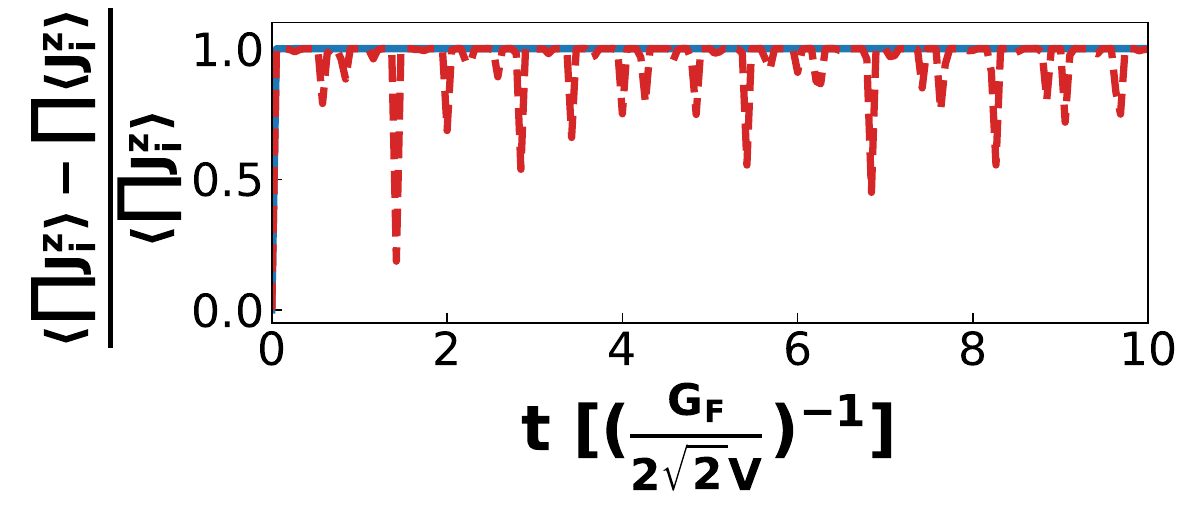}}
 \caption{Multi-angle (solid blue line), 
 single-angle (dashed red line), and mean field (dashed dotted orange line) flavor polarization and correlations for the lattice in Fig.~\ref{fig:84lattice}
 and initial wave-function $|\Psi_0\rangle= | \nu_e^{(8)} \nu_{x}^{(4)}\rangle$ in medium. These results were obtained based on eq.~(\ref{eq:psit}).}
 \label{fig:8410_1}
 \end{figure}

 Figure~\ref{fig:8410} shows the flavor polarization for two neutrinos in the lattice and overall correlation as functions of time in vacuum while fig.~\ref{fig:8410_1} displays the time evolution
 for the same initial conditions in a dense matter medium.
 
 The results depicted in these figures agree with previous section: there is rapid flavor equilibration in the many-body methods, particularly in the multi-angle case, while the mean field shows no such behavior.
 Not surprisingly, correlations are significant when this happens. Especially in a dense matter medium, the qualitative difference is quite visible as the mean field shows subdued oscillations. However, the multi-angle shows rapid flavor equilibration, more so for the flavor initially present in smaller amount. The single-angle shows a result in between the multi-angle and mean field.

\section{Systems of sixteen neutrinos}
\label{sec:16}
In this section I continue the many-body study of the neutrino flavor oscillations by considering two different configurations of sixteen neutrinos.

At first, I consider two cubic lattices of electron and x flavor respectively, with the same momentum ratio as in the previous sections. 
In Fig.~\ref{fig:8810_1} I plot the flavor polarization  and correlation in a dense medium. The mean field treatment shows the usual oscillations due to the one body term. Both many-body calculations show an almost immediate flavor equilibration; qualitatively the opposite outcome. Correlations are significant as the plot shows.

\begin{figure}[ht]
 \includegraphics[scale=0.5]{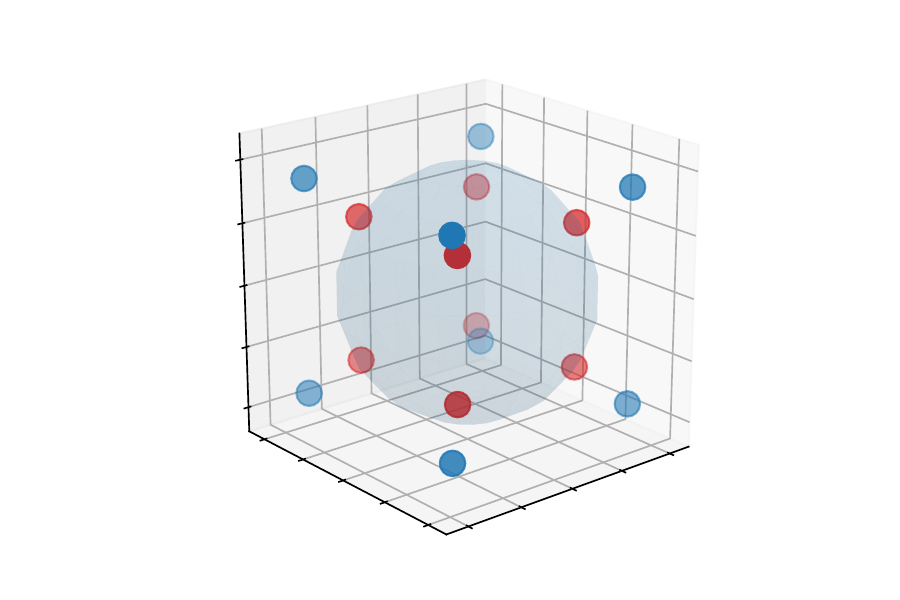}
 \caption{Two cubic lattices with initially electron neutrinos (smaller lattice) and x flavor neutrinos (greater lattice).}
 \label{fig:88lattice}
\end{figure}

\begin{figure}[ht]
 \subfloat [Flavor polarization]
 {\includegraphics[height=0.45\linewidth,width=\linewidth]{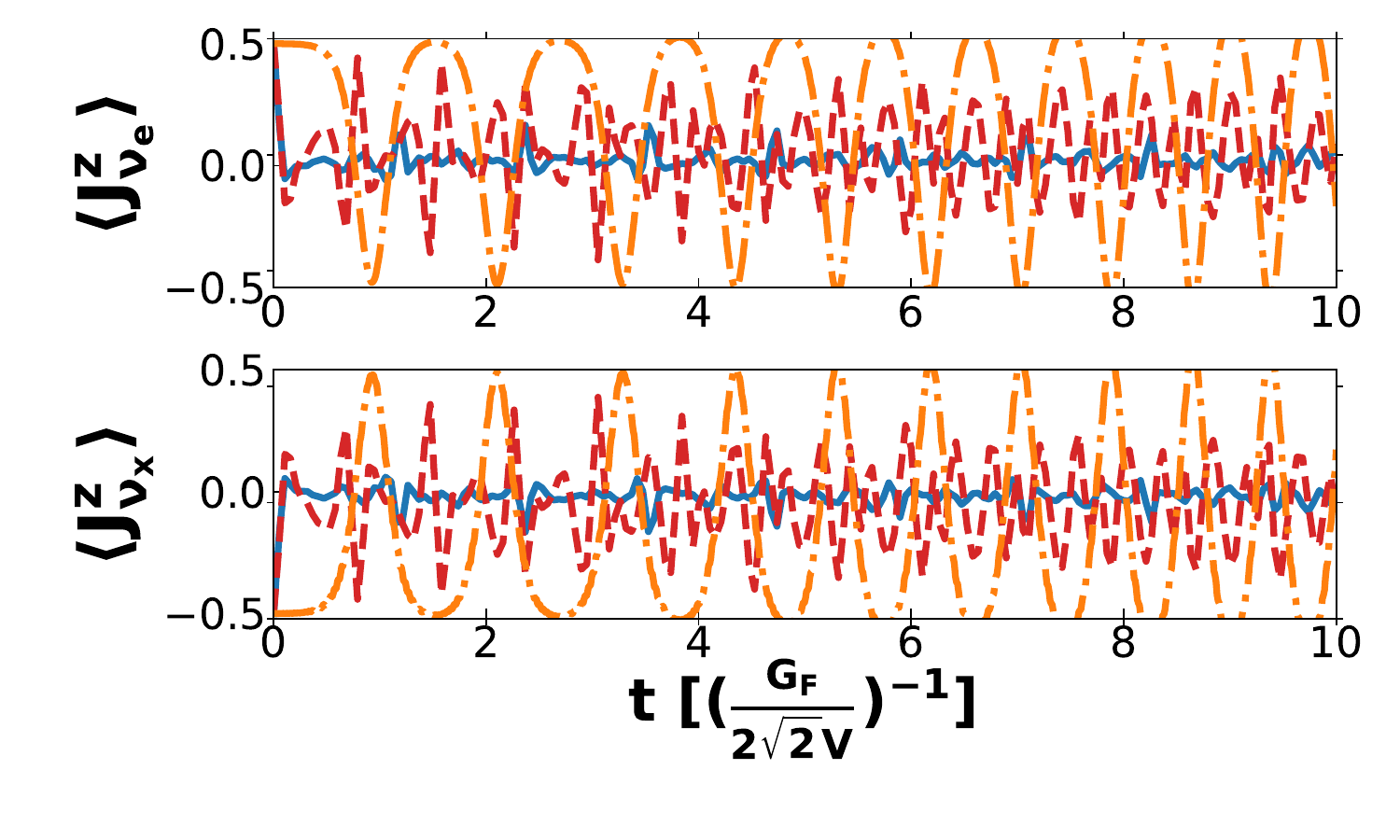}}\\
 \subfloat[Correlations]{\includegraphics[height=0.35\linewidth,width=\linewidth]{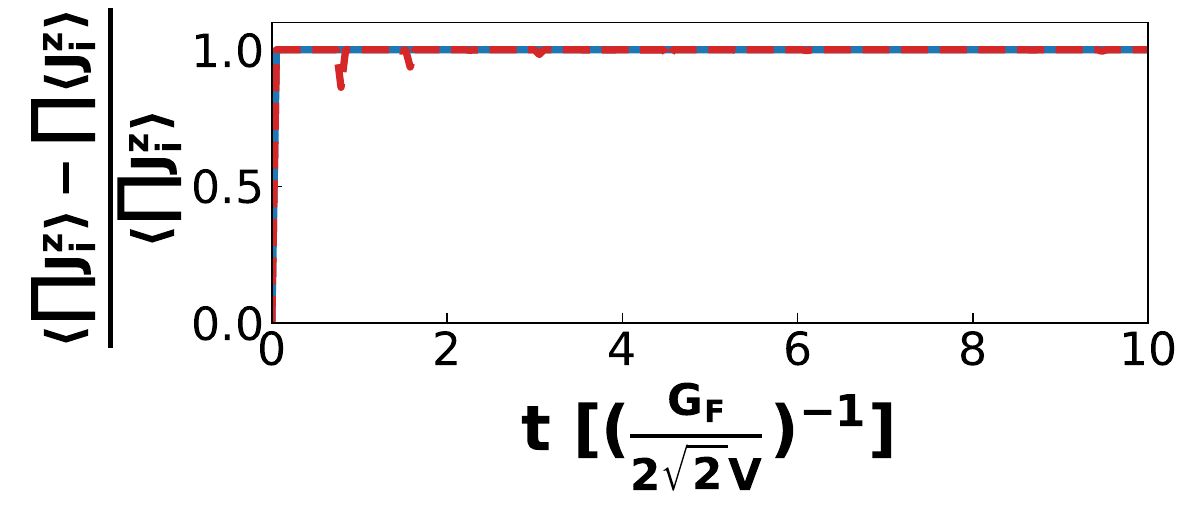}}
 \caption{Multi-angle (solid blue line), 
 single-angle (dashed red line), and mean field (dashed dotted orange line) flavor polarization and correlations for the lattice in Fig.~\ref{fig:88lattice}
 and initial wave-function $|\Psi_0\rangle= | \nu_e^{(8)} \nu_{x}^{(8)}\rangle$ in medium. These results were obtained based on eq.~(\ref{eq:psit}).}
 \label{fig:8810_1}
 \end{figure}
 
\begin{figure}[ht]
 \includegraphics[scale=0.5]{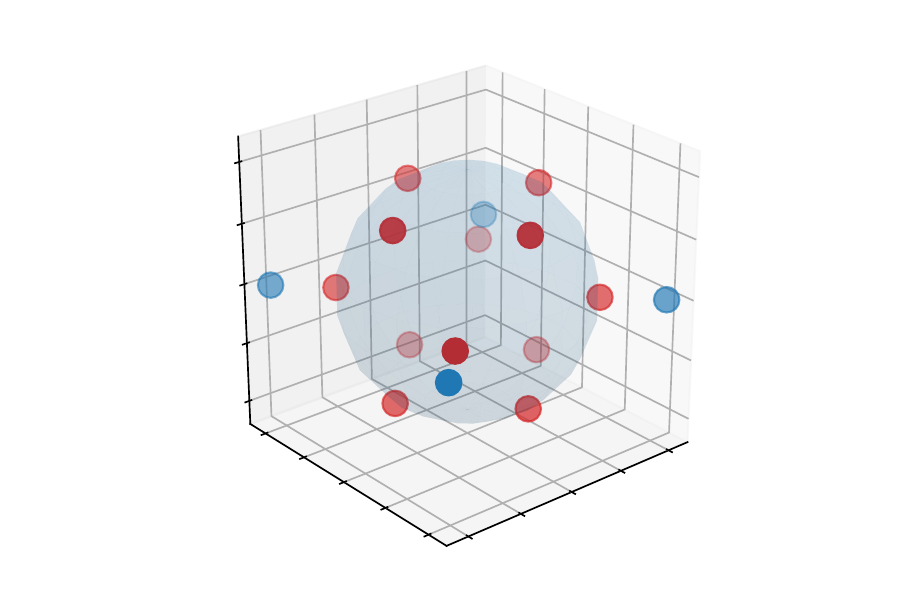}
 \caption{Configuration with initially twelve electron neutrinos (smaller lattice) and four x flavor neutrinos (greater lattice).}
 \label{fig:124lattice}
\end{figure}

\begin{figure}[ht]
 \subfloat [Flavor polarization]
 {\includegraphics[height=0.45\linewidth,width=\linewidth]{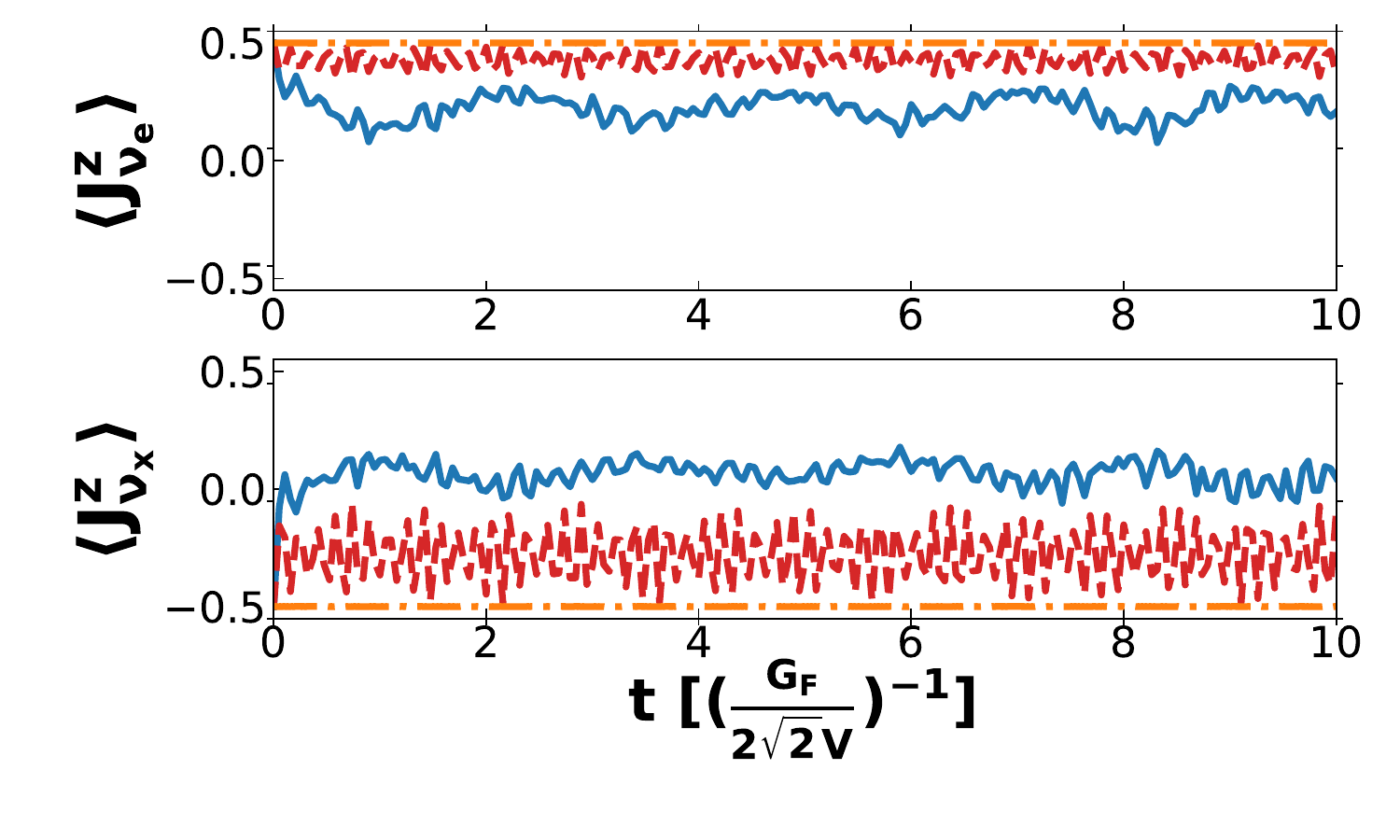}}\\
 \subfloat[Correlations]{\includegraphics[height=0.35\linewidth,width=\linewidth]{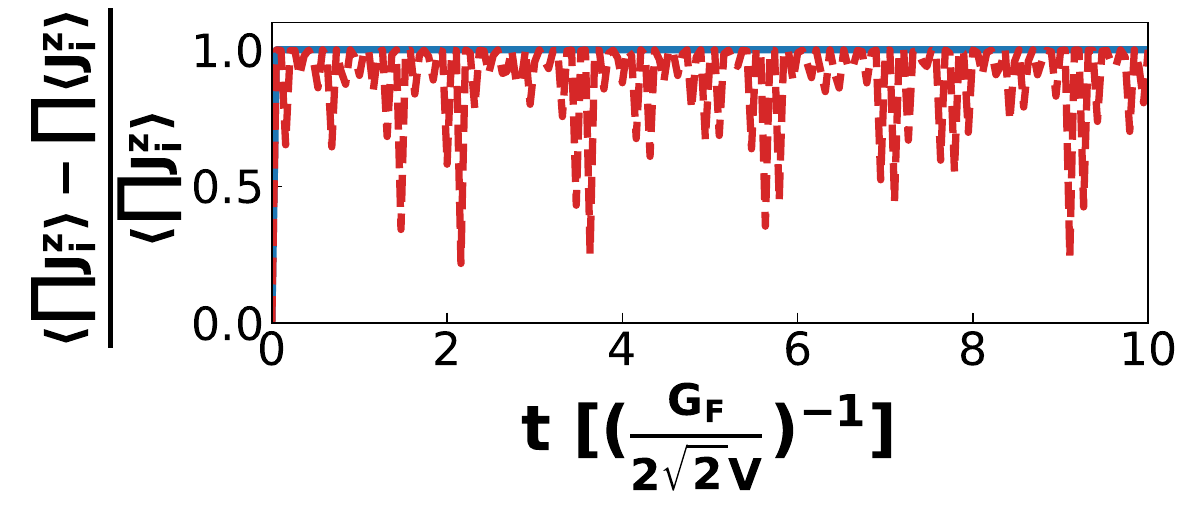}}
 \caption{Multi-angle (solid blue line), 
 single-angle (dashed red line), and mean field (dashed dotted orange line) flavor polarization and correlations for the lattice in Fig.~\ref{fig:124lattice}
 and initial wave-function $|\Psi_0\rangle= | \nu_e^{(12)} \nu_{x}^{(4)}\rangle$ in medium. These results were obtained based on eq.~(\ref{eq:psit}).} 
 \label{fig:124_10}
 \end{figure}
The second configuration has three times more electron neutrinos.
The flavor polarization and correlations depicted in Fig.~\ref{fig:124_10} show a similar trend; with oscillations not present in the mean field approximation 
and flavor equilibration in the many body calculation.

\section{System of twenty neutrinos}
\label{sec:20}
\begin{figure}[ht]
 \includegraphics[scale=0.22]{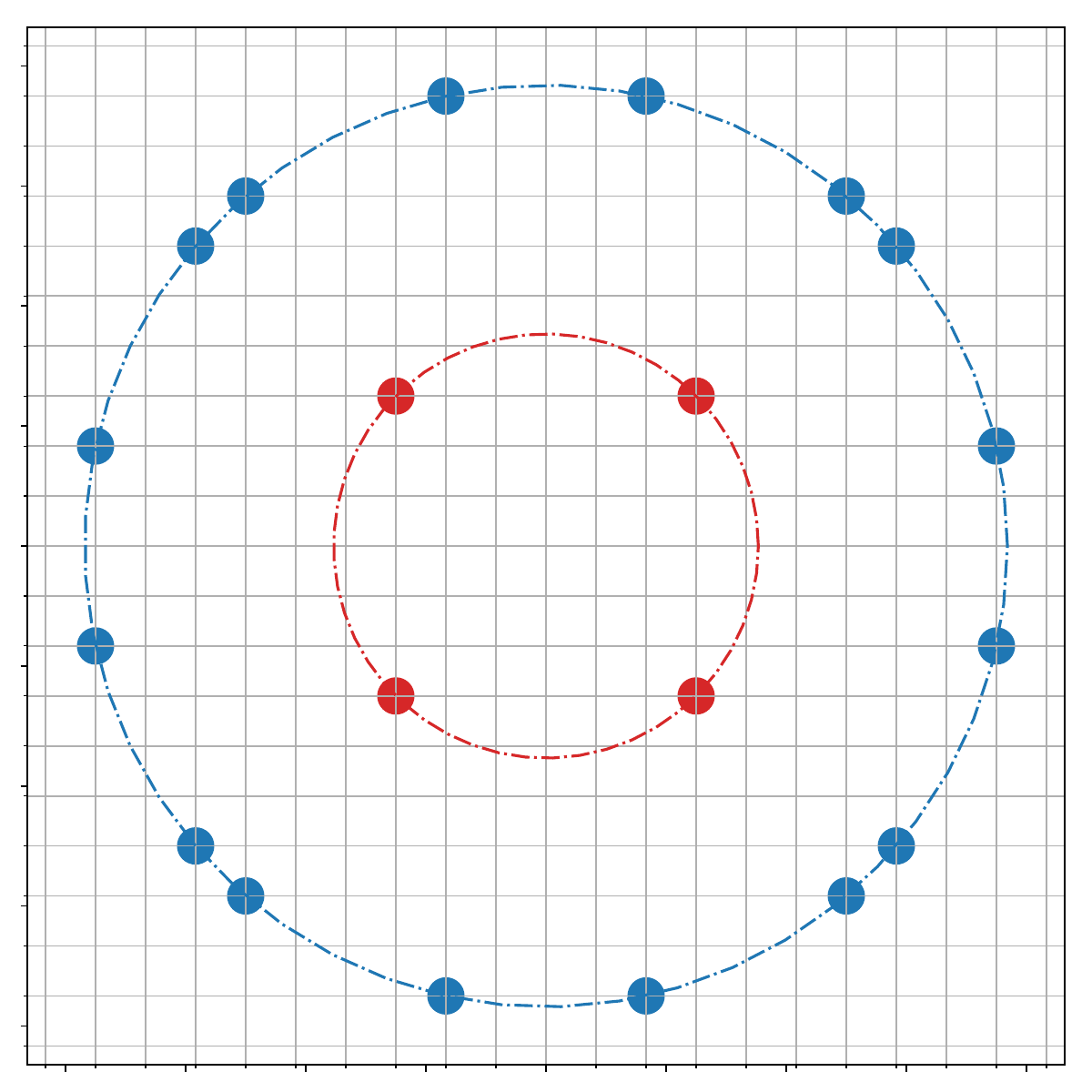}
 \caption{Configuration with initially four electron neutrinos (smaller lattice) and sixteen x flavor neutrinos (greater lattice).}
 \label{fig:20lattice}
\end{figure}
I conclude the many-body study with a system of twenty neutrinos. The configuration is shown in fig.~\ref{fig:20lattice} with four neutrinos on the smaller lattice and twelve on the larger lattice.
Similarly to previous sections, the ratio of the momenta magnitude is three. 

\begin{figure}[ht]
 \subfloat [Flavor polarization]
 {\includegraphics[height=0.45\linewidth,width=\linewidth]{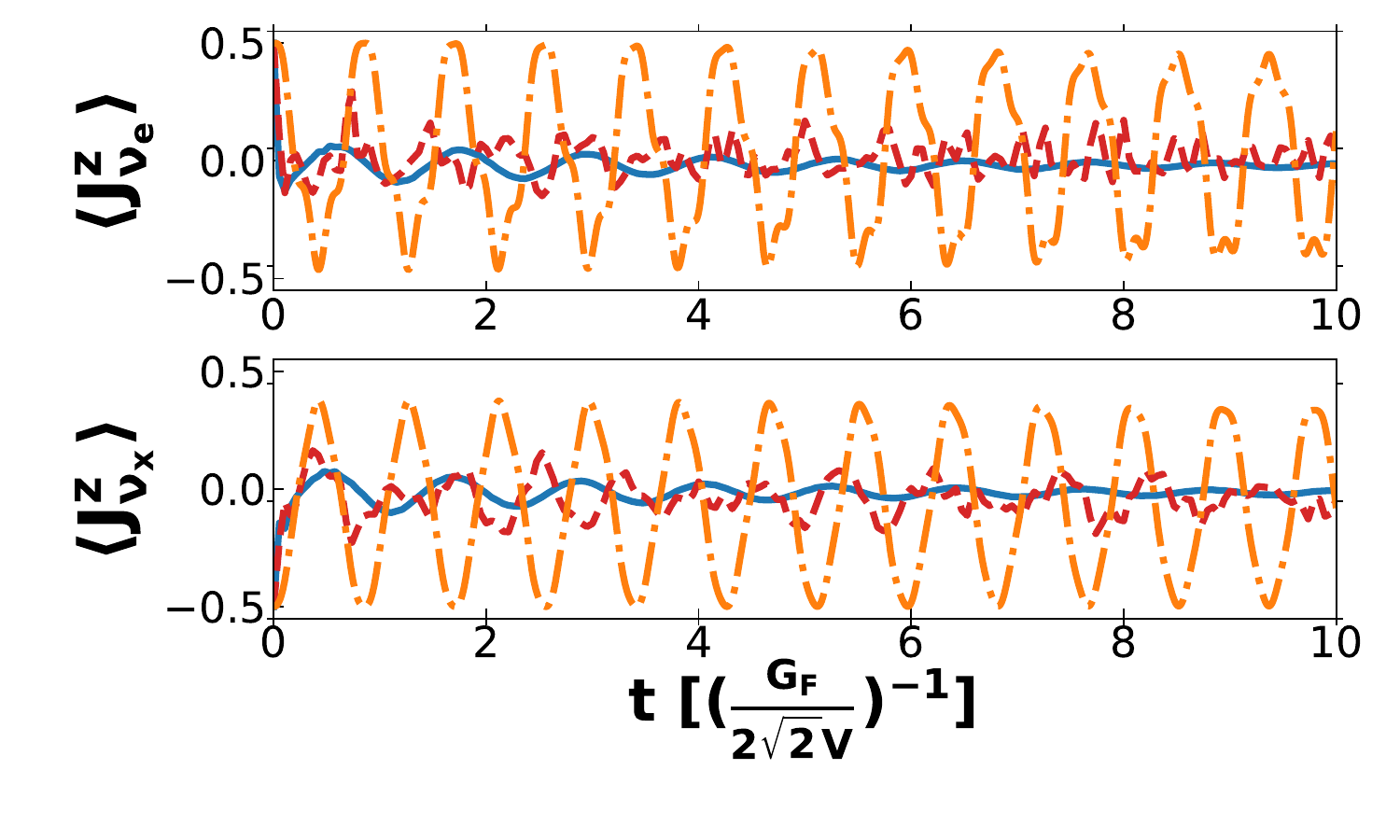}}\\
 \subfloat[Correlations]{\includegraphics[height=0.35\linewidth,width=\linewidth]{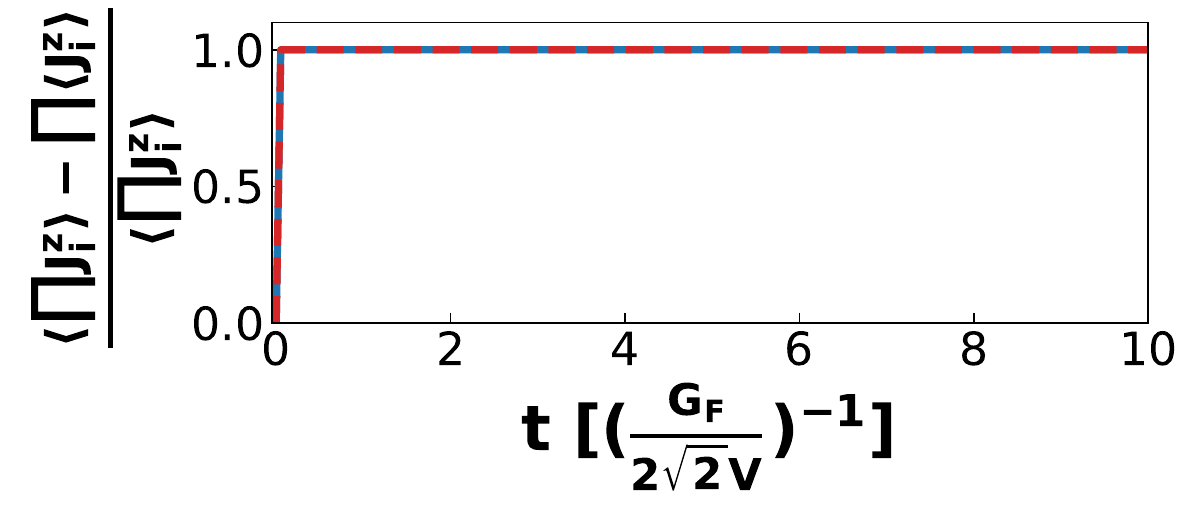}}
 \caption{Multi-angle (solid blue line), 
 single-angle (dashed red line), and mean field (dashed dotted orange line) flavor polarization and correlations for the 
 initial wave-function $|\Psi_0\rangle= | \nu_e^{(4)} \nu_{x}^{(16)}\rangle$ in vacuum with $ {\omega}/2= \mu/4 = \mu_0$.   These results were obtained based on eq.~(\ref{eq:psit}).}
 \label{fig:81210}
 \end{figure}

\begin{figure}[ht]
 \subfloat [Flavor polarization]
 {\includegraphics[height=0.45\linewidth,width=\linewidth]{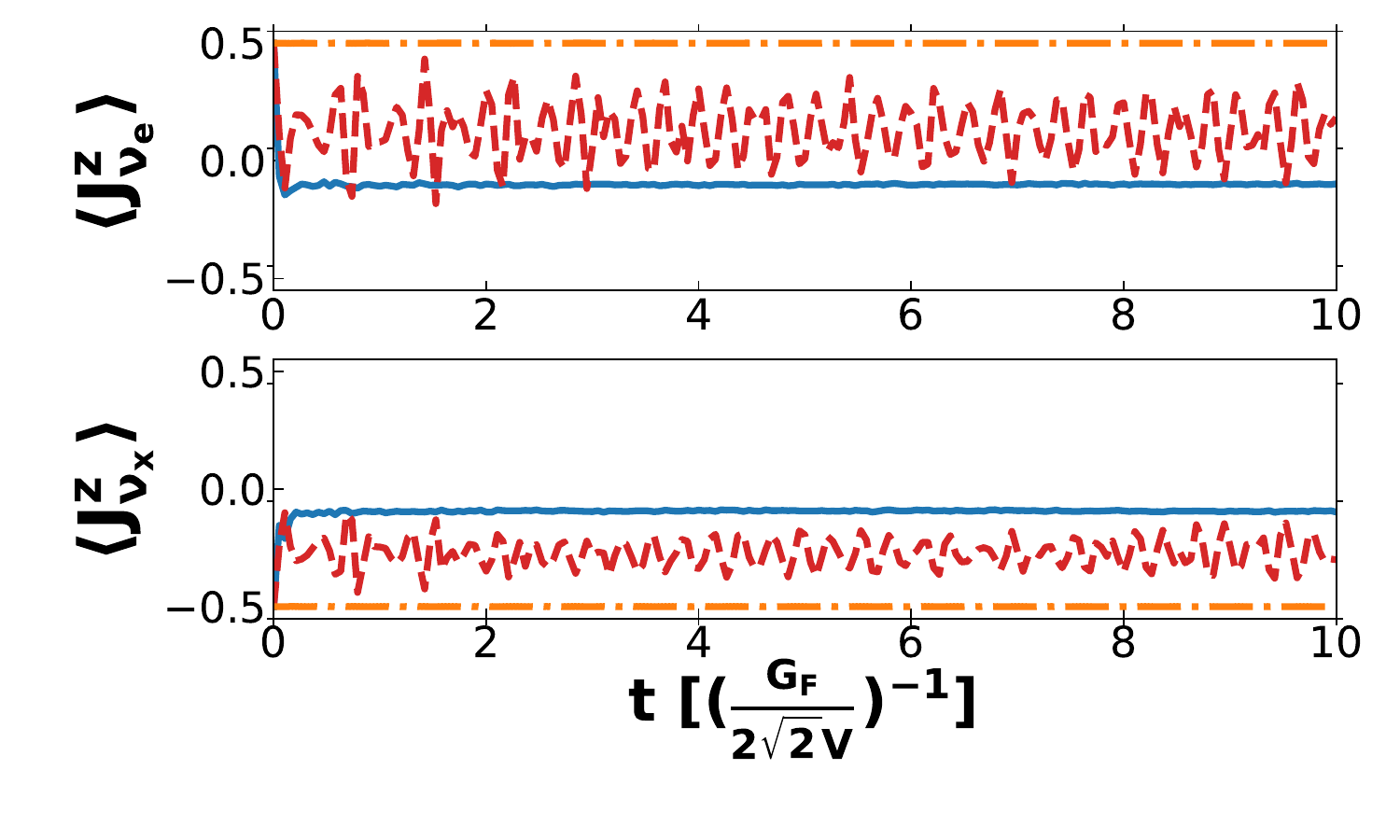}}\\
 \subfloat[Correlations]{\includegraphics[height=0.35\linewidth,width=\linewidth]{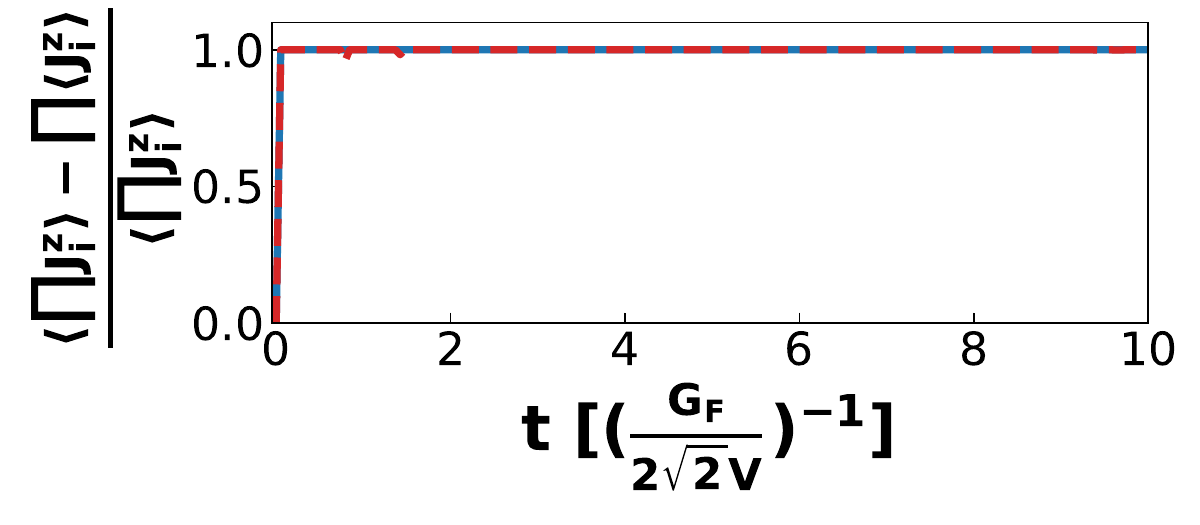}}
 \caption{Multi-angle (solid blue line), 
 single-angle (dashed red line), and mean field (dashed dotted orange line) flavor polarization and correlations for the 
 initial wave-function $|\Psi_0\rangle= | \nu_e^{(4)} \nu_{x}^{(16)}\rangle$ in medium. These results were obtained based on eq.~(\ref{eq:psit}).}
 \label{fig:81210_1}
 \end{figure}

Figure~\ref{fig:81210} shows the flavor polarization for twenty neutrinos in the lattice and overall correlation as functions of time in vacuum while fig.~\ref{fig:81210_1} displays the time evolution
 for the same initial conditions in a dense matter medium. The trend found in previous sections is also present here. One would expect the mean field to approach the many-body 
 results as the particle number is increasing, but this does not happen here. However, given the relatively small number of particles considered in this work, no statements can be made regarding 
 large particle number behavior. Larger lattices will be studied in future work.
 
\section{Entanglement Entropy}
\label{sec:ee}
From the configurations studied in this work a common feature emerges; whenever polarization correlations are significantly present, the mean field approximation fails to capture the time 
evolution described by the many-body method. To have a better understanding, in this section I study the entanglement entropy of each neutrino in the many-body wave-function versus the rest.
Since the initial wave-functions considered in this work are eigenstates of flavor, they are pure states. And, as they time evolve due to the Hamiltonian, they remain pure, 
although not eigenstates of flavor anymore. For such wave-functions,
the Von-Newman entanglement entropy serves as a measure of entanglement between the neutrinos. For a given density matrix it is defined as,
\begin{equation}
\begin{split}
 S=&-\text{Tr}\left( \rho \ln \rho \right),\
 \rho= |\Psi \rangle \langle \Psi |
\end{split}
 \end{equation}
Here I focus on a bipartition of the wave-function with one neutrino in one subspace, and the rest of the neutrinos in the other.
For an in depth description of entanglement and its various measures used in quantum computing, the reader is referred to \cite{Bengtsson:2006, Nielsen:2011, Rangamani:2015, Greenberger:2016}.
The bipartition chosen allows one to study whether one neutrino can be factored out from the rest of the wave-function. If this is possible, the single particle mean field should describe the many-particle 
system. 
The wave-function can be decomposed based on this partition,
\begin{equation}
 \begin{split}
  | \Psi_{\nu^{(N)}} \rangle =  | \nu_e \rangle \otimes \sum_i c_{e}{}_i | \Psi^i_{\nu^{(N-1)}} \rangle +| \nu_x \rangle \otimes \sum_i c_{x}{}_i | \Psi^i_{\nu^{(N-1)}} \rangle
 \end{split}
\end{equation}
The label $i$ denotes some basis of orthonormal wave-functions of $N-1$ neutrinos. 

The reduced density matrix is obtained by tracing out the $N-1$ neutrinos,
$\rho_{\nu}=\text{Tr}_{\Psi^i_{\nu^{(N-1)}}}\left(\rho_{\Psi_{\nu^{(N)}}} \right)$ and the respective entanglement entropy is $
  S_{\nu}=-\text{Tr}\left( \rho_{\nu} \ln \rho_{\nu} \right)$.
  This quantity is zero when there is no entanglement and $\ln(2)$ when the neutrino is maximally entangled with the other neutrinos in the wave-function. Thus, if $S_{\nu}=0$ one expects a single 
particle mean field description to be appropriate. As the entropy increases, the mean field result should get further away from the exact calculation. 
In appendix~\ref{app:cubic} in fig.~\ref{fig:811} I find both correlations and entanglement entropy vanish when the mean field approximation agrees with the many-body result. This happens when only one flavor is initially present.

\begin{figure}[ht]
\subfloat[$|\Psi_0\rangle= | \nu_e^{(4)} \nu_{x}^{(4)}\rangle$]
 {\includegraphics[height=0.44\linewidth,width=0.9\linewidth]{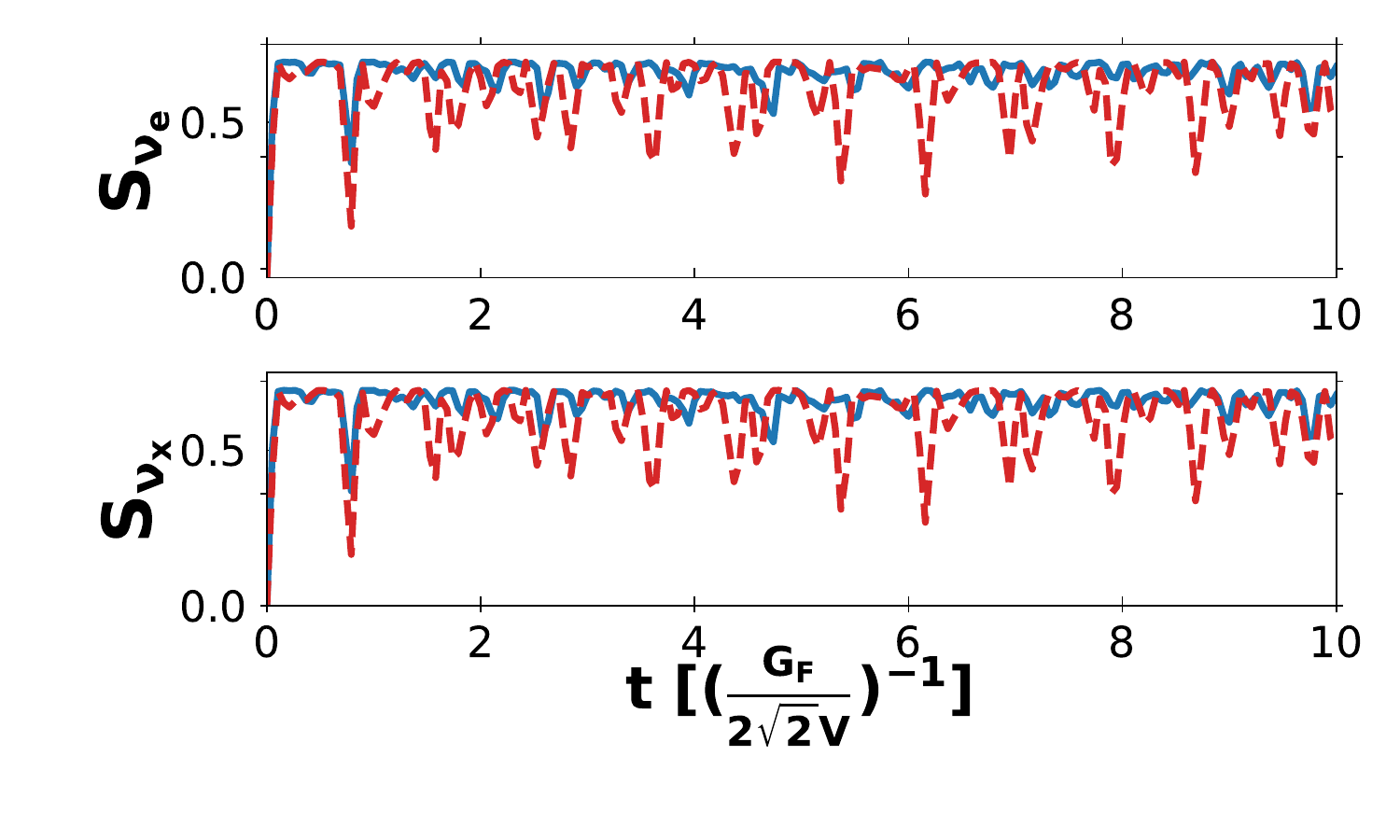}}\\
\subfloat[$|\Psi_0\rangle= | \nu_e^{(8)} \nu_{x}^{(4)}\rangle$]
 {\includegraphics[height=0.44\linewidth,width=0.9\linewidth]{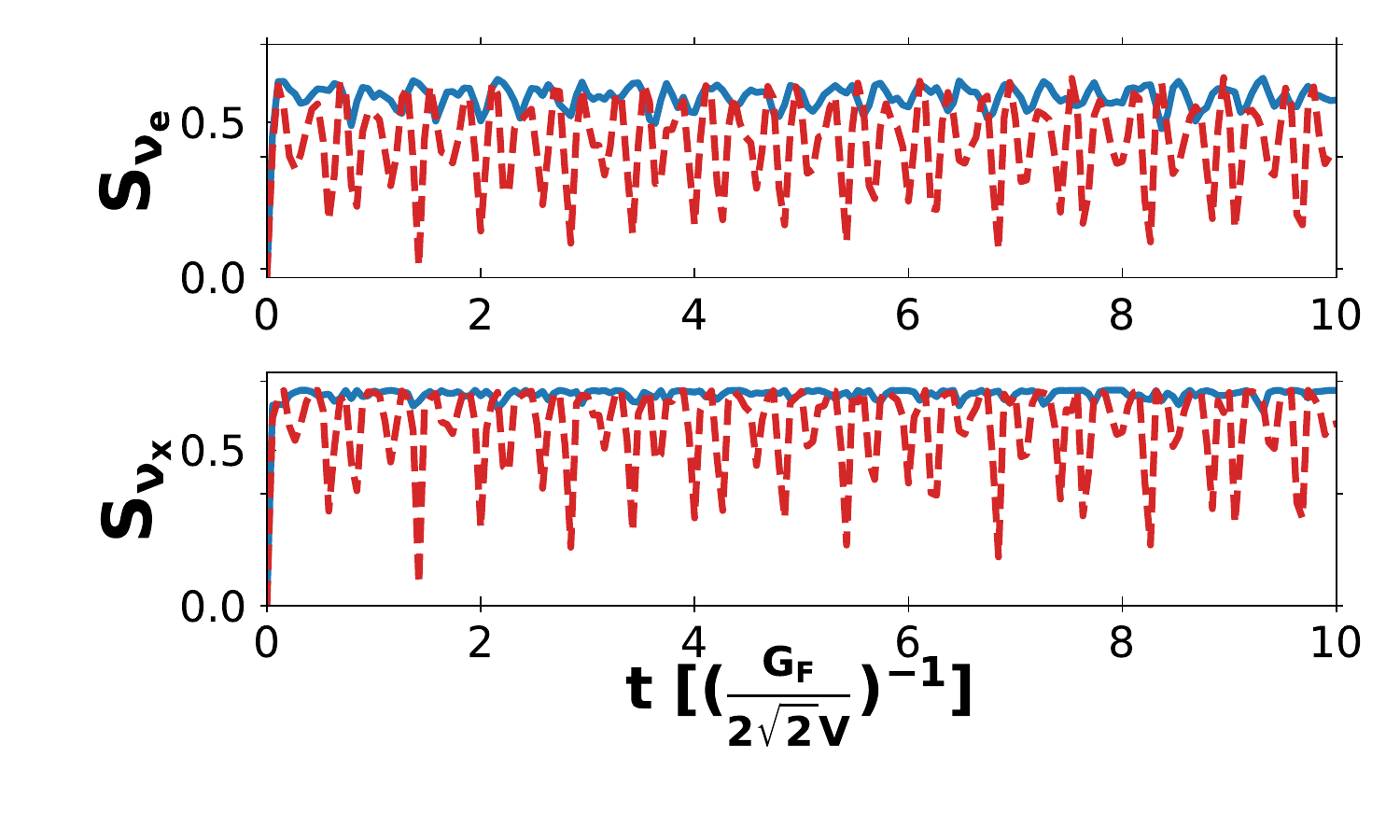}}\\
\subfloat[$|\Psi_0\rangle= | \nu_e^{(8)} \nu_{x}^{(8)}\rangle$]
 {\includegraphics[height=0.44\linewidth,width=0.9\linewidth]{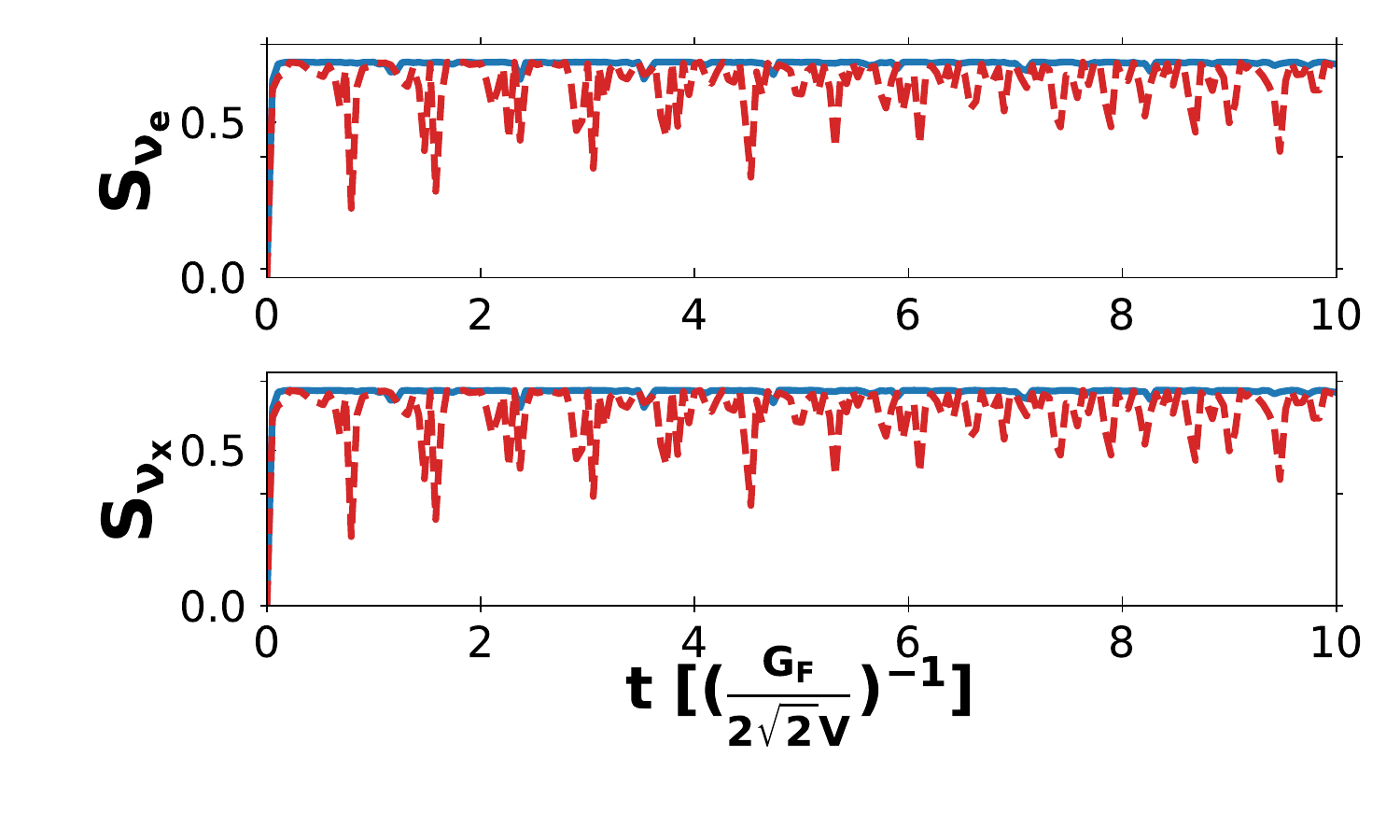}}\\
 \subfloat[$|\Psi_0\rangle= | \nu_e^{(4)} \nu_{x}^{(16)}\rangle$]
{\includegraphics[height=0.44\linewidth,width=0.9\linewidth]{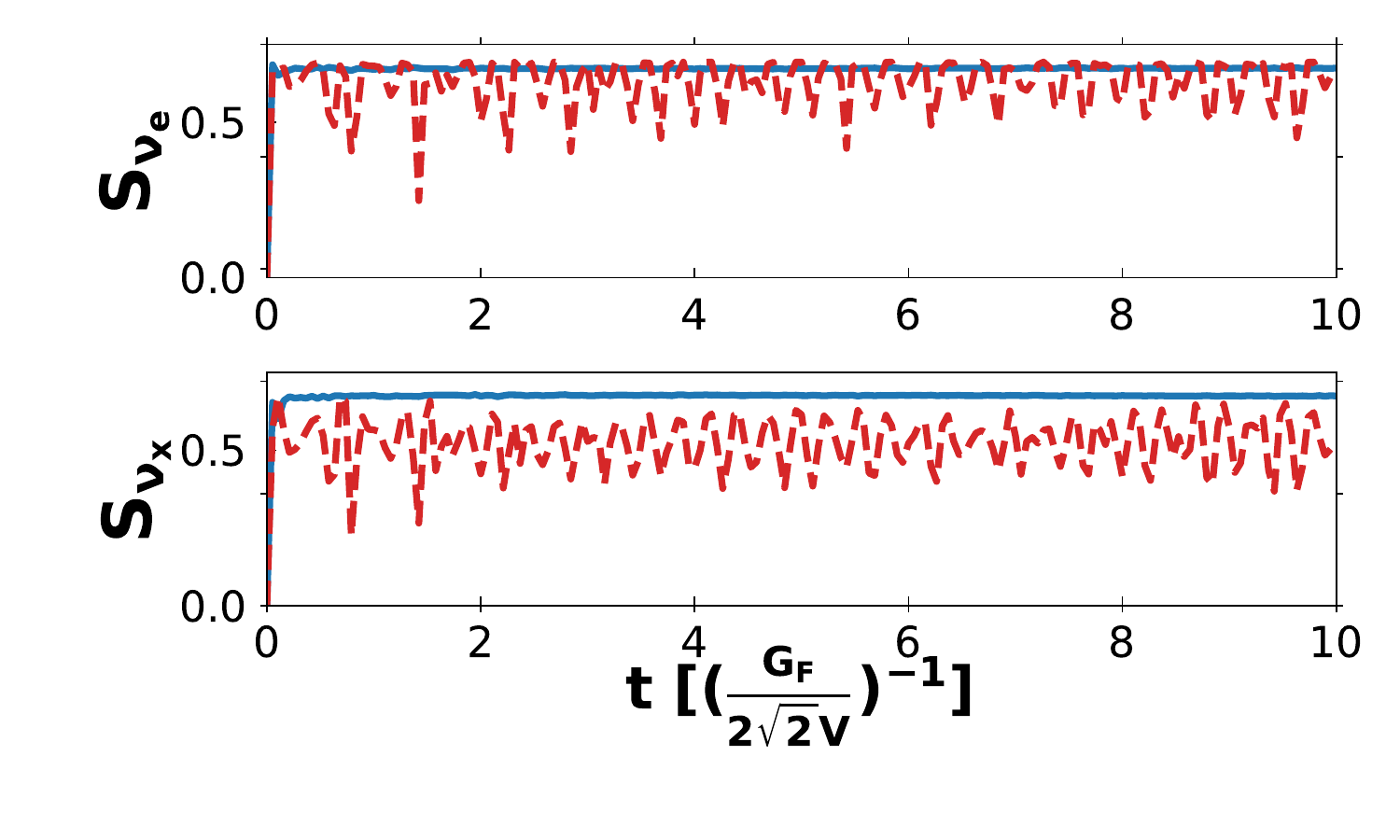}}\\
 \caption{Bipartite entanglement entropy for the individual neutrinos in medium for the configurations described in the previous sections. These results were obtained based on eq.~(\ref{eq:psit}).}
 \label{fig:8e8xee}
\end{figure}

When both flavors are initially present, correlations and entanglement develop. For two neutrino beams, one can obtain analytical results:
\begin{equation}
 \begin{split}
  S_{\nu_e}^{\text{multi-angle}}(t)=& S_{\nu_x}^{\text{multi-angle}}(t)\\
  =&\frac{1}{2} \bigg[ \log \left(\frac{1}{4} \sin ^2(2 \mu (1-\cos \vartheta_{\bold{1\bold{2}}}  ) t)\right)\\
  &+\left| \cos (2 \mu (1-\cos \vartheta_{\bold{1}\bold{2}}) t)\right|  \\
  &\times \log \left(\frac{1+ \left| \cos (2 \mu (1-\vartheta_{\bold{1}\bold{2}}) t)\right|}{1-\left| \cos (2 \mu  (1-\cos \vartheta_{\bold{1}\bold{2}}) t)\right|}\right)\bigg]\\
  S_{\nu_e}^{\text{single-angle}}(t)=& S_{\nu_x}^{\text{single-angle}}(t)\\
  =&\frac{1}{2} \bigg[ \log \left(\frac{1}{4} \sin ^2( \mu (1-\cos \vartheta_{\bold{1\bold{2}}}  ) t)\right)\\
  &+\left| \cos ( \mu (1-\cos \vartheta_{\bold{1}\bold{2}}) t)\right|  \\
  &\times \log \left(\frac{1+ \left| \cos ( \mu (1-\vartheta_{\bold{1}\bold{2}}) t)\right|}{1-\left| \cos ( \mu  (1-\cos \vartheta_{\bold{1}\bold{2}}) t)\right|}\right)\bigg]\\
 \end{split}
\end{equation}
For all lattices studied in this work, when both flavors are initially present, 
I find that neutrinos get rapidly entangled. Figure~\ref{fig:8e8xee} displays
the entanglement entropy for systems of eight, twelve, sixteen and twenty particles as function of time for initial electron and x flavor respectively. 
In \cite{Friedland:2003} the authors find that entanglement does not develop when only the single-angle two body term in the Hamiltonian is considered. Later, in \cite{Bell:2003}, entanglement was found for that 
Hamiltonian when systems of fourteen particles were studied. In the case  at hand, I am studying the full multi-angle Hamiltonian and I find entanglement when both flavors 
are initially present. Due to the fact that the Hamiltonian dimensions increase exponentially with the particle number, it is very computationally expensive to study much larger systems.
While the main focus of this work is the multi-angle Hamiltonian, it is worth pondering what happens for larger and larger systems.
A simple assumption would be that correlations would decrease as the number of particles increases, thus entanglement would decrease as well. 
Unfortunately, if this question is posed for the transverse Ising spin chain the answer contradicts this assumption. In \cite{Calabrese:2005} the authors find that the entanglement entropy
increases with time until it reaches a maximal value. This value of the entanglement depends on the initial state and does not decrease and the particle number increases. 
A conclusive statement about the multi-angle Hamiltonian can be made once larger systems are studied, which is to pursued in the future. In the meantime, I will focus on the effects of the two-body operator only,
and for simplification, I will consider the single-angle approximation and compare it to the mean field result in the next section.

\section{Infinitely large system with only neutrino-neutrino interactions}
\label{sec:inf_two_body}
\subsection{Many-body treatment}
In this section, in order to study infinitely large systems, I will ignore the mass term in the Hamiltonian, and study  the system in the single-angle approximation. 
This is the Hamiltonian studied in~\cite{Friedland:2003eh},
\begin{equation*}
 \begin{split}
  H& = \mu \sum_{i,j} \bold{J}_i \cdot \bold{J}_j
  = \mu (\sum_i \bold{J}_i)^2
  = \mu \bold{J}^2,\\
  H& | J, m_j \rangle = \mu J(J+1) | J, m_j \rangle=E_J| J, m_j \rangle,
 \end{split}
\end{equation*}
where,
\begin{equation*}
 \begin{split}
  | \nu_e \rangle = | \frac{1}{2}, \frac{1}{2} \rangle,
  | \nu_x \rangle = | \frac{1}{2}, -\frac{1}{2}\rangle.
 \end{split}
\end{equation*}
Here I follow the same steps as in~\cite{Friedland:2003eh}, and proceed to calculate the polarization and entanglement entropy as functions of time.
The initial wave-function chosen is 
\begin{equation*}
 | \Psi_0 \rangle = | \nu_e^{(N)} \nu_x^{(N)} \rangle =| \frac{N}{2}, \frac{N}{2}\rangle \otimes| \frac{N}{2}, -\frac{N}{2}\rangle.
\end{equation*}

Based on~\cite{Friedland:2003eh} one can express the wave-function at a given time as a direct product of a single neutrino state and the rest,
\begin{equation*}
 \begin{split}
  &| \Psi(t) \rangle =e^{-i Ht} | \Psi_0 \rangle\\
  =& |\nu_e \rangle \otimes \sum_{J=0}^{N-1} \frac{e^{-i E_Jt}+e^{-i E_{J+1}t}}{2} \eta(N,J) |J+\frac{1}{2}, - \frac{1}{2}; (2N-1) \rangle \\
   +& |\nu_x \rangle \otimes \sum_{J=0}^{N-1} (-1)^{N-J}\frac{e^{-i E_Jt}-e^{-i E_{J+1}t}}{2} \eta(N,J) |J+\frac{1}{2},  \frac{1}{2}; (2N-1)\rangle
 \end{split}
\end{equation*}
where $\eta(N,J)$ is the Clebsch-Gordan coefficient~\cite{Baird:1964},
\begin{equation*}
\begin{split}
 \eta(N,J) =& \langle \frac{N-1}{2},\frac{N-1}{2}; \frac{N}{2}, -\frac{N}{2}| J + \frac{1}{2}, -\frac{1}{2} \rangle\\
 =&\sqrt{2(J+1)}\sqrt{\frac{(N-1)! N!}{(-J+N-1)! (J+N+1)!}}
 \end{split}
\end{equation*}
The reduced density matrix is obtained by tracing over the subspace of $2N-1$ neutrinos,
\begin{equation}
\begin{split}
 \rho_{\nu} =& \left( \sum_{J=0}^{N-1}\frac{\eta^2(N,J)}{2} \big[1+\cos(2 \mu (J+1)t)\big] \right)|\nu_e \rangle \langle \nu_e | \\
 +& \left( \sum_{J=0}^{N-1}\frac{\eta^2(N,J)}{2} \big[1-\cos(2 \mu (J+1)t)\big] \right) |\nu_x \rangle \langle \nu_x |\\
 =& \frac{1}{2}[1+A(N,\mu,t)] |\nu_e \rangle \langle \nu_e | +\frac{1}{2}[1-A(N,\mu,t)] |\nu_x \rangle \langle \nu_x | 
\end{split}
\end{equation}
The function $A(N,\mu,t)$ is given as follows,
\begin{equation}
\begin{split}
 A(N,\mu,t)=& \frac{1}{(N+1)}  \big[e^{-2 i \mu t}\ {}_2{F}_1\left(2,1-N;N+2;-e^{-2 i \mu t}\right)\\
 +&e^{2 i \mu t} \, _2{F}_1\left(2,1-N;N+2;-e^{2 i \mu t}\right)\big]
\end{split}
 \end{equation}
where ${}_2{F}_1\left(a,b;c;z\right)$ is the Gauss hyper-geometric function.
The expression in front of $|\nu_e \rangle \langle \nu_e |$ is the so called survival probability for the neutrino to remain in the initial flavor~\cite{Friedland:2003eh},
while the expression in front of $|\nu_x \rangle \langle \nu_x |$ is the probability for it to change flavor. In the full Hamiltonian $\bold{B} \cdot \bold{J}$ is a constant of motion
while for this Hamiltonian $J_z$ is. This means that the coefficients for  $|\nu_e \rangle \langle \nu_x |$ and $|\nu_x \rangle \langle \nu_e |$ are zero in this case by symmetry, but not in the full
Hamiltonian.
The entanglement entropy is,
\begin{equation}
 \begin{split}
  S_{\nu_e}=&  \ln 2 + \frac{A(N,\mu,t)}{2} \ln \left( \frac{1-A(N,\mu,t)}{1+A(N,\mu,t)} \right).
 \end{split}
\end{equation}
In addition, one can also study the polarization of individual neutrinos from the expressions above,
\begin{equation}
 \begin{split}
  P^z_{\nu_e}(t)=-P^z_{\nu_x}(t)= \frac{A(N,\mu,t)}{2}
  \label{eq:mn1k}
 \end{split}
\end{equation}
The complete derivation of $| \Psi(t) \rangle$ can be found in appendix~\ref{app:mntb}. The derivation of the survival probability has already been performed in~\cite{Friedland:2003eh}.
The authors did not study entanglement entropy or compare the polarization from the exact solution with the mean field in that article. 
Their focus was on the time scale associated with flavor equilibration, and found it to be $\tau^{-1} \propto \mu_0 \sqrt{N}$. As $\mu_0$ depends on the normalization volume $V$, the authors took $V \propto \text{cm}^{3}$,
and concluded that the time scale was too large to be significant for supernovae ($\tau \propto 10^{22}$ s). However, one could argue that this quantity should be related to the neutrino density, $V^{-1} \propto n_{\nu}$. 
If $\frac{n_{\nu}}{n_0}\propto 10^{-10}$, where $n_0=0.16\ \text{fm}^{-3}$ is the nuclear saturation density which is reached at the core of proto-neutron stars, the time scale becomes relevant ($\tau \propto 10^{-6}$ s). As the goal of this work is the mathematical solution for the time evolution 
due to the multi-angle Hamiltonian, I refrain from making any conclusive statements as to which choice is appropriate. I proceed in the next subsection to compare to the mean field result for the same Hamiltonian and 
initial wave-function.

\subsection{Mean Field method}
\begin{figure}[ht]
 \subfloat [Flavor polarization]
 {\includegraphics[height=0.5\linewidth,width=\linewidth]{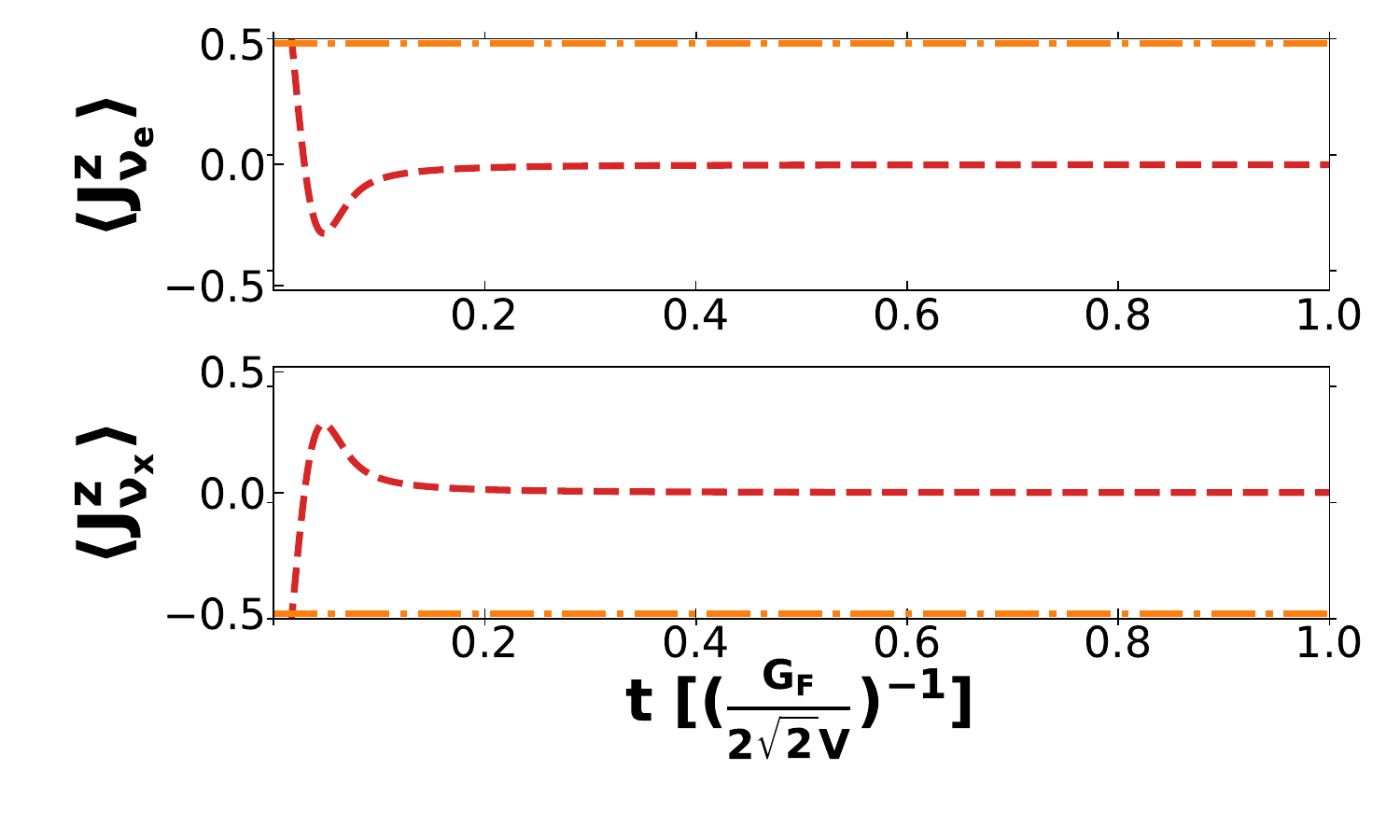}}\\
 \subfloat[Correlations]{\includegraphics[height=0.5\linewidth,width=\linewidth]{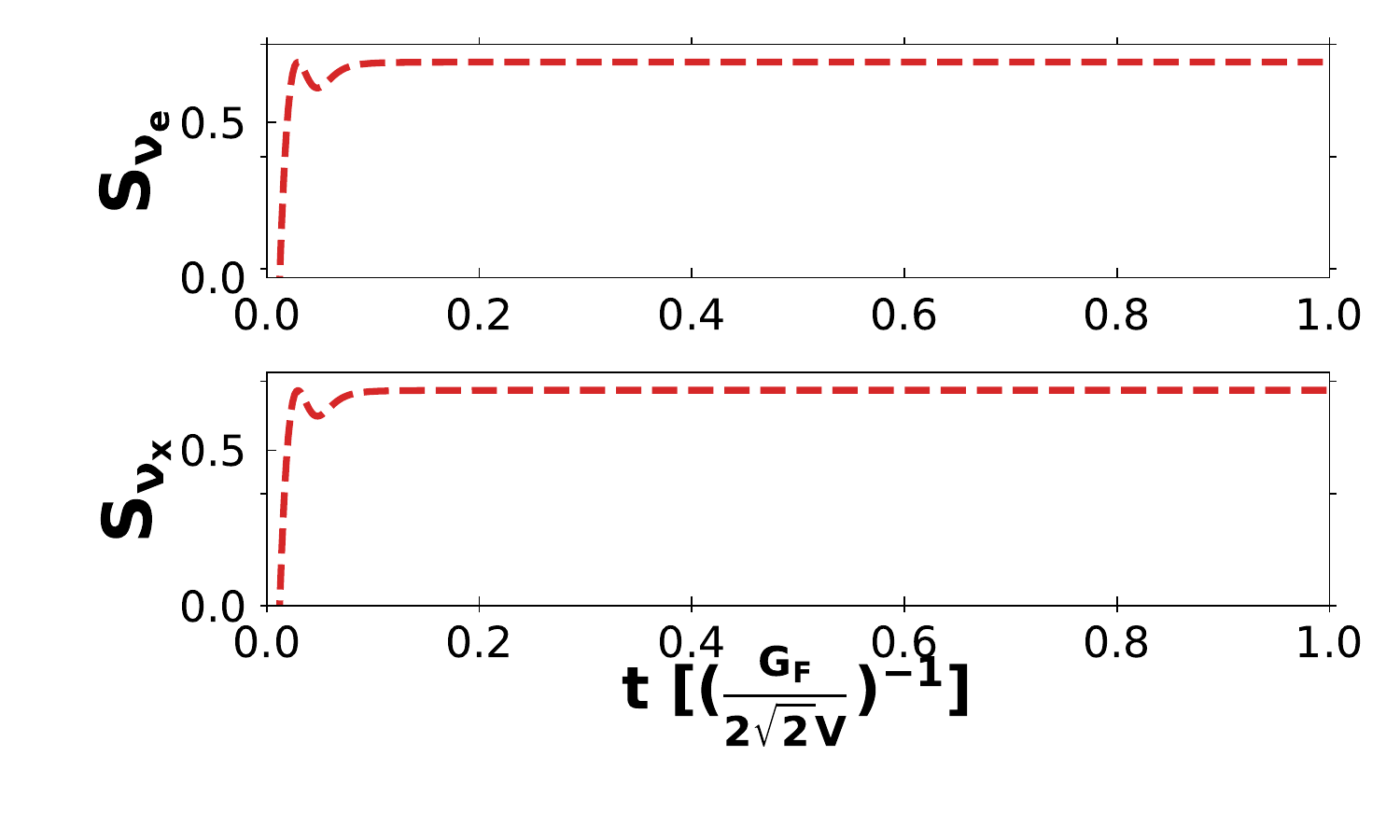}}
 \caption{Single-angle (dashed red line), and mean field (dashed dotted orange line) flavor polarization and bipartite entanglement entropy for the 
 initial wave-function $|\Psi_0\rangle= | \nu_e^{(500)} \nu_{x}^{(500)}\rangle$ in vacuum with $\mu/4 = \mu_0$.}
 \label{fig:500500_01}
 \end{figure}

The equations of motion are derived from eq.~\ref{eq:mf2} by setting $\omega_p =0$,
\begin{equation}
 d_t \bold{P}_{\bold{p}}= \mu_0 \sum_{\bold{q}} \bold{P}_{\bold{q}} \times \bold{P}_{\bold{p}}.
 \label{eq:mf2N}
\end{equation}
In fig.~\ref{fig:500500_01} I plot the polarization and the entanglement entropy as functions of time for the system of 1000 neutrinos. The many body result is provided by the analytical expression if eq.~\ref{eq:mn1k} and the mean field is solved 
numerically. The qualitative difference between the many-body result and the mean field is quite striking. The mean field shows no oscillations at all, while the exact result shows rapid flavor equilibration. This
is agreement with all the configurations studied so far. The lack of oscillations in the vacuum predicted by the mean field can be explained by comparing this system with the two particle system in section~\ref{subec:mf}.
All the 500 initial electron neutrinos are equivalent for the given Hamiltonian, and the same can be said for the x flavor ones. The system of equations in ~\ref{eq:mn1k}, then, can be written in terms of the sum of the polarizations,
and the difference between electron and x flavor. From section~\ref{subec:mf}, the equations of motion for the 1000 neutrino system can be written as,
\begin{equation}
 \begin{split}
 d_t \bold{P}_{\text{tot}}= d_t \bold{P}_{\text{diff}} = 0.
 \end{split}
\end{equation}
The mass term, responsible for the time evolution of these two quantities is zero, so they are both constants of motion. 
No matter how large the system, if the initial wave-function has equal numbers of both flavors, the mean
field will show no oscillation,
\begin{equation}
 P^{\text{(mean-field)}}{}^z_{\nu_e}(t)=-P^{\text{(mean-field)}}{}^z_{\nu_x}(t)=\frac{1}{2}.
\end{equation}
On the other hand, $\lim_{t \rightarrow \infty}F(\sqrt{N}\mu t)=0$, so
\begin{equation}
\lim_{t \rightarrow \infty}P^z_{\nu_e}(t)=\lim_{t \rightarrow \infty}P^z_{\nu_x}(t)=0.
\end{equation}
Actually, the asymptotic value is reached very quickly as fig.~\ref{fig:500500_01} shows. Not surprisingly, neutrinos get rapidly entangled with each other and this quantity reaches its maximum 
rather quickly as shown in the same plot. This result agrees qualitatively with what I have found in the previous sections. 

\section{Conclusion}
\label{sec:conc}
Neutrino flavor oscillations are important for the thermodynamic evolution of core collapse supernovae and neutron star mergers, and also impact the nucleosynthesis of heavy elements in these
environments. Due to the complicated nature of this many-body system, the mean field approximation is widely used in numerical simulations. However, it needs to be
compared to an exact many-body treatment in order to assess its region of validity. By considering uniform matter and discretized momentum space, 
I am able to solve exactly the time evolution of flavor oscillations. 
I provide analytical results for the two neutrino beam scenario and numerical results for cubic and quadratic lattices of eight, twelve, sixteen and twenty particles.
I find that when both flavors are present, fast collective oscillations develop and there is flavor equilibration but these are not observed in the mean field treatment.
This qualitative difference can be ascribed to correlations developing among neutrinos for the mixed flavor scenario as neutrinos rapidly entangle in time units $\mu_0^{-1}=(\frac{G_F}{2\sqrt{2}V})^{-1}$.
When only one flavor is present, correlations vanish, there is no entanglement, oscillations vanish, and the mean field agrees with the exact solution. While 
the single-angle approximation on the lattice does show fast collective oscillations, there are differences from the multi-angle result 
when the two neutrino flavors are initially present but not in equal amounts. To confirm my findings I calculate the flavor polarization and entanglement entropy in the large $N$ limit by studying only the
two-body contribution in the single-angle approximation based on~\cite{Friedland:2003eh}. 

The implications for supernovae might be quite interesting as the neutrino mean free path due to weak reactions could be larger than the mean free path for collective oscillations depending on the 
choice of the normalization volume $V$.
To give a more definitive answer, anti-neutrinos must be included and larger lattices need to be considered. I plan to follow up on this in future work.

\begin{acknowledgments}
I thank Andre Sieverding, Amol Patwardhan, and Manibrata Sen for the useful discussions. Particular thanks go to Yong-Zhong Qian, Joseph Kapusta and A. B. Balantekin for their advice. 
No man is an island, and the opportunity to consult these experts does not come by easily. 
This work was supported in part by the NSF (PHY-1630782), and the Heising-Simons Foundation (2017-228).  
\end{acknowledgments}
\appendix
\onecolumngrid
\section{Time evolution of flavor polarization}
\label{ap:dtjk}
Equation~(\ref{eq:dtJk}) is derived by employing the commutation relations of the Pauli matrices,
\begin{equation}
 \begin{split}
  -i d_t J^{k} =&[H,J^k]\\
  =&[\sum_{\bold{p}_1,\bold{q}_1} \frac{\overline{\omega}_{p_1}}{2N_{\nu}} \left(\sin(\alpha)\sigma^x_{\bold{p}_1} \sigma^0_{\bold{q}_1} - \cos(\alpha) \sigma^z_{\bold{p}_1} \sigma^0_{\bold{q}_1}\right) 
  + \frac{\mu}{4} (1-\cos \vartheta_{\bold{p}_1\bold{q}_1}) (\sigma^x_{\bold{p}_1} \sigma^x_{\bold{q}_1}
  +\sigma^y_{\bold{p}_1} \sigma^y_{\bold{q}_1}+\sigma^z_{\bold{p}_1} \sigma^z_{\bold{q}_1}),\frac{1}{2}\sum_{\bold{p}_2,\bold{q}_2}\sigma^{k}_{\bold{p}_2}\sigma^0_{\bold{q}_2}]\\
  =&\frac{1}{2} \sum_{\bold{p}_1,\bold{q}_1} \frac{\overline{\omega}_{p_1}}{2N_{\nu}} \left(\sin(\alpha)[\sigma^x_{\bold{p}_1},\sigma^k_{\bold{p}_1}] \sigma^0_{\bold{q}_1} 
  - \cos(\alpha) [\sigma^z_{\bold{p}_1},\sigma^k_{\bold{p}_1}] \sigma^0_{\bold{q}_1}\right) + \frac{\mu}{4} (1-\cos \vartheta_{\bold{p}_1\bold{q}_1}) ([\sigma^x_{\bold{p}_1},\sigma^k_{\bold{p}_1}] \sigma^x_{\bold{q}_1}
  +[\sigma^y_{\bold{p}_1},\sigma^k_{\bold{p}_1}] \sigma^y_{\bold{q}_1}+[\sigma^z_{\bold{p}_1},\sigma^k_{\bold{p}_1}] \sigma^z_{\bold{q}_1})\\
  =& i \sum_{\bold{p}_1,\bold{q}_1} \frac{\overline{\omega}_{p_1}}{2N_{\nu}} \left(\sin(\alpha) \epsilon^{1kj}\sigma^j_{\bold{p}_1} \sigma^0_{\bold{q}_1} - \cos(\alpha) \epsilon^{3kj}\sigma^j_{\bold{p}_1} \sigma^0_{\bold{q}_1}\right)
  + \frac{\mu}{4} (1-\cos \vartheta_{\bold{p}_1\bold{q}_1}) \left(\epsilon^{1kj} \sigma^j_{\bold{p}_1} \sigma^x_{\bold{q}_1}
  +\epsilon^{2kj} \sigma^j_{\bold{p}_1} \sigma^y_{\bold{q}_1}+\epsilon^{3kj}\sigma^j_{\bold{p}_1} \sigma^z_{\bold{q}_1}\right)\\
  =& i \sum_{\bold{p}_1,\bold{q}_1} \frac{\overline{\omega}_{p_1}}{2N_{\nu}} \left(\sin(\alpha) \epsilon^{1kj}\sigma^j_{\bold{p}_1} \sigma^0_{\bold{q}_1} - \cos(\alpha) \epsilon^{3kj}\sigma^j_{\bold{p}_1} \sigma^0_{\bold{q}_1}\right)\\
  =&-i \sum_{\bold{p}} \overline{\omega}_{p} \left(\sin(\alpha) \epsilon^{k1j}J^j_{\bold{p}} - \cos(\alpha) \epsilon^{k3j}J^j_{\bold{p}} \right)\\
  =& -i \sum_{\bold{p}} \overline{\omega}_{p} \left( \bold{B} \times \bold{J}_{\bold{p}}\right)^k  \rightarrow \\
 d_t \bold{J} =& \sum_{\bold{p}} \overline{\omega}_{p} \left( \bold{B} \times \bold{J}_{\bold{p}}\right) 
  \end{split}
\end{equation}
where $\sigma^0_{\bold{p}}$ is the identity matrix at momentum $\bold{p}$. Similarly, for the individual lattice point I find
\begin{equation}
 \begin{split}
   d_t J^k_p
  =&i [\sum_{\bold{p}_1,\bold{q}_1} \frac{\overline{\omega}_{p_1}}{2N_{\nu}} \left(\sin(\alpha)\sigma^x_{\bold{p}_1} \sigma^0_{\bold{q}_1} - \cos(\alpha) \sigma^z_{\bold{p}_1} \sigma^0_{\bold{q}_1}\right) 
  + \frac{\mu}{4} (1-\cos \vartheta_{\bold{p}_1,\bold{q}_1}) (\sigma^x_{\bold{p}_1} \sigma^x_{\bold{q}_1}
  +\sigma^y_{\bold{p}_1} \sigma^y_{\bold{q}_1}+\sigma^z_{\bold{p}_1} \sigma^z_{\bold{q}_1}),\frac{1}{2}\sum_{\bold{q}}\sigma^{k}_{\bold{p}}\sigma^0_{\bold{q}}]\\
 =& -\frac{\overline{\omega}_{p}}{N_{\nu}} \sum_{\bold{q}_1}  \left(\sin(\alpha) \epsilon^{1kj}J^j_{\bold{p}} \sigma^0_{\bold{q}_1} - \cos(\alpha) \epsilon^{3kj}J^j_{\bold{p}} \sigma^0_{\bold{q}_1}\right) 
 -\mu \sum_{\bold{q}_1} (1-\cos\vartheta_{\bold{p},\bold{q}_1}) \left(\epsilon^{1kj} J^x_{\bold{p}} J^j_{\bold{q}_1}
  +\epsilon^{2kj} J^y_{\bold{p}} J^j_{\bold{q}_1}+\epsilon^{3kj}J^z_{\bold{p}} J^j_{\bold{q}_1}\right)\\
 =&\overline{\omega}_{p}   \left(\sin(\alpha) \epsilon^{k1j}J^j_{\bold{p}} - \cos(\alpha) \epsilon^{k3j}J^j_{\bold{p}} \right) 
 +\mu \sum_{\bold{q}} (1-\cos\vartheta_{\bold{p},\bold{q}}) \left(\epsilon^{k1j} J^x_{\bold{p}} J^j_{\bold{q}}
  +\epsilon^{k2j} J^y_{\bold{p}} J^j_{\bold{q}}+\epsilon^{k3j}J^z_{\bold{p}} J^j_{\bold{q}}\right)\\
  =&\overline{\omega}_p \left( \bold{B} \times \bold{J}_{\bold{p}}\right)^k+\mu \sum_{\bold{q}}(1-\cos\vartheta_{\bold{p},\bold{q}}) \left( \bold{J}_{\bold{q}} \times \bold{J}_{\bold{p}} \right)^k
 \end{split}
\end{equation}

\section{Exact Time Evolution for two neutrino beams}
\label{ap:twobeams}
The elements of the Dirac basis are Kronecker products of Pauli matrices. 
Using the bilinear property of the Kronecker product~\cite{graham2018kronecker} and the commutation identities of the Pauli matrices,
\begin{equation}
 \begin{split}
  \left[\sigma^{ij},\sigma^{kl}\right]=&\left[ \sigma^i \otimes  \sigma^j,\sigma^k \otimes \sigma^l \right] \\
  =& \sigma^i \sigma^k \otimes \sigma^j \sigma^l -   \\
  =& \left( \sigma^i \sigma^k \otimes \sigma^j \sigma^l- \sigma^i \sigma^k \otimes \sigma^l \sigma^j \right) + \left( \sigma^i \sigma^k \otimes \sigma^l \sigma^j - \sigma^k \sigma^i \otimes \sigma^l \sigma^j \right)\\
  =& \sigma^i \sigma^k \otimes \left[\sigma^j,\sigma^l \right] + \left[\sigma^i, \sigma^k \right] \otimes \sigma^l \sigma^j\\ 
  =& \left( i \sum_r \epsilon^{ikr}\sigma^r \right) \otimes \left( 2i \sum_s \epsilon^{jls} \sigma^s \right) + \left( 2i \sum_r \epsilon^{ikr} \sigma^r \right) \otimes \left( i \sum_s \epsilon^{ljs} \sigma^s \right)\\
  =&-2 \sum_{r,s} \left( \epsilon^{ikr} \epsilon^{jls} + \epsilon^{ikr} \epsilon^{ljs} \right)  \sigma^{rs}
 \end{split}
 \label{eq:diracbasis}
\end{equation}
The Hamiltonian is 
\begin{equation}
 \begin{split}
  H=&\frac{\overline{\omega}}{2} \left[\sin(\alpha)\sigma_{\bold{1}\bold{2}}^{1 0} + \cos(\alpha) \sigma_{\bold{1}\bold{2}}^{30}\right] 
  +\frac{\overline{\omega}}{2} \left[\sin(\alpha) \sigma_{\bold{2}\bold{1}}^{1 0} + \cos(\alpha) \sigma_{\bold{2}\bold{1}}^{30}\right]\\
  +& \frac{\mu}{4} (1-\cos \vartheta_{\bold{1},\bold{2}})
  \left[ \sigma_{\bold{1}\bold{2}}^{11}+ \sigma_{\bold{1}\bold{2}}^{22}+ \sigma_{\bold{1}\bold{2}}^{33}\right]
  + \frac{\mu}{4} (1-\cos \vartheta_{\bold{2},\bold{1}})
  \left[ \sigma_{\bold{2}\bold{1}}^{11}+ \sigma_{\bold{2}\bold{1}}^{22}+ \sigma_{\bold{2}\bold{1}}^{33}\right]
 \end{split}
\end{equation}
The time evolution operator is defined as follows,
\begin{equation}
\begin{split}
 e^{-i H t}
 =& c_1 I + c_2 \sigma_{\bold{1}\bold{2}}^{11}+ c_3 \sigma_{\bold{1}\bold{2}}^{22} + c_4 \sigma_{\bold{1}\bold{2}}^{33} + c_5 \sigma_{\bold{1}\bold{2}}^{10}+ C_6 \sigma_{\bold{1}\bold{2}}^{01}
  + c_7 \sigma_{\bold{1}\bold{2}}^{03}+ c_8 \sigma_{\bold{1}\bold{2}}^{30} + c_9 \sigma_{\bold{1}\bold{2}}^{13}+ c_{10} \sigma_{\bold{1}\bold{2}}^{31}
\end{split}
\end{equation}
Constraints from unitarity,
\begin{equation}
 \begin{split} 
  &\sum_i c_i^2=1\\
  &c_1 c_5+c_2 c_6+c_{10} c_8=0\\
  &c_1 c_7+c_3 C_8+c_6 c_9=0\\
  &c_2 c_5+c_1 c_6+c_7 c_9=0\\
  &c_1 c_2-c_3 c_4+c_5 c_6=0\\
  &c_{10} c_4+c_6 c_7+c_1 c_9=0\\
  &c_{10} c_5+c_3 c_7+c_1 c_8=0\\
  &c_1 c_{10}+c_5 c_8+c_4 c_9=0\\
  &c_1 c_3-c_2 c_4+c_7 c_8=0\\
  &c_1 c_4-c_2 c_3+c_{10} c_9=0
 \end{split}
\end{equation} 
From the Taylor series expansion of $e^{-iHt}$ and the commutator relations in~\ref{eq:diracbasis} one can find the coefficients,
\begin{equation}
 \begin{split}
  c_1=&\frac{1}{4} e^{- i t (\overline{\omega}-\mu/2 (1-\cos(\vartheta_{\bold{1}\bold{2}})))} \left(e^{ i t (\overline{\omega} +2 \mu (1-\cos(\vartheta_{\bold{1}\bold{2}})))}
  +e^{ i \overline{\omega} t}+e^{2 i \overline{\omega} t}+1\right)\\
  c_2=&\frac{1}{8} e^{- i t (\overline{\omega}+\mu/2 (1-\cos(\vartheta_{\bold{1}\bold{2}}))} \left(- e^{ i t (\overline{\omega} + 2 \mu(1-\cos(\vartheta_{\bold{1}\bold{2}}))}
  -\left(1-e^{ i \overline{\omega} t}\right)^2 \cos (2 \alpha )+e^{2 i \overline{\omega} t}+1\right)\\
  c_3=&\frac{1}{4} e^{- i \mu/2 (1+3 \cos(\vartheta_{\bold{1}\bold{2}})) t} \left(e^{2 i \mu \cos(\vartheta_{\bold{1}\bold{2}}) t}-e^{2 i \mu t}\right)\\
  c_4=&\frac{1}{16} e^{- i (t (\overline{\omega}+3/2\mu(1+\cos(\vartheta_{\bold{1}\bold{2}})))+\alpha )} \bigg(\left(1-e^{ i \overline{\omega} t}\right)^2 
  e^{ i (2 \mu \cos(\vartheta_{\bold{1}\bold{2}}) t+4\alpha )}+2 e^{2 i \alpha } \left(e^{ i t (2\overline{\omega}
  +2 \mu \cos(\vartheta_{\bold{1}\bold{2}}))}-2 e^{2 i t (\overline{\omega}+ \mu)}+e^{2 i \mu \cos(\vartheta_{\bold{1}\bold{2}}) t}\right)\\
  &+e^{2 i t (\overline{\omega}+ \mu \cos(\vartheta_{\bold{1}\bold{2}}))}-2 e^{ i t (\overline{\omega}+ 2\mu \cos(\vartheta_{\bold{1}\bold{2}}))}+e^{2 i \mu \cos(\vartheta_{\bold{1}\bold{2}}) t}\bigg)\\
  c_5=&c_6=\frac{1}{4} \left(1-e^{2 i \overline{\omega} t}\right) \sin (\alpha ) e^{- i t (\overline{\omega}+\mu/2 (1-\cos(\vartheta_{\bold{1}\bold{2}})))}\\
  c_7=&c_8=\frac{i}{2} \cos (\alpha ) \sin ( \overline{\omega} t) e^{ -i \mu/2  (1-\cos(\vartheta_{\bold{1}\bold{2}})) t}\\
  c_9=&c_{10}=\frac{1}{2}\sin (2\alpha ) \sin ^2(\overline{\omega}/2 t) e^{- i \mu/2 (1-\cos(\vartheta_{\bold{1}\bold{2}})) t}\\
 \end{split}
\end{equation}
Given this decomposition, one can compute any operator as function of time by working in the Heisenberg picture, $O(t)=e^{i H t}Oe^{-i H t}$. 
In the main article I have provided $J^x(t)$ and $J^z(t)$.  Here I also provide $J^y(t)$,
\begin{equation}
\begin{split}
J^y(t)=& \frac{1}{2} \cos (\overline{\omega} t) (\sigma^{02}_{\bold{1}\bold{2}}+\sigma^{20}_{\bold{1}\bold{2}})\\
-&\frac{1}{2} \cos (\alpha ) \sin ( \overline{\omega} t) (\sigma^{01}_{\bold{1}\bold{2}} +\sigma^{10}_{\bold{1}\bold{2}})\\
-&\frac{1}{2} \sin (\alpha ) \sin (\overline{\omega} t) (\sigma^{03}_{\bold{1}\bold{2}}+\sigma^{30}_{\bold{1}\bold{2}}). \\
\end{split}
\end{equation}
I also provide the flavor polarization for each lattice point,
\begin{equation}
\begin{split}
  {J_Z}_{\bold{1}}(t)=&-\frac{1}{2} \sin ^2(\overline{\omega}/2\ t) \sin (2 \alpha ) 
  \big(\cos^2(\mu (1-\cos\vartheta_{\bold{1}\bold{2}}) t)\sigma^{10}_{\bold{1}\bold{2}}+\sin^2(\mu (1-\cos\vartheta_{\bold{1}\bold{2}}) t)\sigma^{01}_{\bold{1}\bold{2}}\big)\\
  &+\frac{1}{2} \sin (\alpha ) \sin (\overline{\omega} t) \big(\sin ^2( \mu (1-\cos\vartheta_{\bold{1}\bold{2}}) t)\sigma^{02}_{\bold{1}\bold{2}}+\cos^2( \mu (1-\cos\vartheta_{\bold{1}\bold{2}}) t)\sigma^{20}_{\bold{1}\bold{2}}\big)\\
  &+\frac{1}{4} \left(\cos (2 \alpha ) \sin ^2(\overline{\omega}/2\ t)+\cos ^2(\overline{\omega}/2\ t)\right) \big((1-\cos (2  \mu (1-\cos\vartheta_{\bold{1}\bold{2}}) t)) \sigma^{03}_{\bold{1}\bold{2}} +(1+\cos (2  \mu (1-\cos\vartheta_{\bold{1}\bold{2}}) t)) \sigma^{30}_{\bold{1}\bold{2}}\big)\\
  &-\frac{1}{4} \sin (2  \mu (1-\cos\vartheta_{\bold{1}\bold{2}}) t) \left(\cos (2 \alpha ) \sin^2(\overline{\omega}/2\ t)+\cos^2(\overline{\omega}/2\ t)\right) \big( \sigma^{12}_{\bold{1}\bold{2}}-\sigma^{21}_{\bold{1}\bold{2}} \big)\\
  &+\frac{1}{4} \sin (\alpha ) \sin (\mu t) \sin (2  \mu (1-\cos\vartheta_{\bold{1}\bold{2}}) t) \big( \sigma^{13}_{\bold{1}\bold{2}}-\sigma^{31}_{\bold{1}\bold{2}} \big)\\
  &-\frac{1}{4} \sin (2 \alpha ) \sin ^2(\overline{\omega}/2\ t) \sin (2  \mu (1-\cos\vartheta_{\bold{1}\bold{2}}) t) \big( \sigma^{32}_{\bold{1}\bold{2}}-\sigma^{23}_{\bold{1}\bold{2}} \big),\\
  {J_Z}_{\bold{2}}(t)=& 
  -\frac{1}{2} \sin (2 \alpha ) \sin ^2(\overline{\omega}/2\ t) \big( \cos ^2(  \mu (1-\cos\vartheta_{\bold{1}\bold{2}}) t) \sigma^{01}_{\bold{1}\bold{2}} + \sin^2(  \mu (1-\cos\vartheta_{\bold{1}\bold{2}}) t) \sigma^{10}_{\bold{1}\bold{2}}\big)\\
  &+\frac{1}{2} \sin (\alpha ) \sin (\overline{\omega} t) \big(\cos ^2( \mu (1-\cos\vartheta_{\bold{1}\bold{2}}) t)\sigma^{02}_{\bold{1}\bold{2}} + \sin^2(  \mu (1-\cos\vartheta_{\bold{1}\bold{2}}) t)\sigma^{20}_{\bold{1}\bold{2}} \big)\\
  &+\frac{1}{4} \left(\cos (2 \alpha ) \sin ^2(\overline{\omega}/2\ t)+\cos ^2(\overline{\omega}/2\ t)\right) \big((1+\cos (2  \mu (1-\cos\vartheta_{\bold{1}\bold{2}}) t)) \sigma^{03}_{\bold{1}\bold{2}} + (1-\cos (2  \mu (1-\cos\vartheta_{\bold{1}\bold{2}}) t)) \sigma^{30}_{\bold{1}\bold{2}}\big)\\
  &+\frac{1}{4} \sin (2 \mu (1-\cos(\vartheta_{\bold{1}\bold{2}})) t) \left(\cos (2 \alpha ) \sin ^2(\overline{\omega}/2\ t)+\cos ^2(\overline{\omega}/2\ t)\right) \big( \sigma^{12}_{\bold{1}\bold{2}} - \sigma^{21}_{\bold{1}\bold{2}}\big)\\
  &-\frac{1}{4} \sin (\alpha ) \sin (\overline{\omega} t) \sin (2  \mu (1-\cos\vartheta_{\bold{1}\bold{2}}) t) \big( \sigma^{13}_{\bold{1}\bold{2}} - \sigma^{31}_{\bold{1}\bold{2}} \big)\\
  &+\frac{1}{4} \sin (2 \alpha ) \sin ^2(\overline{\omega}/2\ t) \sin (2  \mu (1-\cos\vartheta_{\bold{1}\bold{2}}) t) \big( \sigma^{32}_{\bold{1}\bold{2}} - \sigma^{23}_{\bold{1}\bold{2}} \big).
\end{split}
\end{equation}
To calculate the correlation between individual lattice points, the following operator is needed as well:
\begin{equation}
 \begin{split}
  {J_Z}_{\bold{1}}{J_Z}_{\bold{2}}(t)=&\sin ^2(\alpha ) \cos ^2(\alpha ) \sin ^4(\overline{\omega}/2\ t)\sigma^{11}_{\bold{1}\bold{2}}\\
  &-\cos (\overline{\omega}/2\ t) \sin ^3(\overline{\omega}/2\ t) \sin ^2(\alpha ) \cos (\alpha ) \big(\sigma^{12}_{\bold{1}\bold{2}}+\sigma^{21}_{\bold{1}\bold{2}}\big)\\
  &-\frac{1}{16} \left(2 \sin (4 \alpha ) \sin ^4(\overline{\omega}/2\ t) + \sin (2 \alpha ) \sin ^2(\overline{\omega} t)\right) \big(\sigma^{13}_{\bold{1}\bold{2}}+\sigma^{31}_{\bold{1}\bold{2}}\big)\\
  &+\frac{1}{4} \sin ^2(\alpha ) \sin ^2(\overline{\omega} t) \sigma^{22}_{\bold{1}\bold{2}}\\
  &+\frac{1}{32} \left(2 \sin (\alpha ) \sin (\overline{\omega} t)+3 \sin (\alpha ) \sin (2 \overline{\omega} t)+8 \sin (3 \alpha ) \sin ^3(\overline{\omega}/2\ t) \cos (\overline{\omega}/2\ t)\right) \big(\sigma^{23}_{\bold{1}\bold{2}}+\sigma^{32}_{\bold{1}\bold{2}}\big)\\
  &+\frac{1}{4} \left(\cos (2 \alpha ) \sin ^2(\overline{\omega}/2\ t)+\cos ^2(\overline{\omega}/2\ t)\right)^2 \sigma^{33}_{\bold{1}\bold{2}}.
 \end{split}
\end{equation}
Similar calculations can be performed for the 2 other directions. The two flavor eigenstates in the Pauli basis are,
\begin{equation}
 \begin{split}
  | \nu_e \rangle=&\begin{pmatrix} 1 \\ 0 \end{pmatrix},\\
  | \nu_x \rangle=&\begin{pmatrix} 0 \\ 1 \end{pmatrix}.\\
 \end{split}
\end{equation}
The wave-function for the two neutrino beams is the direct product $|\nu_i \nu_j \rangle=| \nu_i \rangle \otimes | \nu_j \rangle$. 
Given the information above one can calculate the time evolution for the expectation values of the flavor polarizations. Since $\langle {J_Z}_{\bold{1}} \rangle$ has already been provided in the main article, here
I provide the rest,
\begin{equation}
 \begin{split}
  \langle {\nu_e}_{\bold{1}}, {\nu_e}_{\bold{2}} | {J_Z}_{\bold{2}} | {\nu_e}_{\bold{1}}, {\nu_e}_{\bold{2}} \rangle (t) =& -\langle {\nu_x}_{\bold{1}}, {\nu_x}_{\bold{2}} | {J_Z}_{\bold{2}} | {\nu_x}_{\bold{1}}, {\nu_x}_{\bold{2}} \rangle (t)\\
  =& \frac{1}{2} \left(\cos (2 \alpha ) \sin ^2(\overline{\omega}/2\ t)+\cos ^2(\overline{\omega}/2\ t)\right), \\
  \langle {\nu_e}_{\bold{1}}, {\nu_x}_{\bold{2}} | {J_Z}_{\bold{2}} | {\nu_e}_{\bold{1}}, {\nu_x}_{\bold{2}} \rangle (t) =& -\langle {\nu_x}_{\bold{1}}, {\nu_e}_{\bold{2}} | {J_Z}_{\bold{2}} | {\nu_x}_{\bold{1}}, {\nu_e}_{\bold{2}} \rangle (t)\\
  =& -\frac{1}{2} \cos (2 \mu (1-\cos \vartheta_{\bold{1}\bold{2}}) t) \left(\cos (2 \alpha ) \sin ^2(\overline{\omega}/2\ t)+\cos ^2(\overline{\omega}/2\ t)\right)  \\
  \langle {\nu_e}_{\bold{1}}, {\nu_e}_{\bold{2}} | {J_Z}_{\bold{1}}{J_Z}_{\bold{2}} | {\nu_e}_{\bold{1}}, {\nu_e}_{\bold{2}} \rangle (t) =&\langle {\nu_x}_{\bold{1}}, {\nu_x}_{\bold{2}} | {J_Z}_{\bold{1}}{J_Z}_{\bold{2}} | {\nu_x}_{\bold{1}}, {\nu_x}_{\bold{2}} \rangle (t)\\
   =&- \langle {\nu_e}_{\bold{1}}, {\nu_x}_{\bold{2}} | {J_Z}_{\bold{1}}{J_Z}_{\bold{2}} | {\nu_e}_{\bold{1}}, {\nu_x}_{\bold{2}} \rangle (t)\\  
   =&- \langle {\nu_x}_{\bold{1}}, {\nu_e}_{\bold{2}} | {J_Z}_{\bold{1}}{J_Z}_{\bold{2}} | {\nu_x}_{\bold{1}}, {\nu_e}_{\bold{2}} \rangle (t)\\
   =& \frac{1}{4} \left(\cos (2 \alpha ) \sin ^2(\overline{\omega}/2 t)+\cos ^2(\overline{\omega}/2\ t)\right)^2  \\
   \end{split}
\end{equation}
A similar procedure is employed for the single-angle approximation, but for every pair interaction I use the same angle, $\eta=(1+\cos \vartheta_{\bold{1}\bold{2}})/2$. The qualitative difference between
the single and multi-angle results lies in the fact that in the multi-angle case, a particle can not interact with itself, while the single-angle approximation allows for this to happen:
\begin{equation}
 \begin{split}
  H_{\text{SA}}=&\frac{\overline{\omega}}{2} \left[\sin(\alpha)\sigma_{\bold{1}\bold{2}}^{1 0} + \cos(\alpha) \sigma_{\bold{1}\bold{2}}^{30}\right] 
  +\frac{\overline{\omega}}{2} \left[\sin(\alpha)\sigma_{\bold{2}\bold{1}}^{1 0} + \cos(\alpha) \sigma_{\bold{2}\bold{1}}^{30}\right]\\
  +& \frac{\mu}{4} (1-\eta)
  \left[ \sigma_{\bold{1}\bold{2}}^{11}+ \sigma_{\bold{1}\bold{2}}^{22}+ \sigma_{\bold{1}\bold{2}}^{33}\right]
  + \frac{\mu}{4} (1-\eta)
  \left[ \sigma_{\bold{2}\bold{1}}^{11}+ \sigma_{\bold{2}\bold{1}}^{22}+ \sigma_{\bold{2}\bold{1}}^{33}\right]\\
  +& \frac{\mu}{4} (1-\eta)
  \left[ \sigma_{\bold{1}\bold{1}}^{11}+ \sigma_{\bold{1}\bold{1}}^{22}+ \sigma_{\bold{1}\bold{1}}^{33}\right]
  + \frac{\mu}{4} (1-\eta)
  \left[ \sigma_{\bold{2}\bold{2}}^{11}+ \sigma_{\bold{2}\bold{2}}^{22}+ \sigma_{\bold{2}\bold{2}}^{33}\right]\\
  =&\frac{\overline{\omega}}{2} \left[\sin(\alpha)\sigma_{\bold{1}\bold{2}}^{1 0} + \cos(\alpha) \sigma_{\bold{1}\bold{2}}^{30}\right] 
  +\frac{\overline{\omega}}{2} \left[\sin(\alpha)\sigma_{\bold{2}\bold{1}}^{1 0} + \cos(\alpha) \sigma_{\bold{2}\bold{1}}^{30}\right]\\
  +& \frac{\mu}{2} (1-\eta)
  \left[ \sigma_{\bold{1}\bold{2}}^{11}+ \sigma_{\bold{1}\bold{2}}^{22}+ \sigma_{\bold{1}\bold{2}}^{33}\right]
  +\frac{3}{2} \mu (1-\eta) \sigma_{\bold{1}\bold{2}}^{00}\\
  =& H_{\text{MA}} +\frac{3}{2} \mu (1-\eta) \sigma_{\bold{1}\bold{2}}^{00}
 \end{split}
\end{equation}
For the two particle system this means an overall constant difference between the two Hamiltonians as there is only one angle between the two neutrino beams. However, for a greater number of particles, there are many 
angles present, and the difference between the two Hamiltonians is not a simple constant. For initial conditions with partial polarization, as shown in sections~\ref{sec:cubic} and \ref{sec:quadratic}, there are qualitative
differences in the respective flavor evolution.

\section{Two neutrino beams, mean field}
\label{ap:ptot}
The equation of motion for the total flavor polarization of the system is
\begin{equation}
 d_t \bold{P}_{\text{tot}} = \overline{\omega} \bold{B} \times \bold{P}_{\text{tot}}.\\
\end{equation}
The solution is,
\begin{equation}
 \begin{split}
  \bold{P}_{\text{tot}}(t)=& \hat{B} \times \bold{P}_{0} +  \sin \left(\overline{\omega }B t \right) \left(\hat{B}\cdot \bold{P}_{0}\right) \bold{B} - \cos \left(\overline{\omega }B t \right) \hat{B} \times (\hat{B} \times \bold{P}_{0})  
 \end{split}
\end{equation}
where, $\bold{P}_{\text{tot}}(t=0)=\bold{P}_0=({S_x}_0,{S_y}_0,{S_z}_0)=\frac{1}{2}(\langle \Psi_0 |\big( \sigma^x_{\bold{1}}+\sigma^x_{\bold{2}}\big) |\Psi_0 \rangle,
\langle \Psi_0 |\big( \sigma^y_{\bold{1}}+\sigma^y_{\bold{2}}\big)|\Psi_0 \rangle,
\langle \Psi_0 |\big( \sigma^z_{\bold{1}}+\sigma^z_{\bold{2}}\big)|\Psi_0 \rangle )$.

Since $\bold{B}$ is a unit vector, $\bold{B}=\sin(\alpha) \hat{x}-\cos \left(\alpha\right)$:
\begin{equation}
 \begin{split}
  \bold{P}_{\text{tot}}(t)=& \bold{B} \times \bold{P}_{0} +  \sin \left(\overline{\omega } t \right) \left(\bold{B}\cdot \bold{P}_{0}\right) \bold{B} - \cos \left(\overline{\omega } t \right) \bold{B} \times (\bold{B} \times \bold{P}_{0})  \leftrightarrow
 \end{split}
\end{equation}
\begin{equation}
 \begin{split}
  \bold{P}_{\text{tot}}(t)=& \begin{pmatrix} 
                 \left(\sin ^2(\alpha )+\cos ^2(\alpha ) \cos (\overline{\omega} t)\right) {S_x}_0 +\cos (\alpha ) \sin (\overline{\omega} t){S_y}_0 - \frac{1}{2}\sin (2 \alpha ) (1-\cos (\overline{\omega} t)) {S_z}_0 \\
                 -\cos (\alpha ) \sin (\overline{\omega} t) {S_x}_0 +\cos (\overline{\omega} t) {S_y}_0  -\sin (\alpha ) \sin (\overline{\omega} t) {S_z}_0\\
                -\frac{1}{2}\sin (2 \alpha ) (1-\cos (\overline{\omega} t)) {S_x}_0 +\sin (\alpha ) \sin (\overline{\omega} t) {S_y}_0 + \left(\cos ^2(\alpha )+\sin ^2(\alpha ) \cos (\overline{\omega} t)\right) {S_z}_0\\
               \end{pmatrix}
 \end{split}
\end{equation}

\section{Cubic lattice flavor oscillations}
\label{app:cubic}
\subsection{Initial conditions without polarization}
\begin{figure}[ht]
 \includegraphics[height=0.25\linewidth,width=0.5\linewidth]{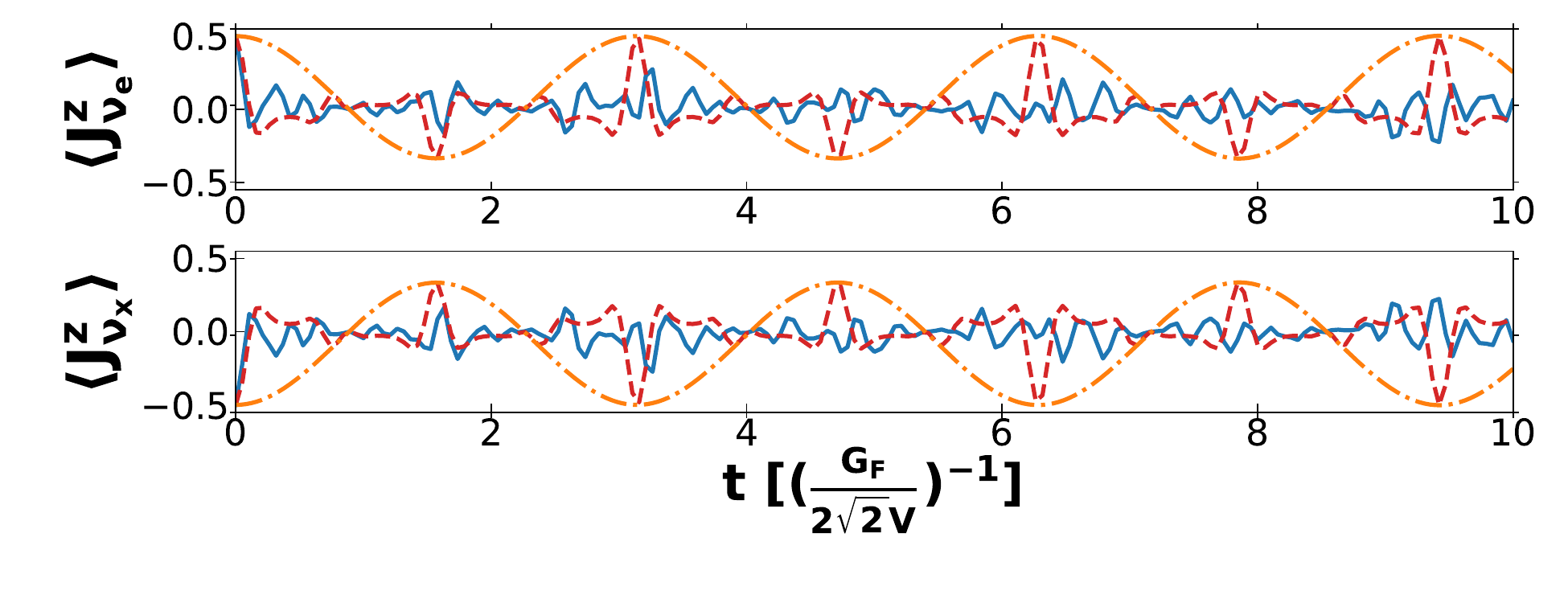}
 \caption{Multi-angle (solid blue line), 
 single-angle (dashed red line), and mean field (dashed dotted orange line) calculations of the flavor polarization as function of time for the cubic lattice in Fig.~\ref{fig:8p}
 and initial wave-function $|\Psi_0\rangle= | \nu_{e}^{(4)} \nu_{x}^{(4)}\rangle$ in vacuum with $ {\omega}/2 = \mu = \mu_0$.   These results were obtained based on eq.~(\ref{eq:psit}).}
 \label{fig:810}
\end{figure}

Based on the insight of the previous sections, at first I consider 
mixed polarization, with four lattice points of electron flavor and the other four of x flavor. 
There are two oscillation modes, each representing one of the two flavors.
In Fig.~\ref{fig:810} I plot the polarization as function of time for two lattice points, each starting with one of the two flavors.
As the figure shows, neutrino interactions dominate; these are fast collective oscillations, in agreement with previous sections.
While there are two oscillation modes that develop, representing the two subgroups in the lattice, the result is qualitatively the same as in the two neutrino beam mode.
Mean field approximation does not display any collective oscillation, but it does distinguish between the two subgroups.

\subsection{Electron flavor initial conditions}
\begin{figure}[ht]
 \subfloat[Multi-angle (solid blue line), 
 single-angle (dashed red line), and mean field (dashed dotted orange line) calculations of the flavor polarization.]
 {\includegraphics[height=0.2\linewidth,width=0.45\linewidth]{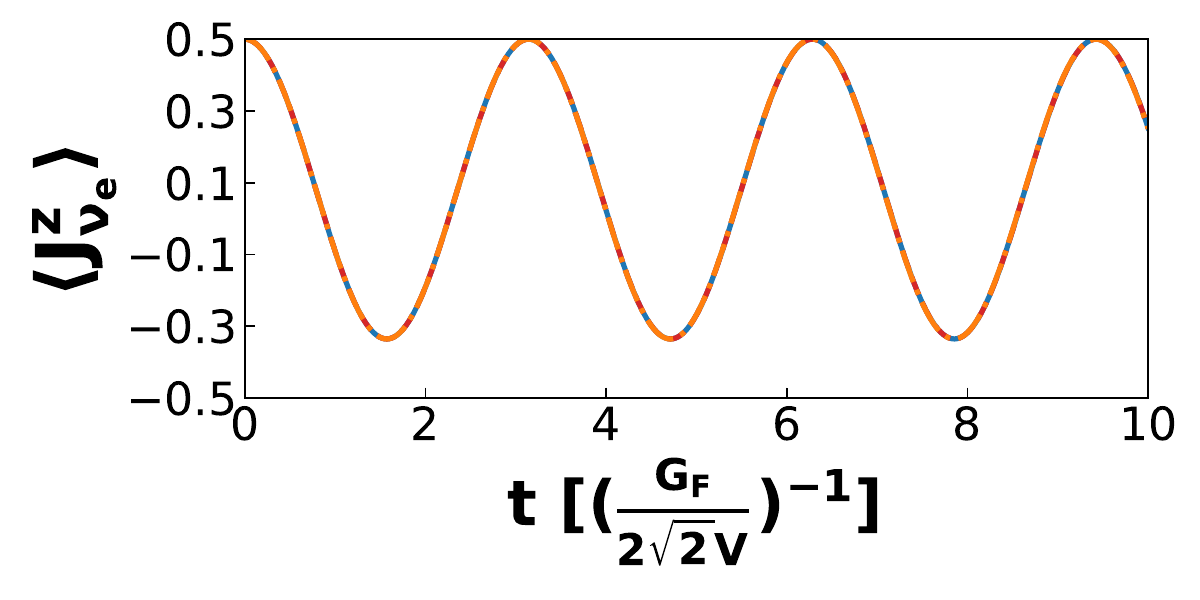}}
  \subfloat[Multi-angle (solid blue line), 
 single-angle (dashed red line) correlations]{\includegraphics[height=0.2\linewidth,width=0.45\linewidth]{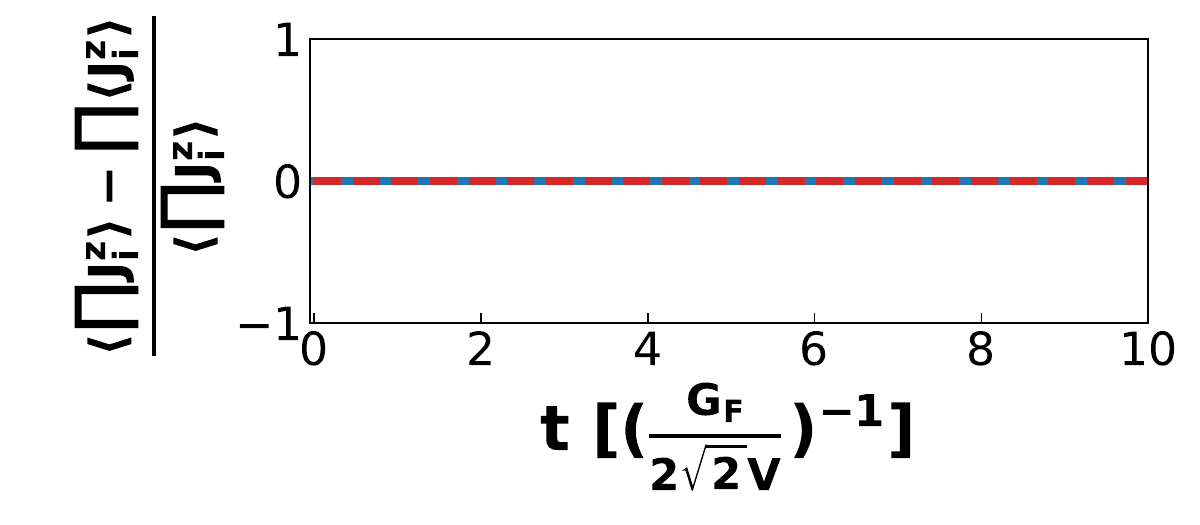}}\\
 \subfloat[Multi-angle (solid blue line), 
 single-angle (dashed red line) entanglement entropy]{\includegraphics[height=0.2\linewidth,width=0.45\linewidth]{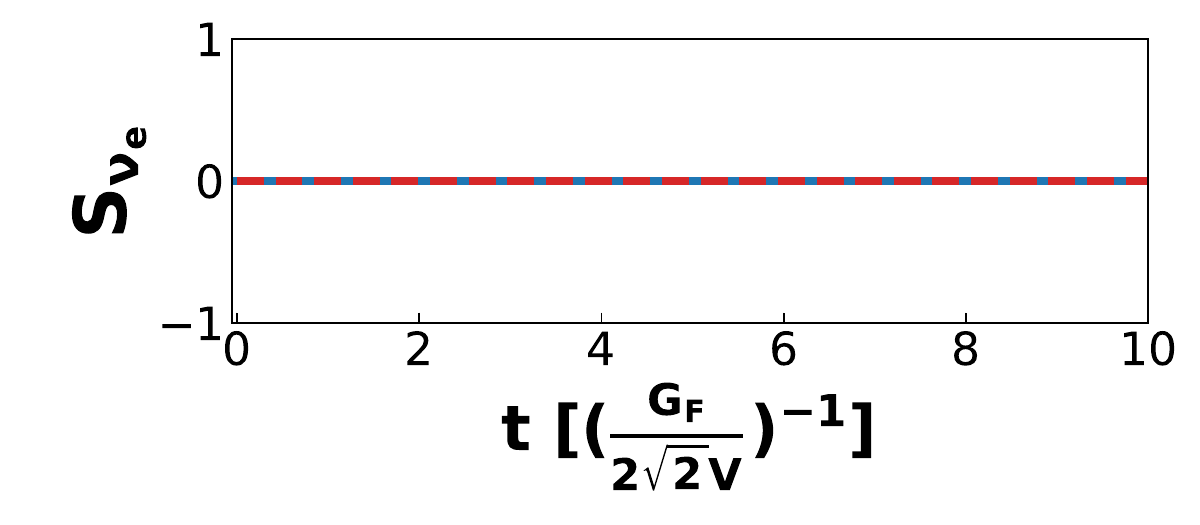}}
 \caption{Flavor polarization, correlations and entanglement entropy for a cubic lattice with an initial wave-function of only electron flavor and $\overline{\omega}_p/2=\mu/4=\mu_0$.   These results were obtained based on eq.~(\ref{eq:psit}).}
 \label{fig:811}
\end{figure}

In the two beam case I found complete agreement among the exact solution and various approximations if only one flavor is present initially. As Fig.~\ref{fig:811} shows, 
the same happens for a cubic lattice and only 
the usual vacuum oscillations are present. 
There are no correlations or entanglement entropy in agreement with the result for the two neutrino beams.

\section{Infinitely large systems}
\label{app:mntb}
The derivation here is based on the work in \cite{Friedland:2003eh} and the reader is referred to that article for an in depth set up of the system at hand.
The initial wave-function can be decomposed by means of Clebsch-Gordan coefficients in the basis of total angular momentum with zero z-projection,  
\begin{equation}
\begin{split}
| \Psi_0 \rangle =& | \nu_e^{(N)} \nu_x^{(N)} \rangle \\
=&| \frac{N}{2}, \frac{N}{2}\rangle \otimes| \frac{N}{2}, -\frac{N}{2}\rangle\\
 =& \sum_{J=0}^{N-1} c(N,J) | J, 0 \rangle_{\nu^{(2N)}}
 \end{split}
\end{equation}
where,
\begin{equation}
\begin{split}
 c(N,J)=&\langle \frac{N}{2}, \frac{N}{2}; \frac{N}{2},\frac{N}{2}| J,0\rangle\\
 =&\frac{N! \sqrt{2J+1}}{\sqrt{(N-J)!(N+J-1)!}}
 \end{split}
\end{equation}
Since,  
\begin{equation}
 H | J,0\rangle = \mu J (J+1) | J,0\rangle = E_J | J,0\rangle,
\end{equation}
one can calculate the time evolution as follows,
\begin{equation}
 \begin{split}
  |\Psi (t) \rangle =& e^{-i H t}| \Psi_0 \rangle\\
  =& \sum_{J=0}^{N} c(N,J) e^{-i E_J t} | J, 0 \rangle_{\nu^{(2N)}}.
  \label{eq:j02n}
 \end{split}
\end{equation}
Since the goal is to understand what happens to a particular neutrino, here I focus on the first one, which happens to be initially of electron flavor. 
\begin{equation}
 \begin{split}
  |\Psi_0 \rangle = |\nu_e\rangle \otimes \sum_{J=0}^{N-1} \eta(N,J) | J+\frac{1}{2}, - \frac{1}{2} ; (2N-1)\rangle
  \label{eq:psi0}
 \end{split}
\end{equation}
This is a bipartition of a system of $2N$ into a subsystem of 1 and a subsystem of $2N-1$ neutrinos.
However, exactly the same 
derivation can be performed for any other neutrino, for instance the last one, which is initially of x flavor. 
In addition, each of the wave-functions in eq.~\ref{eq:j02n} can be decomposed as direct products,
\begin{equation}
 \begin{split}
  | J, 0 \rangle_{\nu^{(2N)}} =&| \nu_e \rangle \otimes \left( a_J | J+\frac{1}{2}, -\frac{1}{2} \rangle_{\nu^{(2N-1)}} +  b_J | J-\frac{1}{2}, -\frac{1}{2} \rangle_{\nu^{(2N-1)}} \right)\\
  +& | \nu_x \rangle \otimes \left( c_J | J+\frac{1}{2}, \frac{1}{2} \rangle_{\nu^{(2N-1)}} +  d_J | J-\frac{1}{2}, \frac{1}{2} \rangle_{\nu^{(2N-1)}} \right)
 \end{split}
\end{equation}
As the minimal value of $J$ is zero, $b_0=d_0=0$, and the maximal value of $J$ is $N$, $a_N=c_N=0$. Due to parity under reflection along the z-axis ($\nu_e \leftrightarrow \nu_x$),
\begin{alignat}{2}
  &| J, 0 \rangle_{\nu^{(2N)}} &&\rightarrow (-1)^{N-J} | J, 0 \rangle_{\nu^{(2N)}}\\
  &| J \pm \frac{1}{2}, \pm \frac{1}{2} \rangle_{\nu^{(2N-1)}} &&\rightarrow | J \pm \frac{1}{2}, \mp \frac{1}{2} \rangle_{\nu^{(2N-1)}}
\end{alignat}
Thus, $c_J=(-1)^{N-J} a_J,\ d_J=(-1)^{N-J} b_J$. 
Using these identities, eq.~\ref{eq:j02n} can be written as,
\begin{equation}
 \begin{split}
  |\Psi \rangle (t) =& \sum_{J=0}^{N} c(N,J) e^{-i E_J t} \bigg[ | \nu_e \rangle \otimes \left( a_J | J+\frac{1}{2}, -\frac{1}{2} \rangle_{\nu^{(2N-1)}} +  b_J | J-\frac{1}{2}, -\frac{1}{2} \rangle_{\nu^{(2N-1)}} \right)\\
  +& | \nu_x \rangle \otimes \left( c_J | J+\frac{1}{2}, \frac{1}{2} \rangle_{\nu^{(2N-1)}} +  d_J | J-\frac{1}{2}, \frac{1}{2} \rangle_{\nu^{(2N-1)}} \right) \bigg]\\
  =& | \nu_e \rangle \otimes \sum_{J=0}^{N-1} \left(  c(N,J) e^{-i E_J t}  a_J | J+\frac{1}{2}, -\frac{1}{2} \rangle_{\nu^{(2N-1)}} +  c(N,J+1) e^{-i E_{J+1} t} b_{J+1} | J+\frac{1}{2}, -\frac{1}{2} \rangle_{\nu^{(2N-1)}}\right)\\
 +&| \nu_x \rangle \otimes \sum_{J=0}^{N-1} (-1)^{N-J}\left(  c(N,J) e^{-i E_J t}  a_J | J+\frac{1}{2}, \frac{1}{2} \rangle_{\nu^{(2N-1)}} -  c(N,J+1) e^{-i E_{J+1} t} b_{J+1} | J+\frac{1}{2}, \frac{1}{2} \rangle_{\nu^{(2N-1)}}\right)\\
 \end{split}
\end{equation}
Since initially the first neutrino is of electron flavor,
\begin{equation}
 \begin{split}
  c(N,J)  a_J | J+\frac{1}{2}, \frac{1}{2} \rangle_{\nu^{(2N-1)}} =  c(N,J+1)  b_{J+1} | J+\frac{1}{2}, \frac{1}{2} \rangle_{\nu^{(2N-1)}}
 \end{split}
\end{equation}
By reflection along the z-axis,
\begin{equation}
 \begin{split}
   c(N,J)  a_J | J+\frac{1}{2}, -\frac{1}{2} \rangle_{\nu^{(2N-1)}} =  c(N,J+1)  b_{J+1} | J+\frac{1}{2}, -\frac{1}{2} \rangle_{\nu^{(2N-1)}}
  \end{split}
\end{equation}
Thus,
\begin{equation}
 \begin{split}
  |\Psi \rangle (t) =& | \nu_e \rangle \otimes \sum_{J=0}^{N-1} \left(e^{-i E_J t} + e^{-i E_{J+1} t} \right) c(N,J)  a_J  | J+\frac{1}{2}, -\frac{1}{2} \rangle_{\nu^{(2N-1)}}\\
  =& | \nu_x \rangle \otimes \sum_{J=0}^{N-1} (-1)^{N-J}\left( e^{-i E_J t} -  e^{-i E_{J+1} t} \right)  c(N,J) a_J | J+\frac{1}{2}, \frac{1}{2} \rangle_{\nu^{(2N-1)}}
  \label{eq:psit2}
 \end{split}
\end{equation}
By setting $t=0$ in eq.~\ref{eq:psit2} one should get the result in eq.~\ref{eq:psi0}, which means
\begin{equation}
 c(N,J)  a_J | J+\frac{1}{2}, -\frac{1}{2} \rangle_{\nu^{(2N-1)}} = \eta(N,J) | J+\frac{1}{2}, - \frac{1}{2} ; (2N-1)\rangle,
\end{equation}
and by reflecting along the z-axis,
\begin{equation}
 c(N,J)  a_J | J+\frac{1}{2}, \frac{1}{2} \rangle_{\nu^{(2N-1)}} = \eta(N,J) | J+\frac{1}{2}, \frac{1}{2} ; (2N-1)\rangle,
\end{equation}
Putting altogether,
\begin{equation}
\begin{split}
  |\Psi (t) \rangle=& |\nu_e \rangle \otimes \sum_{J=0}^{N-1} \frac{e^{-i E_Jt}+e^{-i E_{J+1}t}}{2} \eta(N,J) |J+\frac{1}{2}, - \frac{1}{2}; (2N-1) \rangle \\
   +& |\nu_x \rangle \otimes \sum_{J=0}^{N-1} (-1)^{N-J}\frac{e^{-i E_Jt}-e^{-i E_{J+1}t}}{2} \eta(N,J) |J+\frac{1}{2},  \frac{1}{2}; (2N-1)\rangle,
\end{split}
\end{equation}
which concludes the derivation.

\twocolumngrid

\twocolumngrid

\end{document}